\documentclass[journal, twocolumn, 10pt]{IEEEtran}
\usepackage{url}
\usepackage{graphicx}
\usepackage{epstopdf}
\usepackage{algorithm}
\usepackage{algorithmicx}
\usepackage{algpseudocode}
\usepackage{cite}
\usepackage{color}
\usepackage{amsmath,amssymb,amsfonts,amsthm}
\usepackage{bm}
\usepackage{setspace}
\usepackage{multirow}
\usepackage{mathtools}
\usepackage{wrapfig}
\usepackage{subcaption}
\usepackage{multicol}
\usepackage{lipsum}
\usepackage{fancyhdr}
\usepackage{comment}

\newtheorem{thm} {Theorem}

\newtheorem{cor} {Corollary}

\newcommand{\bbR}{\mathbb{R}}

\newcommand{\asconv}{\overset{\text{a.s.}}{\longrightarrow}}
\newcommand{\ra}{\rightarrow}
\newcommand{\LB}{\left\{}
\newcommand{\RB}{\right\}}
\newcommand{\Lb}{\left[}
\newcommand{\Rb}{\right]}
\newcommand{\lb}{\left(}
\newcommand{\rb}{\right)}

\newcommand{\none}{{n_1}}
\newcommand{\ntwo}{{n_2}}
\newcommand{\nmin}{{n_{\min}}}
\newcommand{\nmax}{{n_{\max}}}
\newcommand{\rhomin}{{\rho_{\min}}}
\newcommand{\rhomax}{{\rho_{\max}}}

\newcommand{\bDelta}{\mathbf{\Delta}}
\newcommand{\bLambda}{\mathbf{\Lambda}}
\newcommand{\bA}{\mathbf{A}}

\newcommand{\bZ}{\mathbf{Z}}
\newcommand{\bAt}{\widetilde{\mathbf{A}}}
\newcommand{\bV}{\mathbf{V}}
\newcommand{\bW}{\mathbf{W}}

\newcommand{\bD}{\mathbf{D}}

\newcommand{\bL}{\mathbf{L}}
\newcommand{\bC}{\mathbf{C}}
\newcommand{\bS}{\mathbf{S}}
\newcommand{\cS}{\mathcal{S}}

\newcommand{\bone}{\mathbf{1}}
\newcommand{\bzero}{\mathbf{0}}
\newcommand{\onenone}{\mathbf{1}_{\none}}
\newcommand{\onentwo}{\mathbf{1}_{\ntwo}}

\newcommand{\yone}{\mathbf{y}_1}
\newcommand{\ytwo}{\mathbf{y}_2}
\newcommand{\bd}{\mathbf{d}}

\newcommand{\bx}{\mathbf{x}}
\newcommand{\by}{\mathbf{y}}

\newcommand{\bz}{\mathbf{z}}
\newcommand{\pUB}{p_{\text{UB}}}
\newcommand{\pLB}{p_{\text{LB}}}
\newcommand{\tUB}{t_{\text{UB}}}
\newcommand{\tLB}{t_{\text{LB}}}

\newcommand{\hpLB}{\widehat{p}_{\text{LB}}}

\newcommand{\htLB}{\widehat{t}_{\text{LB}}}
\newcommand{\hatt}{\widehat{t}}
\newcommand{\hatm}{\widehat{m}}
\newcommand{\hn}{\widehat{n}}

\newcommand{\hp}{\widehat{p}}

\newcommand{\hpij}{\widehat{p}_{ij}}
\newcommand{\hG}{\widehat{G}}

\newcommand{\hL}{\widehat{\mathbf{L}}}

\newcommand{\Wbar}{\overline{W}}
\newcommand{\bWbar}{\overline{\mathbf{W}}}

\newcommand{\hWbar}{\widehat{\Wbar}}

\newcommand{\SK}{S_{2:K}}
\newcommand{\bX}{\mathbf{X}}
\newcommand{\bY}{\mathbf{Y}}
\newcommand{\btY}{\widetilde{\mathbf{Y}}}
\newcommand{\btL}{\widetilde{\mathbf{L}}}
\newcommand{\bM}{\mathbf{M}}
\newcommand{\bI}{\mathbf{I}}
\newcommand{\bO}{\mathbf{O}}
\newcommand{\bU}{\mathbf{U}}
\newcommand{\bv}{\mathbf{v}}
\newcommand{\bu}{\mathbf{u}}
\newcommand{\bvk}{\mathbf{v_k}}
\newcommand{\bnu}{\boldsymbol{\nu}}
\newcommand{\trace}{\textnormal{trace}}
\newcommand{\diag}{\textnormal{diag}}

\newcommand{\convd}{\overset{d}{\longrightarrow}}
\newcommand{\vectorize}{\textbf{vec}}
\newcommand{\cC}{\mathcal{C}}

\begin{document}

\title{Phase Transitions and a Model Order Selection Criterion for Spectral Graph Clustering}

\author{Pin-Yu~Chen and Alfred O. Hero III,~\emph{Fellow},~\emph{IEEE}
	\thanks{P.-Y. Chen is with IBM Thomas J. Watson Research Center, Yorktown Heights, NY 10598, USA. Email: pin-yu.chen@ibm.com.
		A. O. Hero is with the Department of Electrical Engineering and Computer Science, University of Michigan, Ann Arbor, MI 48109, USA. Email : hero@umich.edu.}
	\thanks{This work has been partially supported by the Army Research Office
		(ARO), grants W911NF-15-1-0479 and W911NF-15-1-0241, and by the Consortium for Verification
		Technology under Department of Energy National Nuclear Security Administration,
		award DE-NA0002534. This work was conducted while P.-Y. Chen was at the University of Michigan, Ann Arbor, USA.
		Part of this work is presented at IEEE ICASSP 2017.}
}

\maketitle
\thispagestyle{empty}
\begin{abstract}
One of the longstanding open problems in spectral graph clustering (SGC) is the so-called model order selection problem: automated selection of the correct number of clusters. This is equivalent to the problem of finding the number of connected components or communities in an undirected graph. We propose automated model order selection (AMOS), a solution to the SGC model selection problem under a random interconnection model (RIM) using a novel selection criterion that is based on an asymptotic phase transition analysis. AMOS can more generally be applied to discovering hidden block diagonal structure in symmetric non-negative matrices. Numerical experiments on simulated graphs validate the phase transition analysis, and real-world network data is used to validate the performance of the proposed model selection procedure.
\end{abstract}

\begin{IEEEkeywords}
	community detection, model selection, network analysis, phase transition, spectral clustering 
\end{IEEEkeywords}

\section{Introduction}
Undirected graphs are widely used for network data analysis, where nodes can represent entities or data samples, and the existence and strength of edges can represent relations or affinity between nodes.
For attributional data (e.g., multivariate data samples), such a graph can be constructed by calculating and thresholding  the similarity measure between nodes. For relational data (e.g., friendships), the edges reveal the interactions between nodes. The goal of graph clustering is to group the nodes into clusters of high similarity. Applications of graph clustering, also known as community detection \cite{White05,CPY14spectral}, include but are not limited to graph signal processing  \cite{Sandryhaila13,Bertrand13Mag,Shuman13,Miller13detection,Dong2012clustering,Oselio14,Xu14,SihengChen15signalrecovery,Sandryhaila14,Wang15localset}, 
multivariate data clustering \cite{ng2002spectral,zelnik2004self,Luxburg07}, image segmentation \cite{Shi00,yu2002concurrent},  structural identifiability in physical systems \cite{Radicchi13}, and network vulnerability assessment \cite{CPY14ComMag}. 

Spectral clustering \cite{ng2002spectral,zelnik2004self,Luxburg07} is a popular method for graph clustering, which we refer to as spectral graph clustering (SGC).
 It works by transforming the graph adjacency matrix into a graph Laplacian matrix \cite{Merris94}, computing its eigendecomposition, and performing K-means clustering \cite{hartigan1979algorithm} on the eigenvectors to partition the nodes into clusters. 
  Although heuristic methods have been proposed to automatically select the number of clusters \cite{Polito01grouping,ng2002spectral,zelnik2004self}, rigorous theoretical justifications on the selection of the number of eigenvectors for clustering are still lacking and little is known about the capabilities and limitations of spectral clustering on graphs.

The contributions of this paper are twofold. First, we analyze the performance of spectral clustering on undirected unweighted graphs generated by a random interconnection model (RIM), where each cluster can have arbitrary internal connectivity structure and the  inter-cluster edges are assumed to be random. Under the RIM, we establish
a breakdown condition on the ability to identify correct clusters using SGC. Furthermore, when all of the cluster interconnection probabilities are identical, a model we call the homogeneous RIM, this breakdown condition specifies a critical phase transition threshold $p^* \in [0,1]$ on the inter-cluster connection probability $p$. 
When this interconnection probability is below the critical phase transition threshold,
SGC can perfectly detect the clusters.  On the other hand,  when the interconnection probability is above the critical phase transition threshold, SGC fails to identify the clusters.
This breakdown condition and phase transition analysis apply to weighted graphs as well, where the critical phase transition threshold depends not only on the interconnection probability but also on the weights of the interconnection edges.

Second, we show that the phase transition results for the homogeneous RIM can be used to bound the phase transitions of SGC for the inhomogeneous RIM. This leads to a method for automatically selecting the number of clusters in SGC, which we call automated model order selection (AMOS). 
	AMOS works by sequentially increasing the model order while running multi-stage tests for testing for RIM structure. Specifically, for a given model order and an estimated cluster membership map obtained from  SGC, we first test for local RIM structure for a single cluster pair using a binomial test of homogeneity. This is repeated for all cluster pairs and, if they pass the RIM test, we proceed to the second stage of testing, otherwise we increase the model order and start again. The second stage consists of testing whether the RIM is globally homogeneous or inhomogeneous. This is where the phase transition results are used - if any of the estimated inter-cluster connection probabilities exceed the critical phase transition threshold the model order is increased.     
	In this manner, the outputs from AMOS are the clustering results from SGC of minimal model order that are deemed reliable.
	
Simulation results on both unweighted and weighted graphs generated by different network models validate our phase transition analysis. 
Comparing to other graph clustering methods,
experiments on real-world network datasets show that the AMOS algorithm indeed outputs clusters that are more consistent with the ground-truth meta information. 
For example, when applied to network data with longitude and latitude meta information, such as the Internet backbone map across North American and Europe, and the Minnesota road map, 
	the clusters identified by the AMOS algorithm are more consistent with known geographic separations.

	The rest of this paper is organized as follows. Sec. \ref{sec_related} discusses previous work on phase transition and model order selection for graph clustering. Sec. \ref{sec_RIM} introduces the RIM and the mathematical formulation of SGC. Sec. \ref{sec_phase_spec} describes the breakdown condition and phase transition analysis of SGC, including unweighted and weighted graphs. Sec. \ref{sec_ASGC} summarizes the proposed AMOS algorithm for SGC. Sec. \ref{sec_num} discusses numerical experiments and comparisons on simulated graphs and real-world datasets. Sec. \ref{sec_conclusion} concludes this paper. 

\section{Related Work}
\label{sec_related}
\subsection{Phase transitions in graph clustering} 
In recent years, researchers have established phase transitions
in the accuracy of graph clustering  under a diverse set of network models \cite{Decelle11PRE,Alamgirv11,Nadakuditi12Detecability,abbe2014exact,CPY14spectral,CPY14modularity,Hajek15}. A widely used network model is the stochastic block model (SBM) \cite{Holland83}, where the edge connections within and between clusters are independent Bernoulli random variables. 
Under the SBM, a phase transition on the cluster interconnectivity probability separates clustering accuracy into two regimes: a regime where correct graph clustering is possible, and a regime where correct graph clustering is impossible. The critical values that separate these two regimes are called phase transition thresholds. A summary of  phase transition analysis under the SBM can be found in \cite{abbe2014exact}. 
In this paper, we establish the phase transition analysis of SGC under a more general network model, which we call the random interconnection model (RIM). The RIM does not impose any distributional assumptions on the within-cluster connectivity structure, but assumes the between-cluster edges are generated by a SBM. The formal definition of the RIM is introduced in Sec. \ref{subsec_RIM}.
	The RIM introduced in this paper is a direct generalization of the model introduced in \cite{CPY14spectral}, which is a special case of an unweighted graph with two clusters.

\subsection{Model order selection criterion}
 Most existing model selection algorithms specify an upper bound $K_{\max}$ on the number $K$ of clusters and then select $K$ based on optimizing some objective function, e.g., the goodness of fit of the $k$-cluster model for $k=2,\ldots, K_{\max}$. 
In \cite{ng2002spectral}, the objective is to minimize the sum of cluster-wise Euclidean distances between each data point and the centroid obtained from K-means clustering. In \cite{Polito01grouping}, the objective is to maximize the gap between the $K$-th largest and the $(K+1)$-th largest eigenvalue. In \cite{zelnik2004self}, the authors propose to minimize an objective function that is associated with the cost of aligning the eigenvectors with a canonical coordinate system. In \cite{reichardt2006statistical,arenas2008analysis,schaub2012markov,tremblay2014graph}, model selection is cast as a multiscale  community detection problem.
In \cite{Newman06PNAS}, the authors propose to iteratively divide a cluster based on the leading eigenvector of the modularity matrix until no significant improvement in the modularity measure can be achieved. The Louvain method in \cite{blondel2008fast} uses a greedy algorithm for modularity maximization.
In \cite{Daudin2008}, the authors use the integrated classification likelihood (ICL) criterion \cite{biernacki2000assessing} for graph clustering based a random graph mixture model. 
In \cite{Newman16Estimate}, the authors use the degree-corrected SBM \cite{Karrer11} and Monte Carlo sampling techniques for graph clustering.
In \cite{Krzakala2013,Saade2015spectral}, the authors propose to use the eigenvectors of the nonbacktracking matrix for graph clustering, where the number of clusters is determined by the number of real eigenvalues with magnitude larger than the square root of the largest eigenvalue. Different from these approaches, this paper not only establishes a new model order selection criterion based on the phase transition analysis, but also provides multi-stage statistical tests for determining clustering reliability of SGC.

\section{Random Interconnection Model (RIM) and Spectral Clustering}
\label{sec_RIM}
\subsection{Random interconnection model (RIM)}
\label{subsec_RIM}
Consider an undirected graph where its connectivity structure is represented by an $n \times n$ binary symmetric adjacency matrix $\bA$, where $n$ is the number of nodes in the graph. $[\bA]_{uv}=1$ if there exists an edge between the node pair ($u,v$), and otherwise $[\bA]_{uv}=0$. An unweighted undirected graph is completely specified by its adjacency matrix $\bA$, while a weighted undirected graph is specified by a nonnegative matrix $\bW$, where its nonzero entries denote the weight of an edge. In the next section, 
Theorems \ref{thm_impossible}, \ref{thm_spec} and \ref{thm_principal_angle} apply to unweighted undirected graphs while Theorem \ref{thm_spec_weight} extends these theorems to weighted undirected graphs.

Assume there are $K$ clusters in the graph and denote the size of cluster $k$ by $n_k$. The size of the largest and smallest cluster is denoted by $\nmax$ and $\nmin$, respectively.
Let $\bA_k$ denote the $n_k \times n_k$ adjacency matrix representing the internal edge connections in cluster $k$ and let $\bC_{ij}$ ($i,j \in \{ 1,2,\ldots,K\}$) be an $n_i \times n_j$ matrix representing the adjacency matrix of inter-cluster edge connections between the cluster pair ($i,j$).
The matrix $\bA_k$ is symmetric and $\bC_{ij}=\bC_{ji}^T$ for all $i\neq j$.
 Using these notations, the adjacency matrix of the entire graph can be represented by a block structure, which is
\begin{align}                                                        \label{eqn_network_model_multi}
\bA= \begin{bmatrix}
\bA_1          & \bC_{12} & \bC_{13} & \cdots & \bC_{1K}           \\
\bC_{21}       & \bA_2    & \bC_{23} & \cdots & \bC_{2K} \\
\vdots         & \vdots   & \ddots   & \vdots  & \vdots  \\
\vdots         & \vdots   & \vdots   & \ddots  & \vdots  \\
\bC_{K1}       & \bC_{K2} & \cdots   & \cdots  & \bA_{K}
\end{bmatrix}.
\end{align}
The proposed random interconnection model (RIM) 
assumes that: (1) the adjacency matrix $\bA_k$ is associated with a connected graph of $n_k$ nodes but is otherwise arbitrary; (2) the $K(K-1)/2$ matrices $\{\bC_{ij}\}_{i>j}$ are random and mutually independent, and each $\mathbf C_{ij}$ has i.i.d. Bernoulli distributed entries with Bernoulli parameter $p_{ij} \in [0,1]$. We call this model a \textit{homogeneous} RIM when all random interconnections have equal probability, i.e., $p_{ij}=p$ for all $i \neq j$.  Otherwise, the model is called an inhomogeneous RIM. In the next section, Theorems \ref{thm_impossible} and \ref{thm_principal_angle} apply to general RIM while Theorems \ref{thm_spec} and \ref{thm_spec_weight} are restricted to the homogeneous RIM.

The stochastic block model (SBM) \cite{Holland83} is a special case of the RIM in the sense that the RIM does not impose any distributional constraints on $\bA_k$. In contrast, under the SBM $\bA_k$ is a Erdos-Renyi random graph with some edge connection probability $p_k \in [0,1]$.

\subsection{Spectral clustering}
Let $\bone_n (\bzero_n)$ be the $n$-element column vector of ones (zeros) and let $\bD=\diag(d_1,d_2,\ldots,d_n)$ be the diagonal degree matrix, where $\bd=\bA \bone_n=[d_1,d_2,\ldots,d_n]^T$ is the degree vector of the graph. The graph Laplacian matrix of the entire graph is defined as $\bL=\bD-\bA$, and similarly
the graph Laplacian matrix of $\bA_k$ is denoted by $\bL_k$.
Let $\lambda_i(\bL)$ denote the $i$-th smallest eigenvalue of $\bL$. Then $\lambda_1(\bL)=0$ since $\bL \bone_n=\bzero_n$ by definition, and $\lambda_2(\bL) >0$ if the entire graph is connected. $\lambda_2(\bL)$ is also known as the algebraic connectivity of the graph as it is a lower bound on the node and edge connectivity of a connected graph  \cite{Fiedler73}.

To partition the nodes in the graph into $K$ ($K \geq 2$) clusters, spectral clustering  uses the $K$ eigenvectors associated with the $K$ smallest eigenvalues of $\bL$ \cite{Luxburg07}. Each node can be viewed as a $K$-dimensional vector in the subspace spanned by these eigenvectors. K-means clustering \cite{hartigan1979algorithm} is then implemented on the $K$-dimensional vectors to group the nodes into $K$ clusters. Vector normalization of the obtained $K$-dimensional vectors or degree normalization of the adjacency matrix can be used to stabilize K-means clustering \cite{ng2002spectral,zelnik2004self,Luxburg07}. 

For analysis purposes, throughout this paper we will focus on the case where the observed graph is connected. If the graph is not connected, the connected components can be easily found and
 the proposed algorithm can be applied to each connected component separately.
Since the smallest eigenvalue of $\bL$ is always $0$ and the associated eigenvector is $\frac{\bone_n}{\sqrt{n}}$, only the higher order eigenvectors will affect the clustering results. By the Courant-Fischer theorem \cite{jennings1992matrix}, the $K-1$ eigenvectors associated with the $K-1$ smallest nonzero eigenvalues of $\bL$, represented by the columns of the eigenvector matrix $\bY \in \mathbb{R}^{n \times (K-1)}$, are the solution of the minimization problem 
\begin{align}
\label{eqn_spectral_clustering_multi_formulation}
&\SK(\bL)=\min_{\bX \in \mathbb{R}^{n \times (K-1)}} \trace(\bX^T \bL \bX), \nonumber \\
	&\text{subject~to}~\bX^T \bX= \bI_{K-1},~\bX^T \bone_n=\bzero_{K-1}, 
\end{align}
where the optimal value $\SK(\bL)=\trace(\bY^T \bL \bY)=\sum_{k=2}^K \lambda_{k}(\bL)$ of (\ref{eqn_spectral_clustering_multi_formulation}) is the sum of the second to the $K$-th smallest eigenvalues of $\bL$, and $\bI_{K-1}$ is the $(K-1) \times (K-1)$ identity matrix. The constraints in  (\ref{eqn_spectral_clustering_multi_formulation}) impose  orthonormality and centrality on the eigenvectors.

\section{Breakdown Condition and Phase Transition Analysis}
\label{sec_phase_spec}
In this section we  establish a 
mathematical condition (Theorem \ref{thm_impossible}) under which SGC fails to accurately identify clusters under the RIM. Furthermore, under the homogeneous RIM assumption of identical interconnection probability $p_{ij}=p$ governing the entries of the matrices $\{\bC_{ij}\}$ in (1), the condition leads to (Theorem \ref{thm_spec}) a  critical phase transition threshold $p^*$ where, if $p<p^*$ SGC correctly identifies the communities with probability one while if $p > p^*$ SGC fails.
The phase transition analysis developed in this section will be used to establish an automated model order selection algorithm for SGC in Sec. \ref{sec_ASGC}.
The proofs of the main theorems (Theorems \ref{thm_impossible}, \ref{thm_spec} and \ref{thm_principal_angle}) are given in the appendix, and the proofs of extended theorems and corollaries are given in the supplementary material.

\begin{table}[]
	\centering
	\caption{Notation of limit expressions.}
	\begin{tabular}{c|c}
		\hline
		expression & limit value of \\ \hline
		$\rho_k$          &     $\frac{n_k}{n}$           \\
		$\rho_{\max}$          &     $\frac{\nmax}{n}$           \\		 
		$\rho_{\min}$          &     $\frac{\nmin}{n}$           \\				
		$c$          &           $\frac{\nmin}{\nmax}$     \\ 
		$c^*$       &          $\frac{1}{{n}} \cdot \min_{k \in \{1,2,\ldots,K\}} \SK(\bL_k) $    \\
		$c^*_2$       &          $\frac{1}{{n}} \cdot \min_{k \in \{1,2,\ldots,K\}} \lambda_2(\bL_k) $    \\		
		$c^*_K$       &          $\frac{1}{{n}} \cdot \min_{k \in \{1,2,\ldots,K\}} \lambda_K(\bL_k) $    \\								
		$\bone$ ($\bzero$)         &          $\bone_n$ ($\bzero_n$)    \\		
		$b_p$          &          		$\frac{\| \bL - \btL \|_F}{n}$	   \\														
		\hline		
	\end{tabular}
	\label{table_limit_value}	
		\vspace{-4mm}
\end{table}

In the sequel, there are a number of limit theorems stated about the behavior of random matrices and vectors whose dimensions go to infinity as the sizes $n_k$ of the clusters goes to infinity while their relative sizes $n_k/n_\ell$ are held constant. Throughout this paper, the convergence of a real matrix $\bX \in \bbR^{a \times b}$ is defined with respect to the spectral norm \cite{Tropp_matrix_concentrate}, defined as $\|\bX\|_2 = \max_{\bz \in \mathbb{R}^{b}, \bz^T \bz=1} \|\bX \bz\|_2$, where $\|\bx\|_2$ denotes the Euclidean norm of the vector $\bx$.	Let $\bX=\sum_{i=1}^{r(\bX)} \sigma_i(\bX) \bu_i(\bX) \bv^T_i(\bX)$ denote the singular value decomposition of  $\bX$, where $\sigma_i(\bX)$ denotes the $i$-th largest singular value of $\bX$, $\bu_i(\bX)$ and $\bv_i(\bX)$ are the associated left and right singular vectors, and $r(\bX)$ denotes the rank of $\bX$. For any two matrices $\bX$ and $\widetilde{\bX}$ of the same dimension,  we write $\bX \ra \widetilde{\bX}$ if as $n_k \ra \infty$ for all $k$, the spectral norm $\| \bX-\widetilde{\bX}\|_2$, equivalently $\sigma_1(\bX-\widetilde{\bX})$,  converges to zero. 
	 By Weyl's inequality \cite{weyl1912asymptotische,o2013random}, $\bX \ra \widetilde{\bX}$ implies $\bX$ and $\widetilde{\bX}$ asymptotically have  the same singular values,  i.e.,
 $|\sigma_i(\bX)-\sigma_i(\widetilde{\bX})| \ra 0$
	 for all $i \in \{1,2,\ldots,\min ( r(\bX),r(\widetilde{\bX}) ) \}$, $\sigma_i(\bX) \ra 0$ and $\sigma_i(\widetilde{\bX}) \ra 0$ for all $i > \min ( r(\bX),r(\widetilde{\bX}) )$. Furthermore, the Davis-Kahan theorem \cite{Davis70,o2013random} establishes that under some mild condition on the gap of singular values of $\bX$ and $\widetilde{\bX}$, $\bX \ra \widetilde{\bX}$ implies $\bX$ and $\widetilde{\bX}$ asymptotically have  the same singular vectors (identical up to sign), i.e., $|\bu_i^T(\bX) \bu_i(\widetilde{\bX})| \ra 1$ and  $|\bv_i^T(\bX) \bv_i(\widetilde{\bX})| \ra 1$  for all $i \in \{1,2,\ldots,\min ( r(\bX),r(\widetilde{\bX}) )\}$. 
	  If $\bX$ is a random matrix and $\widetilde{\bX}$ is a given matrix, then $\bX \asconv \widetilde{\bX}$ is shorthand for
	  $\|\bX-\widetilde{\bX}\|_2 \rightarrow 0$ almost surely. In particular, if the dimension of $\bX$ grows with $n_k$, then for simplicity we often write  $\bX \asconv \bM$, where $\bM$ is a matrix of infinite dimension. For example, let $\bI_n$ denote the $n \times n$ identity matrix. If $\|\bX-\bI_n\|_2 \asconv 0$ as $n \ra \infty$, then for simplicity we write $\bX \asconv \bI$, where $\bI$ is the identity matrix of infinite dimension. 
	  While this infinite dimensional notation is non-rigorous,  its use in place of the more cumbersome notation $\|\bX-\bI_n\|_2 \asconv 0$ greatly simplifies the presentation. For vectors, we say $\bx \in \bbR^n$ converges to $\widetilde{\bx} \in \bbR^n$  if $\|\bx-\widetilde{\bx}\|_2 \ra 0$ as $n \ra \infty$. Similarly,  for a vector $\bx$, if $\|\bx-\mathbf{m}_n\|_2 \rightarrow 0$ as $n \ra \infty$, where $\mathbf{m}_n$ is a vector of increasing dimension, we use the notation $\bx \ra \mathbf{m}$, where $\mathbf{m}$ is the infinite dimensional limit of $\mathbf{m}_n$. Table \ref{table_limit_value} summarizes
the limit expressions  presented in this paper.

Based on the RIM (\ref{eqn_network_model_multi}), Theorem \ref{thm_impossible} establishes a general breakdown condition under which SGC fails to correctly identify the clusters. 

\begin{thm}[Breakdown condition]
	\label{thm_impossible}
	Let $\bY=[\bY_1^T,\bY_2^T,\ldots,\bY_K^T]^T$ be the cluster partitioned eigenvector matrix associated with the graph Laplacian matrix $\bL$ obtained by solving (\ref{eqn_spectral_clustering_multi_formulation}),	where  $\bY_k \in \mathbb{R}^{n_k \times (K-1)}$ with its rows indexing the nodes in cluster $k$. 	
	Let $\bAt$ be the $(K-1) \times (K-1)$ matrix with $(i,j)$-th element
	\begin{align}
	[\bAt]_{ij}=\left\{
	\begin{array}{ll}
    \lb  n_i+n_K \rb p_{iK}+\sum_{z=1,z \neq i}^{K-1} n_z p_{iz} , & \text{~if~} i=j , \nonumber \\
	n_i\cdot \lb p_{iK}-p_{ij} \rb &\text{~if~} i \neq j.
	\end{array}
	\right.	
	\end{align}
	The following holds almost surely as $n_k \ra \infty$ and $\frac{\nmin}{\nmax} \ra c >0$. 
	If $\liminf_{n \ra \infty} \frac{1}{n} \min_{i \in \{1,\ldots, K-1\},~j \in \{2,\ldots, K\}} |\lambda_i (\bAt)-\lambda_j ( \bL)|>0$,
	then $\bY_k^T \bone_{n_k} \ra \bzero_{K-1}$, $\forall~k\in \{1,2,\ldots,K\}$, and hence spectral graph clustering cannot be successful.		
\end{thm}
Since the eigenvalues of $\bAt$ depend only on the RIM parameters $p_{ij}$ and $n_k$ whereas the eigenvalues of $\bL$ depend not only on these parameters but also on the internal adjacency matrices $\bA_k$, 
	Theorem \ref{thm_impossible} specifies how the graph connectivity structure  affects the success of SGC.

 For the special case of homogeneous RIM, where $p_{ij}=p$, for all $i\neq j$,
	Theorem \ref{thm_spec} establishes the existence of a phase transition 
	in the accuracy of SGC as the interconnection probability $p$ increases. A similar phase transition likely exists for the inhomogeneous RIM (i.e., $p_{ij}$'s are not identical), but an inhomogeneous extension of Theorem \ref{thm_spec}  is an open problem. Nonetheless, Theorem \ref{thm_principal_angle} shows that the homogeneous RIM phase transition threshold $p^*$ in Theorem \ref{thm_spec} can be used to bound clustering accuracy when the RIM is inhomogeneous.

\begin{thm}[Phase transition]
	\label{thm_spec}
	Let $\bY=[\bY_1^T,\bY_2^T,\ldots,\bY_K^T]^T$ be the cluster partitioned eigenvector matrix associated with the graph Laplacian matrix $\bL$ obtained by solving (\ref{eqn_spectral_clustering_multi_formulation}),
	where  $\bY_k \in \mathbb{R}^{n_k \times (K-1)}$ with its rows indexing the nodes in cluster $k$. 
	Let $c^*=\lim_{n \ra \infty} \frac{1}{{n}} \cdot \min_{k \in \{1,2,\ldots,K\}} \SK(\bL_k) $ and assume $c_2^* = \lim_{n \ra \infty} \frac{1}{n} \min_{k \in \{1,2,\ldots,K\}}\lambda_2(\bL_k)>0$.
	Under the homogeneous RIM in (\ref{eqn_network_model_multi}) with constant interconnection probability $p_{ij}=p$,
	there exists a critical value $p^*$ such that the following holds almost surely as $n_k \ra \infty$ and $\frac{\nmin}{\nmax} \ra c >0$: \\	
	\textnormal{(a)}~$ \left\{
	\begin{array}{ll}
	\textnormal{If~} p \leq p^*,~ \frac{\SK \lb \bL \rb}{n}  \ra (K-1)p; \\
	\textnormal{If~} p > p^*,~ c^* + (K-1) \lb 1-\rho_{\max} \rb p  \leq  \frac{\SK \lb \bL \rb}{n} \\ 
 	~~~~~~~~~~~~\leq c^* + (K-1) \lb 1-\rho_{\min} \rb p .  \\
	\end{array}
	\right.$ \\
	In particular, if $p > p^*$ and $c=1$,~ $\frac{\SK \lb \bL \rb}{n} \ra c^* +\frac{(K-1)^2}{K}p$.	
	\\
	Furthermore, reordering the indices $k$ in decreasing cluster size so that $n_1\geq n_2 \geq \ldots \geq n_K$, we have \\	
	\textnormal{(b)}~$\left\{	
	\begin{array}{ll}
	\textnormal{If~} p < p^*,~\sqrt{n_k} \bY_k \ra \bone \bone_{K-1}^T \bV_k\\
	~~~~~~~~~~~~~~~~~~~~~~~=\Lb v^k_1 \bone,v^k_2 \bone,\ldots,v^k_{K-1} \bone \Rb,~ \\ 
	~~~~~~~~~~~~~~~~~~~~~~~~\forall~k \in \{1,2,\ldots,K\}; \\
	\textnormal{If~} p > p^*,~
	\bY_k^T\bone_{n_k} \ra \bzero_{K-1},~\forall~k \in \{1,2,\ldots,K\}; \\
	\textnormal{If~} p = p^*,~\forall~k \in \{1,2,\ldots,K\},~ \sqrt{n_k} \bY_k \ra \bone\bone_{K-1}^T \bV_k \\
	~~~~~~~~~~~~~\textnormal{~or~} \bY_k^T\bone_{n_k} \ra \bzero_{K-1},
	\end{array}
	\right.$ \\
	where $\bV_k=\diag(v^k_1, v^k_2,\ldots, v^k_{K-1}) \in \mathbb{R}^{(K-1) \times (K-1)}$ is a diagonal matrix.\\
	 Finally, $p^*$ satisfies: \\	
	\textnormal{(c)}~$\pLB \leq p^* \leq \pUB$, where 
	$\pLB=\frac{c^*}{(K-1) \rhomax}$ and
	$\pUB  =\frac{c^*}{(K-1) \rhomin}$.
	In particular, 	$\pLB=\pUB$ when $c=1$.	
\end{thm}

Theorem \ref{thm_spec} (a) establishes a phase transition of the partial eigenvalue sum $\frac{\SK(\bL)}{n}$ at some critical value $p^*$, called the critical phase transition threshold. When $p \leq p^*$ the quantity $\frac{\SK(\bL)}{n}$ converges to  $(K-1)p$. When $p> p^*$ the slope  in $p$ of $\frac{\SK(\bL)}{n}$  changes and the intercept $c^*$ depends on the cluster having the smallest partial eigenvalue sum. When all clusters have the same size (i.e., $\nmax=\nmin=\frac{n}{K}$) so that $c=1$, $\frac{\SK(\bL)}{n}$ undergoes a slope change from $K-1$ to $\frac{(K-1)^2}{K}$.

Theorem \ref{thm_spec} (b) establishes that $p>p^*$ renders the entries of the matrix $\mathbf Y_k$ incoherent, making it impossible for SGC to separate the clusters. On the other hand,  $p < p^*$  makes $\mathbf Y_k$ coherent, and hence the row vectors in the eigenvector matrix $\bY$ possess cluster-wise separability. This is stated as follows.

\begin{cor}[Separability of the row vectors in the eigenvector matrix $\bY$ when $p < p^*$]	
	\label{cor_separability}
~\\	Under the same assumptions as in Theorem \ref{thm_spec}, when $p < p^*$, the following properties of $\bY$ hold  almost surely  as $n_k \ra \infty$ and $\frac{\nmin}{\nmax} \ra c >0$:\\
	\textnormal{(a)} The columns of $\sqrt{n_k}  \bY_k$ are constant vectors. \\
	\textnormal{(b)} Each column of $\sqrt{n} \bY$ has at least two nonzero cluster-wise constant components, and these constants have alternating signs such that their weighted sum equals $0$ due to the property $\sum_{k=1}^K v^k_j = 0,~\forall~j \in\{1,2,\ldots,K-1\}$. \\
	\textnormal{(c)} No two columns of $\sqrt{n} \bY$ have the same sign on the cluster-wise nonzero components.	
\end{cor}
These properties imply that for $p<p^*$ the rows in $\bY_k$ corresponding to different nodes are identical (Corollary \ref{cor_separability} (a)), while the row vectors in $\bY_k$ and $\bY_\ell$, $k \neq \ell$,  corresponding to different clusters are distinct  (Corollary \ref{cor_separability} (b) and (c)).
Therefore, the within-cluster distance between any pair of row vectors in each $\bY_k$ is zero, whereas the between-cluster distance between any two row vectors of different clusters is nonzero. This means that as $n_k \ra \infty$ and $\frac{\nmin}{\nmax} \ra c >0$ the ground-truth clusters become the optimal solution to K-means clustering, and
 hence K-means clustering on these row vectors can group the nodes into correct clusters.
Note that when $p>p^*$, from Theorem \ref{thm_spec} (b) the row vectors of $\bY_k$ corresponding to the same cluster sum to a zero vector. This means that the entries of each column in $\bY_k$ have alternating signs and the centroid of the row vectors of each cluster is the origin.
Therefore, K-means clustering on the rows of $\bY$ yields incorrect clusters.

Furthermore, as a demonstration of the breakdown condition in Theorem \ref{thm_impossible}, observe that when $p_{ij}=p$,
Theorem \ref{thm_impossible} implies that  $\frac{\bAt}{n}$ is a diagonal matrix $p \bI_{K-1}$. From (\ref{eqn_case1}) in the appendix we know that $\frac{\lambda_j (\bL)}{n} \ra p$ for $j=2,3,\ldots,K$ almost surely when $p < p^*$. Therefore, under the homogeneous RIM, SGC can only be successful when $p$ is below  $p^*$.

Theorem  \ref{thm_spec} (c) provides upper and lower bounds on the critical threshold value $p^*$ for the phase transition to occur when $p_{ij}=p$. These bounds are determined by the cluster having the smallest partial eigenvalue sum $\SK(\bL_k)$, the number of clusters $K$, and the size of the largest and smallest cluster ($\nmax$ and $\nmin$). When all cluster sizes are identical (i.e., $c=1$), these bounds become tight. Based on Theorem  \ref{thm_spec} (c), the following corollary specifies the properties of $p^*$ and the connection to algebraic connectivity of each cluster. 

\begin{cor}[Properties of $p^*$ and its connection to algebraic connectivity] 
\label{cor_pstar}
	~\\ 	Let $c_n=\frac{ \min_{k \in \{1,2,\ldots,K\}} \SK(\bL_k) }{n}$, $c_{2,n}=\frac{ \min_{k \in \{1,2,\ldots,K\}} \lambda_2(\bL_k) } {n} $ and $c_{K,n}=\frac{\min_{k \in \{1,2,\ldots,K\}} \lambda_K(\bL_k) } {n}$, and let $c^*$, $c_2^*$ and $c_K^*$ denote their limit value, respectively. 
 Under the same assumptions as in Theorem \ref{thm_spec}, the following statements hold almost surely as $n_k \ra \infty$ and $\frac{\nmin}{\nmax} \ra c >0$:\\
	\textnormal{(a)}	If $c_n=\Omega \lb \frac{\nmax}{n}\rb$, then $p^* >0$. \\
	\textnormal{(b)}	If $c_n=o \lb \frac{\nmin}{n} \rb$, then $p^*=0$. \\
	\textnormal{(c)}  $\frac{ c_2^*}{\rhomax} \leq p^* \leq \frac{c_K^*}{\rhomin}$.
\end{cor}

The following corollary specifies the bounds on the critical value $p^*$ for some special types of clusters. These results provide theoretical justification of the intuition that strongly connected clusters, e.g., complete graphs, have high critical threshold value, and weakly connected clusters, e.g., star graphs, have low critical threshold value.
\begin{cor}[bounds on the critical value $p^*$  for special type of cluster] 
	\label{cor_special_graph}
	~\\ Under the same assumptions as in Theorem \ref{thm_spec}, the following statements hold almost surely as $n_k \ra \infty$ and $\frac{\nmin}{\nmax} \ra c >0$:\\
		\textnormal{(a)} If each cluster is a complete graph, then $c \leq p^* \leq 1$.\\
		\textnormal{(b)} If each cluster is a star graph, then $p^*=0$.
\end{cor}

Furthermore, in the special case of a SBM,  where each adjacency matrix $\bA_k$ corresponds to  a Erdos-Renyi random graph with edge connection probability $p_k$, under the same assumptions as in Theorem \ref{thm_spec} we can show that almost surely,
	\begin{align}
	\label{eqn_SBM_critical}
c \cdot \min_{k \in \{1,2,\ldots,K\}} p_k \leq p^* \leq  \frac{1}{c} \cdot \min_{k \in \{1,2,\ldots,K\}} p_k.	
	\end{align}
	The proof of (\ref{eqn_SBM_critical}) is given in the supplementary material. Similar  results for the SBM can also be deduced from the latent space model \cite{rohe2011spectral}.
	
The next corollary summarizes the results from Theorem \ref{thm_spec} for the case of $K=2$ to elucidate the phase transition phenomenon. Note that it follows from Corollary \ref{cor_K2} (b) that below the phase transition ($p< p^*$) the rows in $\bY$ corresponding to different clusters are constant vectors with entries of opposite signs, and thus K-means clustering is capable of yielding correct clusters. On the other hand, above the phase transition ($p>p^*$) 
the entries corresponding to each cluster have alternating signs and the centroid of each cluster is the origin, and thus K-means clustering fails.

\begin{cor}[Special case of Theorem \ref{thm_spec} when $K=2$] 
	\label{cor_K2}
	~\\
	When $K=2$, let $\bY=[\yone^T~\ytwo^T]^T$, let $c^*=\lim_{n \ra \infty}\frac{\lambda_2(\bL_1) +\lambda_2(\bL_2)-\left| \lambda_2(\bL_1) -\lambda_2(\bL_2) \right|}{2n}$ and assume $c_2^* = \lim_{n \ra \infty} \frac{1}{n} \min \{\lambda_2(\bL_1),\lambda_2(\bL_2)\}>0$. Then there exists a critical value $p^*$ such that the following holds almost surely as $n_1,n_2 \ra \infty$ and $\frac{\nmin}{\nmax} \ra c >0$. \\	
	\textnormal{(a)}~$ \left\{
	\begin{array}{lll}
	\textnormal{If~} p \leq p^*,~\frac{\lambda_2(\bL)}{n} \ra p; \\
	\textnormal{If~} p > p^*,~ c^*+ \frac{c}{1+c}p \leq	\frac{\lambda_2(\bL)}{n} \leq    c^* + \frac{1}{1+c}p.
	\end{array}
	\right.$ 
	\\
	\textnormal{(b)}~$\left\{
	\begin{array}{ll}
	\textnormal{If~} p < p^*,~\sqrt{\frac{n \none}{\ntwo}} \yone \ra \pm \bone ~\textnormal{and}~\sqrt{\frac{n \ntwo }{\none}} \ytwo \ra \mp \bone;\\
	\textnormal{If~} p > p^*,~\yone^T\onenone \ra 0~\textnormal{and}~\ytwo^T\onentwo \ra 0.
	\end{array}
	\right.$ \\
	\textnormal{(c)}~$\pLB \leq p^* \leq \pUB$, where
	$\pLB = \frac{2c^*}{1+|\rho_1-\rho_2|}$ and
	$\pUB  = \frac{2c^*}{1-|\rho_1-\rho_2|}$.
\end{cor}

	The above phase transition analysis can also be applied to the  inhomogeneous RIM for which the $p_{ij}$'s  are not constant.
Let 
$p_{\min}=\min_{i \neq j}p_{ij}$ and $p_{\max}=\max_{i \neq j} p_{ij}$. The corollary below shows that under the inhomogeneous RIM when $p_{\max}$ is below $p^*$, 
which is the critical threshold value specified by Theorem \ref{thm_spec} for the homogeneous RIM,
the smallest $K-1$ nonzero eigenvalues of the graph Laplacian matrix $\frac{\bL}{n}$ lie within the internal $[p_{\min},p_{\max}]$ with probability one.

	\begin{cor}[Bounds on the smallest $K-1$ nonzero eigenvalues of $\bL$ under the inhomogeneous RIM]~\\
		\label{cor_early_breakdown}
	Under the RIM with interconnection probabilities $\{p_{ij}\}$, let 
	$p_{\min}=\min_{i \neq j}p_{ij}$, $p_{\max}=\max_{i \neq j} p_{ij}$, and let $p^*$ be the critical threshold value of the homogeneous RIM specified by Theorem \ref{thm_spec}.
	 If $p_{\max} < p^*$, the following statement holds almost surely as
	  $n_k \ra \infty$ and $\frac{\nmin}{\nmax} \ra c >0$:
	  \begin{align}
	     p_{\min}  \leq \frac{\lambda_j (\bL)}{n} \leq  p_{\max},~\forall~ j=2,3,\ldots,K. \nonumber
	  \end{align}
	\end{cor}
In particular, Corollary \ref{cor_early_breakdown} implies that the normalized algebraic connectivity of the inhomogeneous RIM $\frac{\lambda_2(\bL)}{n}$ is between $p_{\min}$ and $p_{\max}$ almost surely as  $n_k \ra \infty$ and $\frac{\nmin}{\nmax} \ra c >0$.

For graphs following the inhomogeneous RIM, Theorem \ref{thm_principal_angle} below establishes that accurate clustering is possible if it can be determined that $p_{\max}<p^*$.		
As defined in Theorem \ref{thm_spec}, let $\bY \in \mathbb{R}^{n \times (K-1)} $ be the eigenvector matrix of $\bL$ under the inhomogeneous RIM, and let $\btY \in \mathbb{R}^{n \times (K-1)}$ be the eigenvector matrix of the graph Laplacian $\btL$ of another random graph, independent of $\bL$, generated by a homogeneous RIM with cluster interconnectivity parameter $p$. 
We can specify the distance between the subspaces spanned by the columns of $\bY$ and $\btY$ by inspecting their principal angles \cite{Luxburg07}. Since $\bY$ and $\btY$ both have orthonormal columns, the vector $\bv$ of $K-1$ principal angles between their column spaces is $\bv=[\cos^{-1}\sigma_1(\bY^T \btY),\ldots,\cos^{-1}\sigma_{K-1}(\bY^T \btY)]^T$, where $\sigma_k(\bM)$ is the $k$-th largest singular value of a real rectangular matrix $\bM$.
Let $\mathbf{\Theta}(\bY,\btY)=\diag(\bv)$, and let $\sin\mathbf{\Theta}(\bY,\btY)$ be defined entrywise. When $p<p^*$, the following theorem provides an upper bound on the 
Frobenius norm of $\sin\mathbf{\Theta}(\bY,\btY)$, which is denoted by $\| \sin\mathbf{\Theta}(\bY,\btY) \|_F$.

\begin{thm}[Distance between column spaces spanned by $\bY$ and $\btY$]
	\label{thm_principal_angle}
Under the RIM with interconnection probabilities $\{p_{ij}\}$, let $p^*$ be the critical threshold value for the homogeneous RIM specified by Theorem \ref{thm_spec}, and define $\delta_{p,n}=\min\{p,|\frac{\lambda_{K+1}(\bL)}{n}-p|\}$. 
For a fixed $p$, let $b_p$ denote the limit of $\frac{\| \bL - \btL \|_F}{n}$.	
If $p < p^*$ and $\delta_{p,n} \ra \delta_p > 0$ as $n_k \ra \infty$,
the following statement holds almost surely as
$n_k \ra \infty$ and $\frac{\nmin}{\nmax} \ra c >0$:
\begin{align}
\label{eqn_principal_angle_bound}
\|\sin\mathbf{\Theta}(\bY,\btY)\|_F \leq \frac{b_p}{ \delta_p}.
\end{align}
Furthermore, let $p_{\max}=\max_{i \neq j} p_{ij}$. If $p_{\max} < p^*$,
$\|\sin\mathbf{\Theta}(\bY,\btY)\|_F \leq \min_{p \leq p_{\max}} \frac{b_p}{ \delta_p}.$
\end{thm}
As established in Corollary \ref{cor_separability}, under the homogeneous RIM when $p<p^*$ the row vectors of the eigenvector matrix $\btY$ 
 are perfectly cluster-wise separable as
 $n_k \ra \infty$ and $\frac{\nmin}{\nmax} \ra c >0$. Under the inhomogeneous RIM, Theorem \ref{thm_principal_angle} establishes that cluster separability can still be expected provided that $\|\sin\mathbf{\Theta}(\bY,\btY)\|_F $ is small and $p<p^*$. As a result, we can bound the clustering accuracy under the inhomogeneous RIM by inspecting the upper bound (\ref{eqn_principal_angle_bound}) on $\|\sin\mathbf{\Theta}(\bY,\btY)\|_F$. Note that if $p_{\max} < p^*$, we can obtain a tighter upper bound on (\ref{eqn_principal_angle_bound}).

Next we extend Theorem \ref{thm_spec} to undirected weighted random graphs obeying the homogeneous RIM. The edges within each cluster are assumed to have nonnegative weights and the weights of inter-cluster edges are assumed to be independently drawn from a common nonnegative bounded distribution. Let $\bW$ denote the $n \times n$ symmetric nonnegative weight matrix of the entire graph. Then the corresponding graph Laplacian matrix is defined as $\bL=\bS-\bW$, where $\bS =\diag( \bW \bone_n )$ is the diagonal matrix of nodal strengths of the weighted graph. Similarly, the symmetric graph Laplacian matrix $\bL_k$ of each cluster can be defined. The following theorem establishes a phase transition phenomenon for such weighted graphs. Specifically, the critical value depends not only on the inter-cluster edge connection probability but also on the mean of inter-cluster edge weights.

\begin{thm}[Phase transition in weighted graphs]
	\label{thm_spec_weight}
	Under the same assumptions as in Theorem \ref{thm_spec}, further assume 
	the weight matrix $\bW$ is symmetric, nonnegative and bounded, and
	the weights of the upper triangular part of $\bW$ 
	are independently drawn from a common nonnegative bounded distribution with mean $\Wbar$. Let $t=p \cdot \Wbar$ and $c^*=\lim_{n \ra \infty }\frac{1}{{n}} \cdot \min_{k \in \{1,2,\ldots,K\}} \SK(\bL_k) $.
	Then there exists a critical value $t^*$ such that the following holds almost surely as $n_k \ra \infty$ and $\frac{\nmin}{\nmax} \ra c >0$: \\	
	\textnormal{(a)}~$ \left\{
	\begin{array}{ll}
	\textnormal{If~} t \leq t^*,~\frac{\SK(\bL)}{n} \ra (K-1)t; \\
	\textnormal{If~} t > t^*,~ c^* + (K-1) \lb 1-\rhomax \rb t \leq \frac{\SK(\bL)}{n} \\
   ~~~~~~~~~~~\leq  c^* + (K-1) \lb 1-\rhomin \rb t.
	\end{array}	
	\right.$ \\
	\textnormal{(b)}~$\left\{
	\begin{array}{ll}
	\textnormal{If~} t < t^*,~\sqrt{n_k} \bY_k \ra \bone \bone_{K-1}^T \bV_k \\
	~~~~~~~~~~~~~~~~~~~~~~=\Lb v^k_1 \bone,v^k_2 \bone,\ldots,v^k_{K-1} 	 \bone \Rb,~\\
	~~~~~~~~~~~~~~~~~~~~~~~\forall~k \in \{1,2,\ldots,K\}; \\
	\textnormal{If~} t > t^*,~\bY_k^T\bone_{n_k} \ra \bzero_{K-1}~\forall~k \in \{1,2,\ldots,K\}; \\
	\textnormal{If~} t \ra t^*,~\forall~k \in \{1,2,\ldots,K\},~ \sqrt{n_k}\bY_k \ra \bone \bone_{K-1}^T \bV_k	\\
		~~~~~~~~~~~~~\textnormal{~or~} \bY_k^T\bone_{n_k} = \bzero_{K-1},
	\end{array}
	\right.$ \\
	where $\bV_k=\diag(v^k_1, v^k_2,\ldots, v^k_{K-1}) \in \mathbb{R}^{(K-1) \times (K-1)}$ is a diagonal matrix. \\	
	\textnormal{(c)}~$\tLB \leq t^* \leq \tUB$, where 
	$\tLB = \frac{c^*}{(K-1)\rhomax}$ and 
	$\tUB  = \frac{c^*}{(K-1)\rhomin}$.
\end{thm}

Theorem \ref{thm_impossible} and Theorem \ref{thm_principal_angle}  
can be  extended to weighted graphs under the inhomogeneous RIM. Moreover, Theorem \ref{thm_spec_weight}  reduces to Theorem \ref{thm_spec}  when $\Wbar=1$.

\section{Automated Model Order Selection (AMOS) Algorithm for Spectral Graph Clustering}
\label{sec_ASGC}
\begin{figure}[t!]		
	\centering
	\includegraphics[scale=0.42]{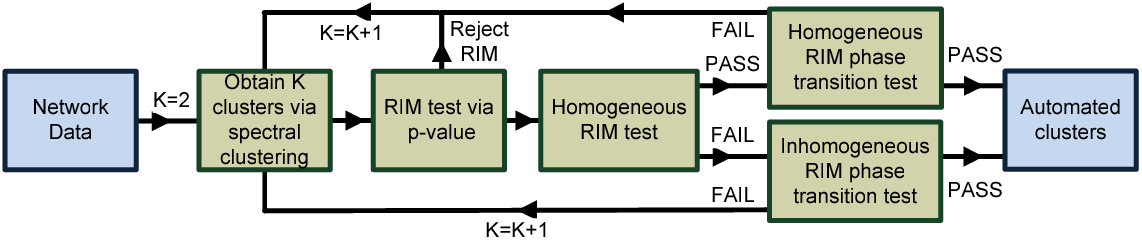}
	\caption{Flow diagram of the proposed automated model order selection (AMOS) scheme in spectral graph cluster (SGC).}
	\label{Fig_automated_clustering}     
	\vspace{-4mm}
\end{figure}

Based on the phase transition analysis in Sec. \ref{sec_phase_spec}, we propose an automated model order selection (AMOS) algorithm for selecting the number of clusters in
spectral graph clustering (SGC). 
 This algorithm  produces p-values of hypothesis tests for testing the RIM and phase transition.
 	In particular, under the homogeneous RIM, we can estimate the critical phase transition threshold for each putative cluster found and use this estimate to construct a test of reliability of the cluster. The statistical tests in the AMOS algorithm are implemented in two phases. The first phase is to test the RIM assumption based on the interconnectivity pattern of each cluster (Sec. \ref{subsec_RIM_test}), and the second phase is to test the homogeneity and variation of the interconnectivity parameter $p_{ij}$ for every cluster pair $i$ and $j$  in addition to making comparisons to the critical phase transition threshold (Sec. \ref{subsec_phase_transition_estimator}). The flow diagram of the proposed algorithm is displayed in Fig. \ref{Fig_automated_clustering}, and the algorithm is summarized in Algorithm \ref{algo_automated_clustering}. The AMOS package is publicly available for download\footnote{https://github.com/tgensol/AMOS}.
 	 	 	Next we explain the functionality of each block in the diagram.

\subsection{Input network data and spectral clustering}
The input network data is a matrix that can be a symmetric adjacency matrix $\bA$, a degree-normalized symmetric adjacency matrix $\bD^{-\frac{1}{2}} \bA \bD^{-\frac{1}{2}}$, a symmetric weight matrix $\bW$, or a normalized symmetric weight matrix $\bS^{-\frac{1}{2}} \bW \bS^{-\frac{1}{2}}$, where $\bD=\diag(\bA \bone_n)$ and $\bS=\diag(\bW \bone_n)$ are assumed invertible. Spectral clustering is then implemented on the input data to produce $K$ clusters $\{\hG_k\}_{k=1}^K$, where  $\hG_k$ is the $k$-th identified cluster with number of nodes $\hn_k$ and number of edges $\hatm_k$. Initially $K$ is set to $2$. The AMOS algorithm works by iteratively increasing $K$ and performing spectral clustering on the data until the output clusters meet
a level of significance 
 criterion specified by the RIM test and phase transition estimator.

\subsection{RIM test via p-value for local homogeneity testing}
\label{subsec_RIM_test}

Given clusters $\{\hG_k\}_{k=1}^K$ obtained from spectral clustering with model order $K$, let $\widehat{\bC}_{ij}$ be the $\hn_i \times \hn_j$ interconnection matrix of edges connecting clusters $i$ and $j$. Our goal is to compute a p-value to test the hypothesis that the matrix $\bA$ in (\ref{eqn_network_model_multi}) satisfies the RIM. More specifically, 
we are testing the null hypothesis that \emph{$\widehat{\bC}_{ij}$ is a realization of a random matrix with i.i.d. Bernoulli entries} (RIM) and the alternative hypothesis that \emph{$\widehat{\bC}_{ij}$ is not a realization of a random matrix with i.i.d. Bernoulli entries entries} (not RIM), for all $i \neq j$, $i>j$. Since the RIM homogeneity model for the interconnection matrices $\{\bC_{ij}\}$ will only be valid when the clusters have been correctly identified, this RIM test can be used to test the quality of a graph clustering algorithm.

\begin{algorithm}[t]
	\caption{p-value computation of V-test for the RIM test}
	\label{algo_RIM_pvalue}
	\begin{algorithmic}
		\State \textbf{Input:} An $n_i \times n_j$ interconnection matrix $\widehat{\bC}_{ij}$
		\State \textbf{Output:} \text{p-value}$(i,j)$
		\State $\bx= \widehat{\bC}_{ij} \bone_{n_j}$~(\# of nonzero entries of each row in $\widehat{\bC}_{ij}$)
		\State $\by= n_j \bone_{n_i}-\bx$~(\# of zero entries of each row in $\widehat{\bC}_{ij}$)	
		\State $X=\bx^T \bx - \bx^T \bone_{n_i}$ and  $Y=\by^T \by - \by^T \bone_{n_i}$.
		\State $N=n_i n_j (n_j-1)$ and $V=\lb \sqrt{X} + \sqrt{Y} \rb^2$.
		\State Compute test statistic $Z=\frac{V-N}{\sqrt{2N}}$
		\State Compute \text{p-value}$(i,j)$$=2 \cdot \min \{ \Phi(Z),1-\Phi(Z) \}$
	\end{algorithmic}
\end{algorithm} 

\begin{algorithm}[t]
	\caption{Automated model order selection (AMOS) algorithm for spectral graph clustering (SGC)}
	\label{algo_automated_clustering}
	\begin{algorithmic}
		\State \textbf{Input:} a connected undirected weighted graph,  p-value significance level $\eta$, homogeneous and inhomogeneous RIM confidence interval parameters $\alpha$, $\alpha^\prime$
		\State \textbf{Output:} number of clusters $K$ and identity of $\{\hG_k\}_{k=1}^K$
		\State Initialization: $K=2$. Flag $=1$.
		\While{Flag$=1$}
		\State Obtain $K$ clusters $\{\hG_k\}_{k=1}^K$ via spectral clustering ($*$)
		\For{$i=1$ to $K$}
		\For{$j=i+1$ to $K$}
		\State Calculate p-value($i,j$) from Algorithm \ref{algo_RIM_pvalue}.
		\If{p-value($i,j$) $\leq \eta$}{~Reject RIM}
		\State Go back to ($*$) with $K=K+1$.
		\EndIf
		\EndFor
		\EndFor 
		\State Estimate $\hp$, $\hWbar$, $\{ \hp_{ij}\}$, and $\htLB$ specified in Sec. \ref{subsec_phase_transition_estimator}.
		\If{$\hp$ lies within the confidence interval in (\ref{eqn_spectral_multi_confidence_interval})}
		\State \# \emph{Homogeneous RIM phase transition test} \#
		\If{$\hp \cdot \hWbar$$  < \htLB$}				 
		Flag$=0$.
		\Else~ 
		Go back to ($*$) with $K=K+1$.		
		\EndIf    
		\ElsIf{$\hp$ does not lie within (\ref{eqn_spectral_multi_confidence_interval})}
		\State \# \emph{Inhomogeneous RIM phase transition test} \#
		\If{$\prod_{i=1}^K \prod_{j=i+1}^K F_{ij}\lb \frac{\htLB}{\hWbar},\hpij \rb \geq 1-\alpha^\prime$}	\\	
		~~~~~~~~~~~Flag$=0$.
		\Else
		~Go back to ($*$) with $K=K+1$.			
		\EndIf
		
		\EndIf
		\EndWhile
		\State Output  $K$ clusters $\{\hG_k\}_{k=1}^K$.		
	\end{algorithmic}
\end{algorithm}

	To compute a p-value for the RIM we use the V-test \cite{potthoff1966testing} for homogeneity testing of the row sums or column sums of $\widehat{\bC}_{ij}$. Specifically, given $s$ independent binomial random variables, the V-test tests that they are all identically distributed. For concreteness, here we apply the V-test to the row sums. Given a candidate set of clusters, the  V-test is applied independently to each of the $K \choose 2$ interconnection matrices $\{\widehat{\bC}_{ij}\}$.
	
	For any interconnection matrix $\widehat{\bC}_{ij}$ the test statistic $Z$ of the V-test converges to a standard normal distribution as $n_i,n_j \ra \infty$, and the p-value for the hypothesis that the row sums of $\widehat{\bC}_{ij}$ are i.i.d. is
	\text{p-value}$(i,j)=2 \cdot \min \{ \Phi(Z),1-\Phi(Z) \}$, where $\Phi(\cdot)$ is the cumulative distribution function (cdf) of the standard normal distribution.
	The proposed V-test procedure is summarized in Algorithm \ref{algo_RIM_pvalue}. The RIM test on $\widehat{\bC}_{ij}$ rejects the null hypothesis if \text{p-value}$(i,j) \leq \eta$, where $\eta$ is the desired single comparison significance level.
	Since the $\bC_{ij}$'s are independent, the p-value threshold parameter $\eta$ can be easily translated into a multiple comparisons significance level for detecting homogeneity of all $\bC_{ij}$'s. It can also be translated into a threshold for testing the homogeneity of at least one of these matrices using family-wise error rate Bonferroni corrections or false discovery rate analysis \cite{simes1986improved,benjamini1995controlling}.

\subsection{A cluster quality measure for RIM}
\label{subsec_phase_transition_estimator}
Once the identified clusters $\{\hG_k\}_{k=1}^K$ pass the RIM test, 
one can empirically determine the reliability of the clustering using the phase transition analysis introduced in the previous section. In a nutshell, if the estimate of $p_{\max}= \max_{i>j} p_{ij}$ falls below the critical phase transition threshold $p^*$ then, by Theorem \ref{thm_principal_angle}, the results of the clustering algorithm can be declared reliable if the clustering quality measure $\|\sin \Theta(\bY,\btY) \|_F$ is small. This is the basis for the proposed AMOS procedure under the assumption of inhomogeneous RIM. For homogeneous RIM models an alternative procedure is proposed. 
The AMOS algorithm (Fig. \ref{Fig_automated_clustering}) runs a serial process of homogeneous and inhomogeneous RIM phase transition tests. 
Each of these is considered separately in what follows.

\noindent{\it $\bullet$ Homogeneous RIM phase transition test}:\\
The following plug-in estimators are used to evaluate the RIM parameters and the critical phase transition threshold under  the homogeneous RIM.   Let $\hatm_{ij}=\bone_{n_i}^T \widehat{\bC}_{ij} \bone_{n_j}$ be the number of inter-cluster edges between clusters $i$ and $j$ (i.e., the number of nonzero entries in $\widehat{\bC}_{ij}$). Then under the inhomogeneous RIM  $\hp_{ij}=\frac{\hatm_{{ij}}}{\hn_i \hn_j}$ is the maximum likelihood estimator (MLE) of $p_{ij}$.
Under the homogeneous RIM, $p_{ij}=p$, and the MLE of $p$ is 
$\hp=\frac{\sum_{i=1}^K \sum_{j=i+1}^K \hatm_{ij}}{\sum_{i=1}^K \sum_{j=i+1}^K \hn_i \hn_j}=\frac{2(m-\sum_{k=1}^{K}\hatm_k)}{n^2-\sum_{k=1}^{K} \hn_k^2}$, where $m$ is the number of edges in the graph. We use the estimates $\hp$ and $\{ \hp_{ij} \}$ to carry out a test for the homogeneous RIM and utilize the estimated critical phase transition threshold developed in this paper to evaluate the clustering quality  when it passes the test. Intuitively, if $\{\hp_{ij}\} $ are close to $\hp$ and $\hp$ is below the estimated phase transition threshold, then the output clusters are regarded homogeneous and reliable. On the other hand, if there is a large variation in $\{ \hp_{ij} \}$, the homogeneity test fails.

A generalized log-likelihood ratio test (GLRT) is used to test the validity of the homogeneous RIM. The details are given in the supplementary material.
By the Wilk's theorem \cite{wilks1938large}, an asymptotic $100(1-\alpha) \%$ confidence interval for $p$ in an assumed homogeneous RIM is 
{\small
	\begin{align}
	\label{eqn_spectral_multi_confidence_interval}
	&\Bigg\{ p:
	\xi_{\binom{K}{2}-1,1-\frac{\alpha}{2}} \leq 
	2\sum_{i=1}^K \sum_{j=i+1}^K  \mathbb{I}_{\{\hp_{ij} \in (0,1)\}} \Lb \hatm_{ij} \ln \hp_{ij} \right.  \nonumber \\
	&\left. \left. +(\hn_i \hn_j -\hatm_{ij} ) \ln (1-\hp_{ij}) \Rb -2 \lb m-\sum_{k=1}^{K}\hatm_{k} \rb  \ln p  \right. \nonumber \\
	& \left.
	- \Lb n^2-\sum_{k=1}^{K} \hn_k^2  
	-2\lb m-\sum_{k=1}^{K}\hatm_{k} \rb \Rb  \ln (1-p) \right.  \leq \xi_{\binom{K}{2}-1,\frac{\alpha}{2}} 
	\Bigg\},
	\end{align} 
}  
where $\xi_{q,\alpha}$ is the upper $\alpha$-th quantile of the central chi-square distribution with degree of freedom $q$.
	
The clusters pass the homogeneous RIM test if $\hp$ is within the confidence interval (\ref{eqn_spectral_multi_confidence_interval}), and by Theorem \ref{thm_spec} the clusters are deemed reliable if $\hp < \hpLB$, an estimate of the lower bound on the critical phase transition threshold value, which is denoted by $\hpLB = \frac{\min_{k \in \{1,2,\ldots,K\}} \SK( \hL_k)}{(K-1)\hn_{\max}}$.

\noindent{\it $\bullet$ Inhomogeneous RIM phase transition test}:\\
As established in  Theorem \ref{thm_principal_angle}, if $\max_{i > j} p_{ij}  < p^*$, we can obtain a tight bound on the clustering quality measure 
$\|\sin \Theta(\bY,\btY) \|_F$, and by the perfect separability in $\btY$ from
Theorem \ref{thm_spec}, we can conclude that the clusters identified by SGC are reliable.  
We use the maximum of MLEs of $p_{ij}$'s, denoted by $\hp_{\max}=\max_{i > j}\hpij$, as a test statistic for testing the null hypothesis $H_0$: \emph{$\max_{i > j} p_{ij}  < p^*$} against the alternative hypothesis $H_1$: \emph{$\max_{i > j} p_{ij}  \geq p^*$}.
The test accepts $H_0$ if $\hp_{\max}<p^*$, and rejects $H_0$ otherwise. 
Using the Anscombe transformation on the $\hpij$'s for variance stabilization \cite{anscombe1948transformation}, let $A_{ij}(x)=\sin^{-1} \sqrt{\frac{x+\frac{c^\prime}{\hn_i \hn_j}}{1+\frac{2c^\prime}{\hn_i \hn_j}}}$, where $c^\prime=\frac{3}{8}$.
 By the central limit theorem,
 $\sqrt{4\hn_i \hn_j+2} \cdot \lb A_{ij}(\hpij)- A_{ij}(p_{ij}) \rb
 \convd N(0,1)$
for all $p_{ij} \in (0,1)$ as $\hn_i,\hn_j \ra \infty$, 
where $\convd$ denotes convergence in distribution and $N(0,1)$ denotes the standard normal distribution \cite{anscombe1948transformation}.
Therefore, under the null hypothesis that $\max_{i>j} p_{ij}<p^*$, from \cite[Theorem 2.1]{chang2000generalized}
an asymptotic $100(1-\alpha^\prime)\%$ confidence interval for $\hp_{\max}$ is $[0,\psi]$,
 where  $\psi(\alpha^\prime,\{\hpij\})$ is a function of the precision parameter $\alpha^\prime \in [0,1]$ and $\{\hpij\}$, which satisfies $\prod_{i=1}^K \prod_{j=i+1}^K \Phi \lb  \sqrt{4\hn_i \hn_j+2} \cdot \lb A_{ij}(\psi)- A_{ij}(\hpij)\rb \rb=1-\alpha^\prime$,
and $\Phi(\cdot)$ is the cdf of the standard normal distribution. Therefore, if $\psi < p^*$, then $\hp_{\max}<p^*$ with probability at least $1-\alpha^\prime$. Note that verifying $\psi < p^*$ is equivalent to checking the condition
\begin{align}
\label{eqn_spectral_multi_confidence_interval_ingomogeneous_RIM}
\prod_{i=1}^K \prod_{j=i+1}^K F_{ij}(p^*,\hpij) \geq 1-\alpha^\prime,
\end{align}
where $F_{ij}(p^*,\hpij)= \Phi \lb  \sqrt{4\hn_i \hn_j+2} \cdot \lb A_{ij}(p^*)- A_{ij}(\hpij)\rb \rb \cdot \mathbb{I}_{\{\hpij \in (0,1)\}}+ \mathbb{I}_{\{\hpij<p^*\}}\mathbb{I}_{\{\hpij \in \{0,1\}\}}$, and $\mathbb{I}_E$ is the indicator function of an event $E$.
For implementation of the inhomogeneous RIM phase transition test, we 
replace $F_{ij}(p^*,\hpij)$  in (\ref{eqn_spectral_multi_confidence_interval_ingomogeneous_RIM}) with $F_{ij}(\hpLB,\hpij)$, and 
check whether $\prod_{i=1}^K \prod_{j=i+1}^K F_{ij}(\hpLB,\hpij) \geq 1-\alpha^\prime$  or not. Since $\pLB \leq p^*$, by the monotonicity of $\Phi(\cdot)$ and $\sin^{-1}(\cdot)$, $\prod_{i=1}^K \prod_{j=i+1}^K F_{ij}(\pLB,\hpij) \geq 1-\alpha^\prime$ implies $\prod_{i=1}^K \prod_{j=i+1}^K F_{ij}(p^*,\hpij)  \geq 1-\alpha^\prime$.

In the phase transition test stage of Algorithm \ref{algo_automated_clustering}, 
the inhomogeneous RIM phase transition test is adopted if the clusters fail the
homogeneous RIM test. Increasing $\eta$ or decreasing $\alpha$ and $\alpha^\prime$ tightens the clustering reliability constraint and may increase the number of output clusters.
These phase transition estimators are extended to weighted graphs by defining the parameter $t_{ij}=p_{ij} \cdot \Wbar$ and using the empirical estimators $\hatt_{ij}=\hp_{ij} \cdot \hWbar$ and  $\htLB = \frac{\min_{k \in \{1,2,\ldots,K\}} \SK( \hL_k)}{(K-1)\hn_{\max}}$ in the AMOS algorithm, where $\hWbar$ is the average weight of the inter-cluster edges. The details are given in the supplementary material.

\subsection{Computational complexity analysis}
The overall computational complexity of the proposed AMOS algorithm is $O(K^3(m+n))$, where $K$ is the number of output clusters, $n$ is the number of nodes, and $m$ is the number of edges. 
Fixing a model order $K$ (i.e., the number of clusters) in the AMOS iteration as displayed in Fig. \ref{Fig_automated_clustering}, there are three contributions to the computational complexity of AMOS: The first contribution is the incremental eigenpair computation - acquiring an additional smallest eigenvector for SGC takes $O(m+n)$ operations via power iteration \cite{CPY_16KDDMLG}, since the number of nonzero entries in the graph Laplacian matrix $\bL$ is $m+n$. The second contribution is RIM parameter estimation - estimating the RIM parameters $\{p_{ij}\}$ and $\Wbar$ takes $O(m)$ operations since they only depend on the number of edges and edge weights. Estimating  $\pLB$ takes $O(K(m+n) \cdot K)=O(K^2(m+n))$ operations for computing the least partial eigenvalue sum among $K$ clusters. The third contribution is  K-means clustering -  $O(nK^2)$ operations \cite{Zaki.Jr:14} for clustering $n$ data points of  dimension $K-1$ into $K$ groups.
As a result, if the AMOS algorithm outputs $K$ clusters, then the iterative process leads to total computational complexity of $O(K^3(m+n))$ operations. For large graphs one can use fast graph Laplacian linear solvers for  efficient eigenvector computation and implementation of AMOS \cite{spielman2010algorithms,livne2012lean}.

\section{Numerical Experiments}
\label{sec_num}
\subsection{Validation of phase transition in simulated graphs}
\begin{figure}[t]
	\centering
	\begin{subfigure}[b]{0.7\linewidth}
		\centering
		\includegraphics[width=\textwidth]{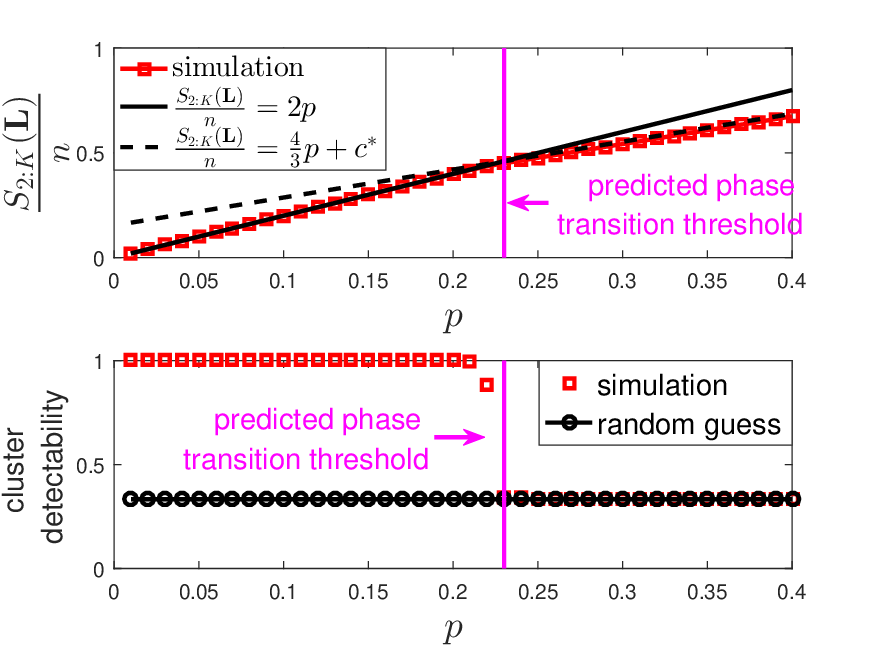}
		\caption{Phase transition in normalized partial sum of eigenvalues $\frac{\SK(\bL)}{n}$ and cluster detectability.}
		\label{Fig_K3_SBM_8000_2000}
	\end{subfigure}%
	\hspace{0.1cm}
	\\
	\centering
	\begin{subfigure}[b]{0.72\linewidth}
		\centering
		\includegraphics[width=\textwidth]{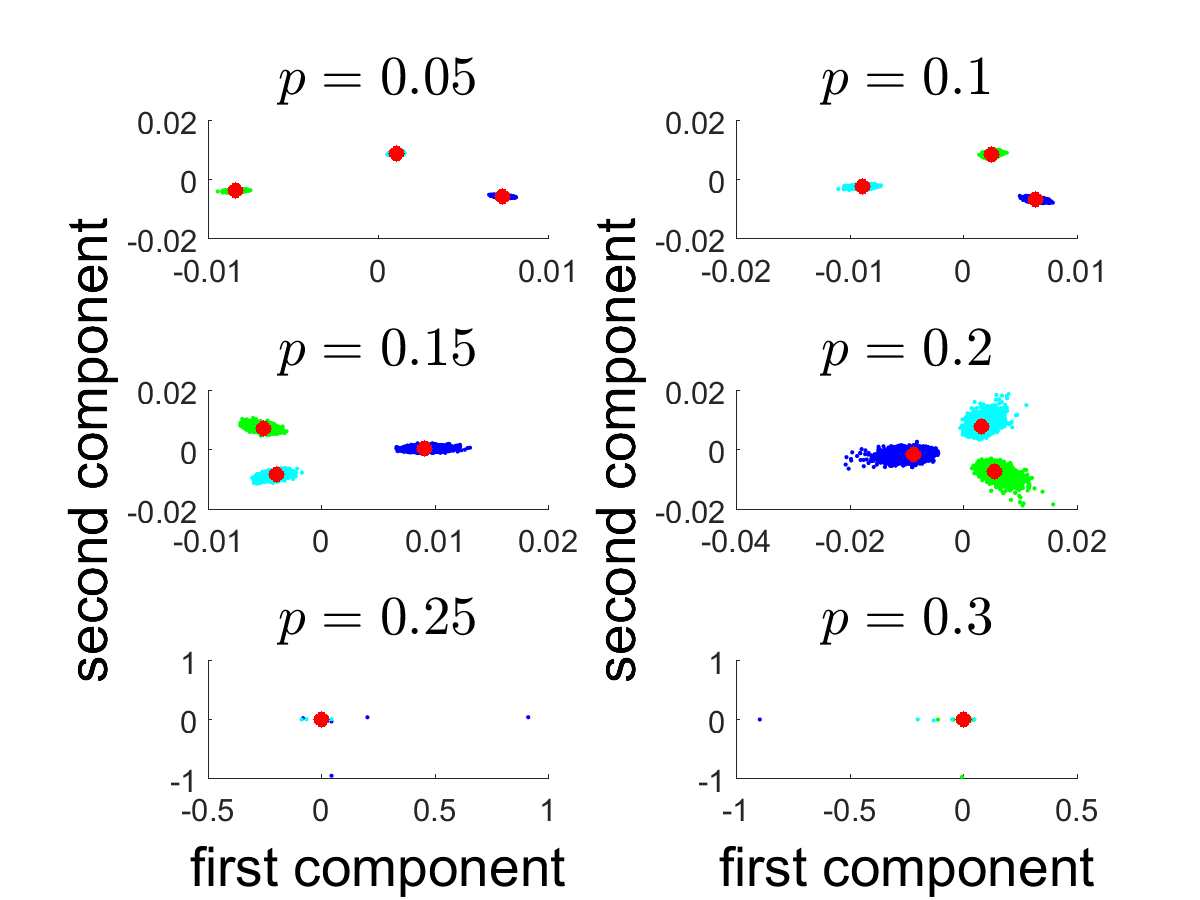}
		\caption{Row vectors in $\bY$ with respect to different $p$. Colors and red solid circles represent clusters and  cluster-wise centroids.}
		\label{Fig_8000_2000_K3_eigenvector}
	\end{subfigure}
	\caption{Phase transition of clusters generated by Erdos-Renyi random graphs. $K=3$, $n_1=n_2=n_3=8000$, and $p_1=p_2=p_3=0.25$.  The empirical critical phase transition threshold value predicted by Theorem \ref{thm_spec} is $p^*=0.2301$.}
	\label{Fig_K3_SBM_8000_2000_all}
	\vspace{-4mm}
\end{figure}

\begin{figure}[t]
	\centering
	\includegraphics[width=0.72\linewidth]{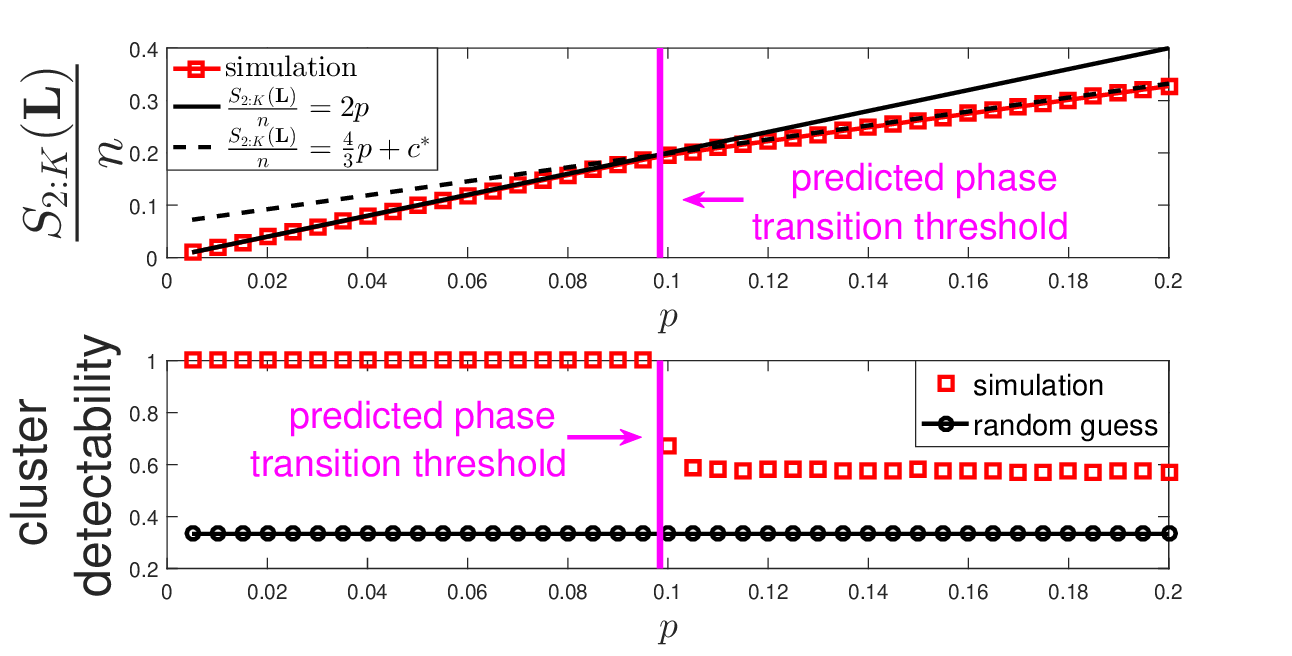}
	\caption{Phase transition of clusters generated by the Watts-Strogatz small world network model. $K=3$, $n_1=n_2=n_3=1000$, average number of neighbors $=200$, and rewire probability for each cluster is $0.4$, $0.4$, and $0.6$.
		The empirical critical threshold value predicted by Theorem \ref{thm_spec} is $p^*=0.0985$.}
	\label{Fig_K3_WS_1000_4_4_6}
	\vspace{-5mm}
\end{figure}

\begin{figure}[t]
	\centering
	\includegraphics[width=0.72\linewidth]{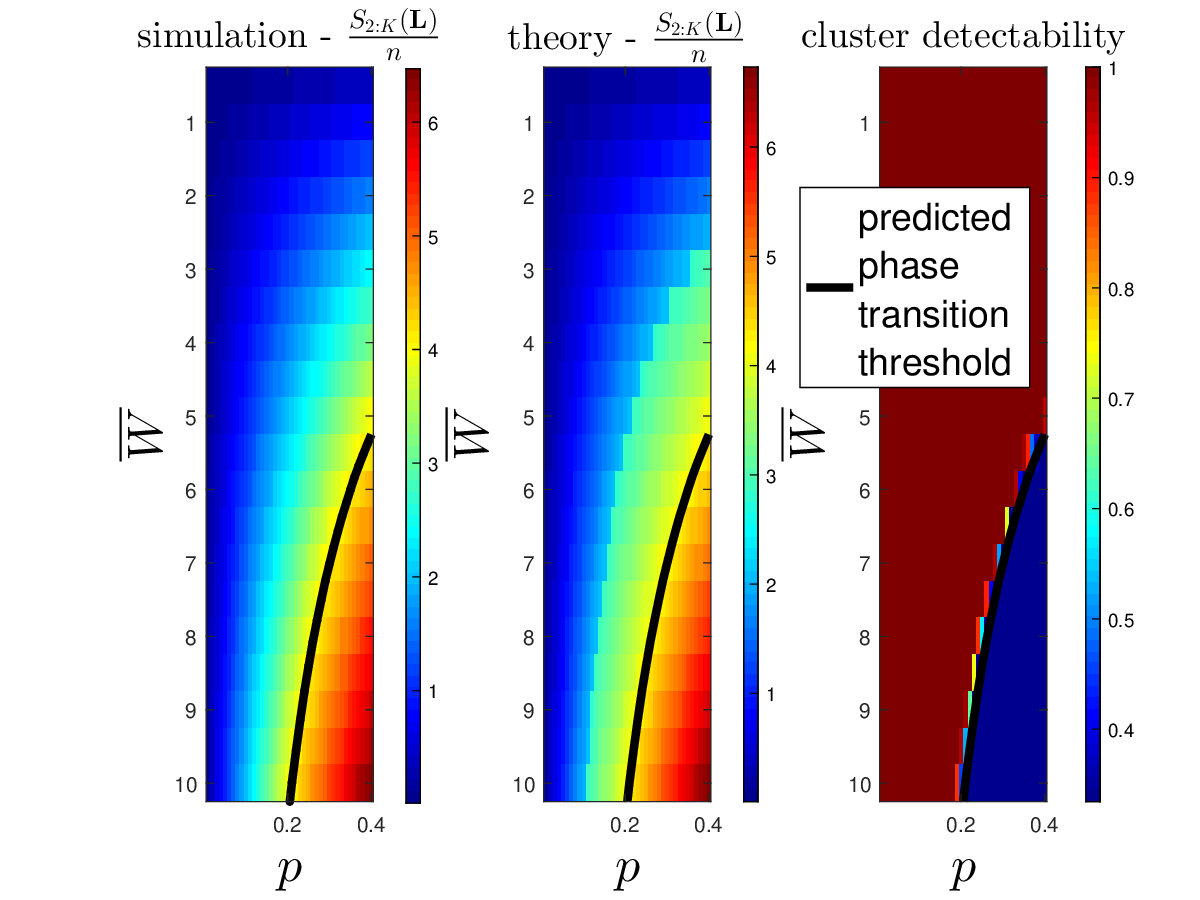}
	\caption{Phase transition of clusters generated by Erdos-Renyi random graphs with exponentially distributed edge weight with mean $10$. $K=3$, $n_1=n_2=n_3=4000$, and $p_1=p_2=p_3=0.25$. The predicted phase transition threshold curve from Theorem \ref{thm_spec_weight} is $p \cdot \Wbar= \frac{K \min_{k \in \{1,2,\ldots,K\}} \SK(\bL_k)}{(K-1)n}$.}
	\label{Fig_SBM_weighted_4000_2000}  
	\vspace{-6mm}  
\end{figure}

We simulate graphs generated by the homogeneous RIM to validate the phase transition analysis. Fig. \ref{Fig_K3_SBM_8000_2000_all} (a) shows the phase transition in partial eigenvalue sum $\SK (\bL)$ and cluster detectability (i.e., the fraction of correctly identified nodes) for clusters generated by Erdos-Renyi random graphs with varying inter-cluster edge connection probability $p$. Random guessing leads to baseline cluster detectability $\frac{1}{K}$. The simulation results verify Theorem \ref{thm_spec} that the simulated graphs transition from almost perfect detectability to low detectability and undergo a change of slope in $\SK (\bL)$ when $p$ exceeds the critical value $p^*$. In addition, the separability of the row vectors of $\bY$ in Corollary \ref{cor_separability} is demonstrated in  Fig. \ref{Fig_K3_SBM_8000_2000_all} (b). Similar phase transitions can be found for clusters generated by the Watts-Strogatz small world network model \cite{Watts98} in Fig. \ref{Fig_K3_WS_1000_4_4_6}. Fig. \ref{Fig_SBM_weighted_4000_2000} shows phase transition of weighted graphs where the inter-cluster edge weights are independently drawn from a common exponential distribution with mean $\Wbar$, which verifies the results in Theorem \ref{thm_spec_weight}. 
The effect of different cluster sizes and sensitivity to the inhomogeneous RIM are discussed in the supplementary material.

\subsection{Automated model order selection (AMOS) on real-world network data}
\label{subsec_SGC_data}

We implement the proposed AMOS algorithm (Algorithm \ref{algo_automated_clustering}) on several real-world network datasets with $\alpha=\alpha^\prime=0.05$, $\eta=10^{-5}$ and compare the clustering results with the self-tuning spectral clustering method proposed in \cite{zelnik2004self} with $K_{\max}=\lceil n / 4 \rceil$. 
Clustering results of the nonbacktracking matrix method \cite{Krzakala2013,Saade2015spectral}, the Louvain method \cite{blondel2008fast}, and the Newman-Reinert method\footnote{For the  Newman-Reinert method, we set the maximum number of clusters to be $K_{\max}=100$ and the number of Monte Carlo samples to be 10000. The final cluster assignment is obtained from the majority vote of Monte Carlo ensembles of the most probable number of clusters, as suggested in \cite{Newman16Estimate}.} \cite{Newman16Estimate} are given in the supplementary material. The details of the network datasets are summarized in Table \ref{table_data_SGC}.
Note that no information beyond network topology is used for clustering. The meta information provided by these datasets are used \emph{ex post facto} to validate the clustering results as presented in Table \ref{table_single_layer}.

\begin{table*}[t]
	\centering 
	\caption{Summary of datasets.}
	\begin{tabular}{l|l|l|l}
		\hline
		Dataset                                                   & Node               & Edge              & Ground-truth meta information                                                                \\ \hline
		IEEE reliability test system \cite{Grigg99}             & 73 power stations  & 108 power lines   & \begin{tabular}[c]{@{}l@{}}3 interconnected \\ power subsystems\end{tabular}    \\ \hline
		Hibernia Internet backbone map \cite{Knight11}          & 55 cities          & 162 connections   & \begin{tabular}[c]{@{}l@{}}city names and  \\ geographic locations
		\end{tabular}                                                        \\ \hline
		Cogent Internet backbone map \cite{Knight11}            & 197 cities         & 243 connections   & \begin{tabular}[c]{@{}l@{}}city names and \\ geographic locations\end{tabular}  \\ \hline
		Minnesota road map \cite{MATLAB_BGL}                   & 2640 intersections & 3302 roads        & geographic locations                                                            \\ \hline
	\end{tabular}
		\vspace{-4mm}
	\label{table_data_SGC}
\end{table*}

\begin{figure}[t]
	\centering
	\begin{subfigure}[b]{0.24\textwidth}
		\includegraphics[width=\textwidth]{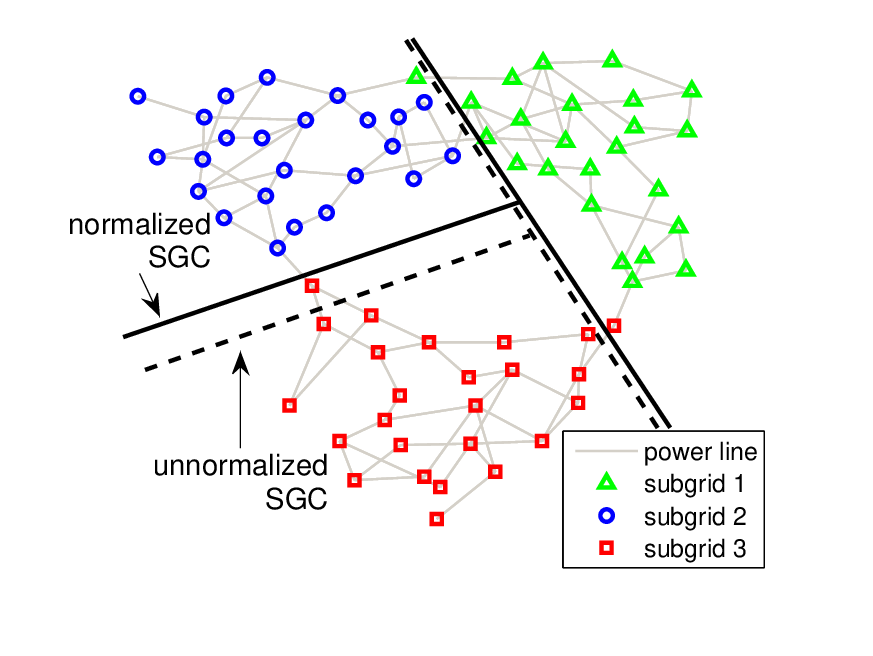}
		\vspace{-0.9cm}
		\caption{Proposed AMOS algorithm. The number of clusters is $3$.}
	\end{subfigure}%
	\hspace{0.01cm}
	\centering
	\begin{subfigure}[b]{0.24\textwidth}
		\includegraphics[width=\textwidth]{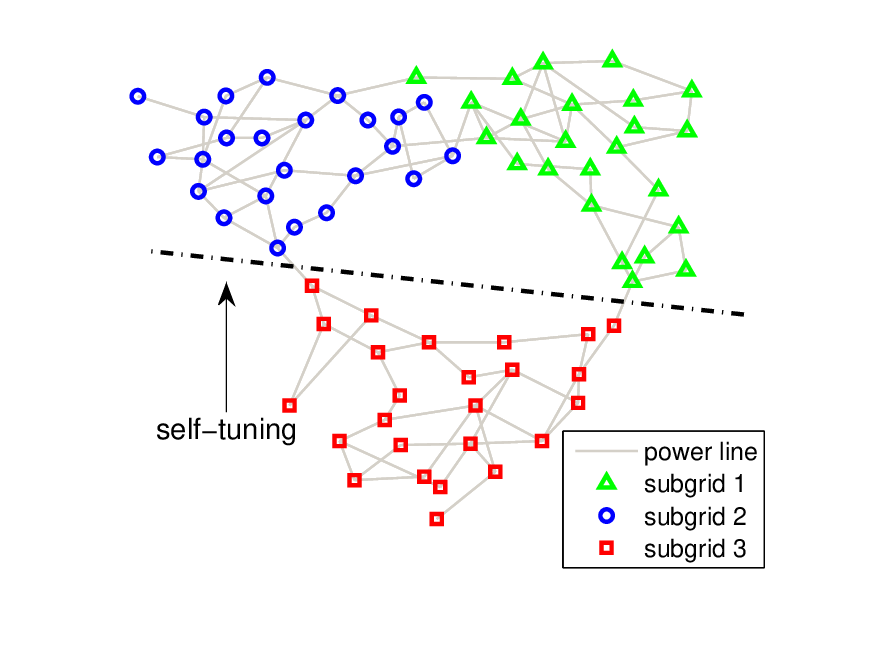}
		\vspace{-0.9cm}
		\caption{Self-tuning spectral clustering \cite{zelnik2004self}. The number of clusters is $2$.}
	\end{subfigure}
	\caption{IEEE reliability test system \cite{Grigg99}. Normalized (unnormalized) spectral graph clustering (SGC) misidentifies $2$ ($3$) nodes, whereas self-tuning spectral clustering fails to identify the third cluster.}
	\label{Fig_IEEE_RTS_SGC}
	\vspace{-2mm}
\end{figure}

Fig. \ref{Fig_IEEE_RTS_SGC} shows the clustering results of IEEE reliability test system for power system. Marker shapes represent different power subsystems. It is observed that AMOS correctly selects the number of true clusters (subsystems), and
unnormalized SGC (taking adjacency matrix as the input data) misidentifies $3$ nodes while normalized SGC (taking degree-normalized adjacency matrix as the input data) only misidentifies $2$ nodes. Self-tuning spectral clustering fails to identify the third cluster.

\begin{figure}[t!]
	\vspace{-8mm}		
	\centering
	\begin{subfigure}[b]{0.24\textwidth}
		\includegraphics[width=\textwidth]{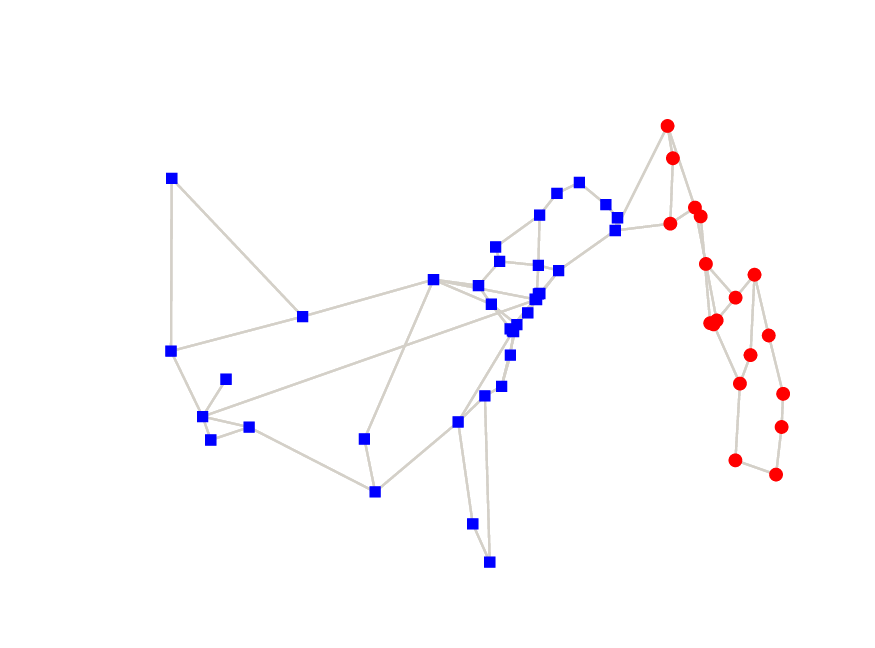}
		\vspace{-0.9cm}
		\caption{Proposed AMOS algorithm. The number of clusters is $2$.}
	\end{subfigure}%
	\hspace{0.01cm}
	\centering
	\begin{subfigure}[b]{0.24\textwidth}
		\includegraphics[width=\textwidth]{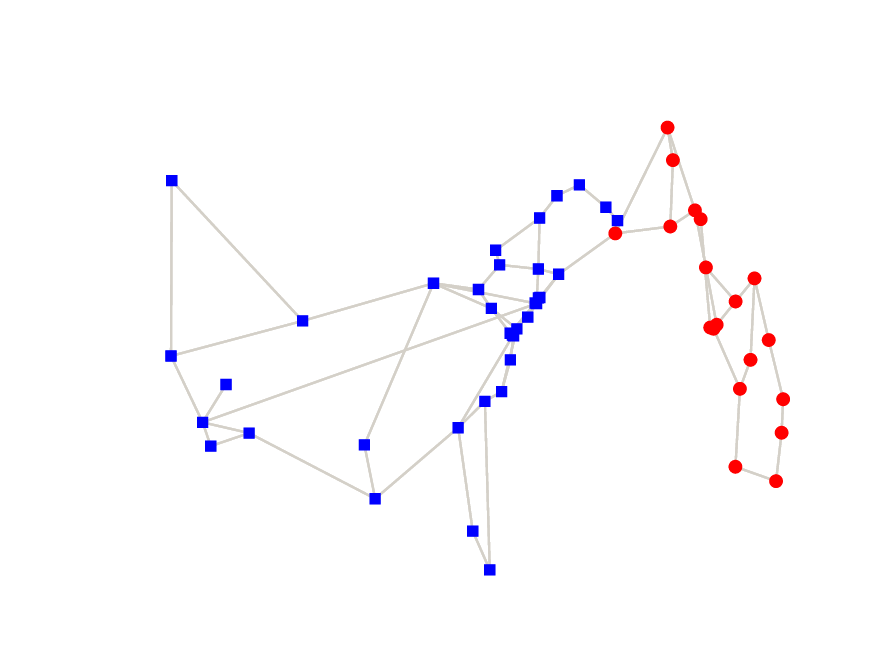}
		\vspace{-0.6cm}
		\caption{Self-tuning spectral clustering \cite{zelnik2004self}. The number of clusters is $2$.}
	\end{subfigure}
	\caption{The Hibernia Internet backbone map across Europe and North America \cite{Knight11}. Cities of different continents are perfectly clustered via automated SGC, whereas one city in North America is clustered with the cities in Europe via self-tuning spectral clustering. Automated clusters found by AMOS, including city names, can be found in the supplementary material.}
	\label{Fig_hibernia}
	\vspace{-6mm}
\end{figure}

We implement AMOS with normalized SGC for the rest of datasets, and in what follows different colors represent different automated clusters.  Fig. \ref{Fig_hibernia} shows the automated clusters of the Hibernia Internet backbone map. AMOS outputs two clusters that perfectly separates the cities in North America and Europe, whereas one city in North America is clustered with the cities in Europe via self-tuning spectral clustering.
Similar consistent clustering results  using AMOS are observed in the Cogent and Minnesota road datasets, which are discussed in the supplementary material.

\begin{table}[t]
	\centering	
	\caption{
		Clustering performance comparison. The number in the parenthesis of the Dataset (Method) column shows the number of ground-truth (identified) clusters. ``NB'' refers to the nonbacktracking matrix method, ``ST'' refers to the self-tuning method, and ``NR'' refers to the Newman-Reinert method. The notation	``-'' means ``not available'' due to lack of ground-truth cluster labels. For each dataset, the method that has the best clustering metric is highlighted in bold face. AMOS has the best or second best performance among all datasets (rows) under all  clustering metrics (columns) studied.}	
	\begin{tabular}{lllllll}
		\hline
		\multicolumn{1}{c|}{Dataset}                                                 & \multicolumn{1}{c}{Method}                                                                                & \multicolumn{1}{c}{NMI}                                                                       & \multicolumn{1}{c}{RI}                                                                        & \multicolumn{1}{c}{F}                                                                 & \multicolumn{1}{c}{C}                                                               & \multicolumn{1}{c}{NC}                                                                        \\ \hline
		\multicolumn{1}{c|}{\begin{tabular}[c]{@{}c@{}}IEEE RTS\\ (3)\end{tabular}}       & \multicolumn{1}{c}{\begin{tabular}[c]{@{}c@{}}AMOS (\textbf{3})\\ Louvain (6)\\ NB (\textbf{3})\\ ST (2) \\NR (6)\end{tabular}}      & \multicolumn{1}{c}{\begin{tabular}[c]{@{}c@{}} \textbf{.89}\\ .74\\ .75\\ .74 \\ .72\end{tabular}} & \multicolumn{1}{c}{\begin{tabular}[c]{@{}c@{}}\textbf{.96}\\ .84\\ .88\\ .78 \\.82\end{tabular}} & \multicolumn{1}{c}{\begin{tabular}[c]{@{}c@{}}\textbf{.94}\\ .67\\ .81\\ .75 \\.64\end{tabular}} & \multicolumn{1}{c}{\begin{tabular}[c]{@{}c@{}}.046\\ .144\\ .070\\ \textbf{.021} \\ .680\end{tabular}} & \multicolumn{1}{c}{\begin{tabular}[c]{@{}c@{}}.068\\ .169\\ .100\\ \textbf{.041}\\.804\end{tabular}} \\ \hline
		\multicolumn{1}{c|}{\begin{tabular}[c]{@{}c@{}}Hibernia \\ (2)\end{tabular}} & \multicolumn{1}{c}{\begin{tabular}[c]{@{}c@{}}AMOS (\textbf{2})\\ Louvain (6)\\ NB (\textbf{2})\\ ST (\textbf{2})\\ NR (\textbf{2})\end{tabular}}      & \multicolumn{1}{c}{\begin{tabular}[c]{@{}c@{}}\textbf{1.0}\\ .27\\ .73\\ .88 \\ .73\end{tabular}} & \multicolumn{1}{c}{\begin{tabular}[c]{@{}c@{}}\textbf{1.0}\\ .51\\ .89\\ .96\\ .89 \end{tabular}} & \multicolumn{1}{c}{\begin{tabular}[c]{@{}c@{}}\textbf{1.0}\\ .33\\ .90\\ .97 \\ .90 \end{tabular}} & \multicolumn{1}{c}{\begin{tabular}[c]{@{}c@{}}.030\\ .222\\ \textbf{.027}\\ .028 \\ \textbf{.027}\end{tabular}} & \multicolumn{1}{c}{\begin{tabular}[c]{@{}c@{}}.057\\ .263\\ .053\\ \textbf{.050} \\.053\end{tabular}} \\ \hline
		\multicolumn{1}{c|}{\begin{tabular}[c]{@{}c@{}}Cogent\\ (2)\end{tabular}}    & \multicolumn{1}{c}{\begin{tabular}[c]{@{}c@{}}AMOS (4)\\ Louvain (11)\\ NB (3)\\ ST (14) \\NR (3)\end{tabular}}    & \multicolumn{1}{c}{\begin{tabular}[c]{@{}c@{}}.42\\ .25\\ .26\\ .34 \\ \textbf{.48}\end{tabular}} & \multicolumn{1}{c}{\begin{tabular}[c]{@{}c@{}}.63\\ .54\\ .54\\ .55 \\\textbf{.68}\end{tabular}} & \multicolumn{1}{c}{\begin{tabular}[c]{@{}c@{}}.53\\ .26\\ .58\\ .29 \\ \textbf{.63}\end{tabular}} & \multicolumn{1}{c}{\begin{tabular}[c]{@{}c@{}}.036\\ .186\\ .073\\ .148 \\\textbf{.029}\end{tabular}} & \multicolumn{1}{c}{\begin{tabular}[c]{@{}c@{}}.049\\ .204\\ .109\\ .164 \\ \textbf{.043}\end{tabular}} \\ \hline
		\multicolumn{1}{c|}{\begin{tabular}[c]{@{}c@{}}Minnesota\\ (-)\end{tabular}} & \multicolumn{1}{c}{\begin{tabular}[c]{@{}c@{}}AMOS (46)\\ Louvain (33)\\ NB (35)\\ ST (100) \\ NR(58) \end{tabular}} & \multicolumn{1}{c}{-}                                                                         & \multicolumn{1}{c}{-}                                                                         & \multicolumn{1}{c}{-}                                                                         & \multicolumn{1}{c}{\begin{tabular}[c]{@{}c@{}}\textbf{.074}\\ .290\\ .140\\ .119 \\ .645\end{tabular}} & \multicolumn{1}{c}{\begin{tabular}[c]{@{}c@{}}\textbf{.076}\\ .299\\ .144\\ .120 \\ .661\end{tabular}} \\ \hline                                                                 
	\end{tabular}
	\label{table_single_layer}
	\vspace{-4mm}
\end{table}

In addition, comparing to the nonbacktracking matrix method \cite{Krzakala2013,Saade2015spectral}, the Louvain method \cite{blondel2008fast}, and the Newman-Reinert method \cite{Newman16Estimate},  the output clusters from the proposed AMOS algorithm are shown to be more consistent with the ground-truth meta information (see supplementary material).


For further clustering quality assessment, we use the following external and internal clustering metrics to evaluate the performance of the aforementioned automated graph clustering methods. The metrics are: (1) normalized mutual information (NMI) \cite{zaki2014data}; (2) Rand index (RI) \cite{zaki2014data}; (3) F-measure (F) \cite{zaki2014data}; (4) conductance (C) \cite{Shi00}; and (5) normalized cut (NC) \cite{Shi00}. External metrics (i.e., NMI, RI and F-measure) can be computed only when ground-truth cluster labels are known, whereas internal metrics (i.e., C and NC) can be computed in the absence of ground-truth cluster labels. For NMI, RI and F, larger value means better clustering quality. For C and NC, smaller value means better clustering quality. The definitions of the five clustering metrics are given in the supplementary material.
Table \ref{table_single_layer} summarizes the external and internal clustering metrics of the aforementioned clustering methods for the graph datasets listed in Table \ref{table_data_SGC}. 
It is observed from Table \ref{table_single_layer} that AMOS outperforms most clustering metrics for all datasets except for the Cogent dataset. For the  Cogent dataset, AMOS has comparable clustering performance to the best method for all clustering metrics.

\section{Conclusion}
\label{sec_conclusion}
This paper establishes a framework for automated model order selection (AMOS) for spectral graph clustering of dense graphs, including unweighted and weighted undirected graphs. The proposed AMOS algorithm is based on a novel
phase transition analysis of spectral clustering on graphs generated by the random interconnection model (RIM) and
an empirical estimator of the critical phase transition threshold established in our theory.
Simulated graphs validate the phase transition analysis, and the output clusters of real-world network data are shown to be consistent with ground-truth meta information. 
Extensions to the cases of sparse graphs, growing clusters and connectivity models beyond the RIM will be future work.

\section*{Appendix}
\label{sec_appendix}
\appendices
\subsection{Proof of Theorem \ref{thm_impossible}}
\label{appen_unsuccess}
Based on the partitioned matrix representation of $\bA$ in 
(\ref{eqn_network_model_multi}),  define the induced graph Laplacian matrix $\bL=\bD-\bA$. In particular, the $(i,j)$-th block is an  $n_i \times n_j$ matrix $\bL_{ij}$ satisfying  
\begin{align}
\label{eqn_Laplacian_multi_simple}
\bL_{ij}=
\left\{
\begin{array}{ll}
\bL_i+\sum_{z=1,~z \neq i}^K \bD_{iz}   , & \hbox{if}~i=j, \\
-\bC_{ij}, & \hbox{if}~i \neq j,
\end{array}
\right.
\end{align}
where $\bL_i$ is the graph Laplacian matrix of $\bA_i$, $\bD_{ij}= \textnormal{diag}(\bC_{ij}\bone_{n_j})$ is the diagonal degree matrix contributed by the inter-cluster edges between clusters $i$ and $j$. 
Applying (\ref{eqn_Laplacian_multi_simple}) to (\ref{eqn_spectral_clustering_multi_formulation}), 
let $\bnu \in \mathbb{R}^{(K-1)}$ and $\bU \in \mathbb{R}^{(K-1) \times (K-1)}$ with $\bU=\bU^T$ be the Lagrange multiplier of the constraints $\bX^T \bone_n=\bzero_{K-1}$ and $\bX^T \bX= \bI_{K-1}$, respectively.  The Lagrangian function is
\begin{align}
\label{eqn_Lagrangian_multi}
\Gamma(\bX)&=\trace(\bX^T \bL \bX)-\bnu^T \bX^T \bone_n \nonumber \\
&~~~ - \trace \lb \bU (\bX^T \bX-\bI_{K-1}) \rb.
\end{align}
Let $\bY \in \mathbb{R}^{n \times (K-1)}$ be the solution of (\ref{eqn_spectral_clustering_multi_formulation}) and let $\bO$ be a matrix of zeros.
Differentiating (\ref{eqn_Lagrangian_multi}) with respect to $\bX$ and substituting $\bY$ into the equations, we obtain the optimality condition
\begin{align}
\label{eqn_Lagrangian_multi_nu}
2 \bL \bY - \bone_n \bnu^T - 2 \bY \bU = \bO,
\end{align}
Left multiplying (\ref{eqn_Lagrangian_multi_nu}) by $\bone_n^T$, we obtain
\begin{align}
\label{eqn_Lagrangian_multi_nu_2}
\bnu=\bzero_{K-1}.
\end{align}
Left multiplying (\ref{eqn_Lagrangian_multi_nu}) by $\bY^T$ and using (\ref{eqn_Lagrangian_multi_nu_2}) we have
\begin{align}
\label{eqn_Lagrangian_multi_mu}
\bU = \bY^T \bL \bY=\diag(\lambda_2(\bL),\lambda_3(\bL),\ldots,\lambda_K(\bL)),
\end{align}
which we denote by the diagonal matrix $\bLambda$.
By (\ref{eqn_spectral_clustering_multi_formulation}) we have
\begin{align}
\label{eqn_Lagrangian_multi_mu_2}
\SK(\bL)=\trace (\bU).
\end{align}

Now let $\bX=[\bX_1^T,\bX_2^T,\ldots,\bX_K^T]^T$ and $\bY=[\bY_1^T,\bY_2^T,\ldots,\bY_K^T]^T$, where $\bX_k \in \mathbb{R}^{{n_k} \times (K-1)}$ and $\bY_k \in \mathbb{R}^{{n_k} \times (K-1)}$. With (\ref{eqn_Lagrangian_multi_mu}), the Lagrangian function in (\ref{eqn_Lagrangian_multi}) can be written as
\begin{align}
\label{eqn_Lagrangian_multi_K}
\Gamma(\bX)&=\sum_{k=1}^{K} \trace( \bX_k^T \bL_k \bX_k) + \sum_{k=1}^{K} \sum_{j=1,j \neq k}^{K} \trace(\bX_k^T  \bD_{kj} \bX_k) \nonumber \\
&~~~-\sum_{k=1}^{K} \sum_{j=1,j \neq k}^{K} \trace(\bX_k^T  \bC_{kj} \bX_j)-\sum_{k=1}^{K} \trace(\bU \bX_k^T \bX_k) \nonumber \\
&~~~+\trace(\bU).
\end{align}
Differentiating (\ref{eqn_Lagrangian_multi_K}) with respect to $\bX_k$ and substituting $\bY_k$ into the equation, we obtain the optimality condition that for all $k \in \{1,2,\ldots,K\}$,
\begin{align}
\label{eqn_Lagrangian_multi_K_diff}
\bL_k \bY_k + \sum_{j=1,j \neq k}^{K} \bD_{kj} \bY_k-\sum_{j=1,j \neq k}^{K} \bC_{kj} \bY_j- \bY_k \bU=\bO.
\end{align}
Note that the matrix $\bD_{ij}$, $i  \neq j$, $i,j \in \{1,2,\ldots,K\}$, has the following property:
	\begin{align}
	\label{eqn_Degree_matrix_concentrate}
	\frac{\bD_{ij}}{n_j}= \frac{\textnormal{diag}(\bC_{ij}\bone_{n_j})}{n_j}
	\asconv p_{ij} \bI
	\end{align}
	as $n_i,n_j \ra \infty$ and $\frac{\nmin}{\nmax} \ra c>0$, where $\bI$ is the identity matrix of infinite dimension and $\textnormal{diag}(\bone_{n_i})=\bI_{n_i} \ra \bI$ as $n_i \ra \infty$. The convergence result in (\ref{eqn_Degree_matrix_concentrate}) can be proved using the fact that each entry of the vector $\bC_{ij} \bone_{n_j}$ is the sum of i.i.d. Bernoulli random variables and  $\|\frac{\bD_{ij}}{n_j} - p_{ij} \bI_{n_i}\|_2=\max_{z \in \{1,2,\ldots,n_i\}}|[\frac{\bD_{ij}}{n_j} - p_{ij} \bI_{n_i}]_{zz}|$. Specifically, by Bernstein's concentration inequality \cite{Resnick13}, $|[\frac{\bD_{ij}}{n_j} - p_{ij} \bI_{n_i}]_{zz}|$ has an exponentially decaying tail and hence by the union bound, $\|\frac{\bD_{ij}}{n_j} - p_{ij} \bI_{n_i}\|_2 \asconv 0$ as $n_i,n_j \ra \infty$. 
Using (\ref{eqn_Degree_matrix_concentrate}) and left multiplying (\ref{eqn_Lagrangian_multi_K_diff}) by $\frac{\bone_{n_k}^T}{n}$ gives
\begin{align}
\label{eqn_Lagrangian_multi_K_one}
&\frac{1}{n} \Lb \sum_{j=1,j \neq k}^{K} n_j p_{kj} \bone_{n_k}^T \bY_k-\sum_{j=1,j \neq k}^{K} n_k p_{kj} \bone_{n_j}^T\bY_j- \bone_{n_k}^T\bY_k \bU \Rb \nonumber \\
&\asconv \bzero_{K-1}^T,~\forall~k.
\end{align}

Using the relation $\bone_{n_K}^T \bY_{K}=-\sum_{j=1}^{K-1} \bone_{n_j}^T \bY_{j}$, 
(\ref{eqn_Lagrangian_multi_K_one}) can be represented as an asymptotic form of  Sylvester's equation 
\begin{align}
\label{eqn_Sylvester_form}
\frac{1}{n} \lb \bAt \bZ - \bZ \bLambda \rb \asconv \bO,
\end{align}
where $\bZ=[ \bY_{1}^T\bone_{n_1}, \bY_{2}^T \bone_{n_2},\ldots, \bY_{{K-1}}^T \bone_{n_{K-1}} ]^T \in \mathbb{R}^{(K-1) \times (K-1)}$, $\bAt$ is the matrix specified in Theorem \ref{thm_impossible}, and we use the relation $\bU=\bLambda=\diag(\lambda_2(\bL),\lambda_3(\bL),\ldots,\lambda_K(\bL))$ from (\ref{eqn_Lagrangian_multi_mu}). 
Let $\otimes$ denote the Kronecker product and let $\vectorize (\bZ) $ denote the vectorization operation of $\bZ$ by stacking the columns of $\bZ$ into a vector. (\ref{eqn_Sylvester_form}) can be represented as  
\begin{align}
\label{eqn_Sylvester_vector_form}
\frac{1}{n}(\bI_{K-1} \otimes \bAt - \bLambda \otimes \bI_{K-1}) \vectorize(\bZ) \asconv \bzero,
\end{align}
where the matrix $\bI_{K-1} \otimes \bAt - \bLambda \otimes \bI_{K-1}$ is the Kronecker sum, denoted by $\bAt \oplus -\bLambda$.
Observe that $\vectorize(\bZ) \asconv \bzero$ is always a trivial solution  to (\ref{eqn_Sylvester_vector_form}), and 
if $\bAt \oplus -\bLambda$ is non-singular (i.e., its determinant is nonzero), $\vectorize(\bZ) \asconv \bzero$ is the unique solution to (\ref{eqn_Sylvester_vector_form}). Since  $\vectorize(\bZ) \asconv \bzero$ and $\sum_{k=1}^K \bone_{n_k}^T \bY_{k}=\bzero^T_{K-1}$ imply $\bone_{n_k}^T \bY_{k} \asconv \bzero^T_{K-1}$ for all $k=1,2,\ldots,K$, the centroid $\frac{\bone_{n_k}^T \bY_{k}}{n_k}$ of each cluster in the eigenspace is asymptotically centered at the origin such that the clusters are not perfectly separable, and hence accurate clustering is impossible. Therefore, a sufficient condition for SGC under the RIM to fail is that the matrix $\bI_{K-1} \otimes \bAt - \bLambda \otimes \bI_{K-1}$ be non-singular. Moreover, using the property of the Kronecker sum that the eigenvalues of $\bAt \oplus -\bLambda$ satisfy
$\{ \lambda_\ell (\bAt \oplus -\bLambda) \}_{\ell=1}^{(K-1)^2}=\{\lambda_i(\bAt)-\lambda_j(\bLambda)\}_{i,j=1}^{K-1}$, the sufficient condition on the failure of SGC under the RIM is $\liminf_{n \ra \infty} \frac{1}{n} \min_{i,j}|\lambda_i ( \bAt )-\lambda_j (\bL)|>0$
 for all $i = 1,2,\ldots,K-1$ and $j =2,3,\ldots,K$.

\subsection{Proof of Theorem  \ref{thm_spec}}
\label{proof_thm_spec}


Following the derivations in Appendix\ref{appen_unsuccess}, since $ \bone_{n_k}^T \bY_k=-\sum_{j=1,j \neq k}^{K}\bone_{n_j}^T\bY_j$, under the homogeneous RIM (i.e., $p_{ij}=p$),
equation (\ref{eqn_Lagrangian_multi_K_one}) can be simplified to
\begin{align}
\label{eqn_Lagrangian_multi_K_one_homo}
\lb p \bI_{K-1} - \frac{\bU}{n} \rb \bY_k^T \bone_{n_k} 
\asconv \bzero_{K-1},~\forall~k.
\end{align}
Below we further divide the optimality condition in (\ref{eqn_Lagrangian_multi_K_one_homo}) into two cases based on whether $\bY_k^T \bone_{n_k} \asconv \bzero_{n_k}$ for all $k$ or not: 
\begin{align}
\label{eqn_case1}
&\text{Case 1:~} \lb p \bI_{K-1} - \frac{\bU}{n} \rb \bY_k^T \bone_{n_k} 
\asconv \bzero_{K-1},~\forall~k  \nonumber \\ &~~~~~~~~~~\textnormal{~and~} \exists~k \textnormal{~s.t.~} \lim_{n \ra \infty} \|\bY_k^T \bone_{n_k}\|_2 > 0; 
\\
\label{eqn_case2}
&\text{Case 2:~}  \bY_k^T \bone_{n_k}  \asconv  \bzero_{K-1},~\forall~k.
\end{align}
 Note that Case 1 immediately implies $ \frac{\bU}{n}  \asconv p \bI_{K-1}$, which is proved as follows. In Case 1, take a $k$ such that $\lb p \bI_{K-1} - \frac{\bU}{n} \rb \bY_k^T \bone_{n_k} 
	\asconv \bzero_{K-1}$ and $\lim_{n \ra \infty} \|\bY_k^T \bone_{n_k}\|_2 > 0$.
Left multiplying $\lb p \bI_{K-1} - \frac{\bU}{n} \rb \bY_k^T \bone_{n_k} $ by $(\bY_k^T \bone_{n_k})^T$ gives $p \|\bY_k^T \bone_{n_k}\|_2^2 - \frac{1}{n}(\bY_k^T \bone_{n_k})^T \bU \bY_k^T \bone_{n_k} \asconv 0$. Since $\lim_{n \ra \infty} \|\bY_k^T \bone_{n_k}\|_2 > 0$ and $\lim_{n \ra \infty} \frac{1}{n}(\bY_k^T \bone_{n_k})^T \bU \bY_k^T \bone_{n_k}  \geq 0$ using (\ref{eqn_Lagrangian_multi_mu}), we obtain  $\frac{\lambda_{j+1}(\bL)}{n}  \asconv p$ if $\lim_{n \ra \infty} |[\bY_k^T \bone_{n_k}]_j|>0$ for $j \in \{1,2,\ldots,K-1\}$. Moreover, to show $\frac{\bU}{n} \asconv p \bI_{K-1}$, it suffices to show $\frac{\lambda_2(\bL)}{n} \asconv p$ and $\frac{\lambda_K(\bL)}{n} \asconv p$ since from (\ref{eqn_Lagrangian_multi_mu}) $\bU$ is a diagonal matrix and its main diagonal are the second to the $K$-th smallest eigenvalue of $\bL$. Using the fact that $\sum_{k=1}^K \bY_k^T \bone_{n_k} = \bzero_{K-1}$, under Case 1 there must exist at least two asymptotically nonzero vectors in $\{\bY_k^T \bone_{n_k} \}_{k=1}^K$. Furthermore, the fact that $\sum_{k=1}^{K} \bY_k^T \bY_k=\bI_{K-1}$ ensures that for each column $j \in \{1,2,\ldots,K-1\}$ of $\bY$, there must exist some $k$ such that the $j$-th column of $\bY_k$ has some nonzero entries and hence $\lim_{n \ra \infty} |[\bY_k^T \bone_{n_k}]_j|>0$, which then implies $ \frac{\bU}{n}  \asconv p \bI_{K-1}$.
As a result, we also obtain
\begin{align}
\label{eqn_partial_eig_sum}
\frac{\SK(\bL)}{n} =\frac{\trace(\bU)}{n}\asconv (K-1)p. 
\end{align}


In Case 1, left multiplying (\ref{eqn_Lagrangian_multi_K_diff}) by $\frac{\bY_k^T}{n}$, using the fact \cite{CPY14spectral} $\frac{\|\bC_{ij}-\overline{\bC}_{ij}\|_2}{\sqrt{n_i n_j}} \asconv 0$ as $n_i,n_j \ra \infty$ and $\frac{\nmin}{\nmax} \ra c>0$, where $\overline{\bC}_{ij}=p \bone_{n_i} \bone_{n_j}^T$ when $p_{ij}=p$, and
 using (\ref{eqn_Degree_matrix_concentrate}) gives
\begin{align}
\label{eqn_spec_multi_eigvector}
&\frac{1}{n} \Lb \bY_k^T \bL_k \bY_k + \sum_{j=1,j \neq k}^{K} n_j p \bY_k^T\bY_k \right. \nonumber \\
&~~~\left.  -\sum_{j=1,j \neq k}^{K} p \bY_k^T \bone_{n_k}\bone_{n_j}^T  \bY_j - \bY_k^T\bY_k \bU \Rb \asconv \bO,~\forall~k.
\end{align}
Since $ \bone_{n_k}^T \bY_k=-\sum_{j=1,j \neq k}^{K}\bone_{n_j}^T\bY_j$,
(\ref{eqn_spec_multi_eigvector}) can be simplified as 
\begin{align}
\label{eqn_spec_multi_eigvector_2}
&\frac{1}{n} \Lb \bY_k^T \bL_k \bY_k + (n-n_k) p \bY_k^T\bY_k + 
p \bY_k^T \bone_{n_k}\bone_{n_k}^T  \bY_k \right. \nonumber \\
&~~~\left.-   \bY_k^T\bY_k \bU \Rb \asconv \bO,~\forall~k.
\end{align}
Taking the trace of (\ref{eqn_spec_multi_eigvector_2}) and using (\ref{eqn_case1}), we have
\begin{align}
\label{eqn_spec_multi_eigvector_trace}
&\frac{1}{n} \Lb \trace(\bY_k^T \bL_k \bY_k) \Rb + \frac{p}{n} \Lb \trace(\bY_k^T \bone_{n_k}\bone_{n_k}^T  \bY_k)  \right.  \nonumber \\ 
&~~~\left.  -n_k  \trace(\bY_k^T\bY_k)
\Rb \asconv 0,~\forall~k.
\end{align}
Rearranging (\ref{eqn_spec_multi_eigvector_trace}), we obtain
\begin{align}
\label{eqn_spec_multi_eigvector_trace_3}
&\frac{1}{n} \Lb \trace(\bY_k^T [\bL_k + p\bone_{n_k}\bone_{n_k}^T - p n_k \bI_{n_k}  ] \bY_k) \Rb  \asconv 0,~\forall~k.
\end{align} 
The optimality condition in (\ref{eqn_spec_multi_eigvector_trace_3}) implies that every column of $\bY_k$ is a constant vector, which is proved as follows. Let $\bz$ be a column of $\bY_k$ and decompose $\bz$ as $\bz=a_n \bone_{n_k} + b_n \bar{\bone}_{n_k}$, where $a_n,b_n \in \bbR$ and  $\bar{\bone}_{n_k} \neq \bzero_{n_k}$ is a linear combination of  all eigenvectors of $\bY_k$ except $\bone_{n_k}$. Since $\bL_k \bone_{n_k}=\bzero_{n_k}$, $\frac{1}{n}   \bz^T [\bL_k + p\bone_{n_k}\bone_{n_k}^T - p n_k \bI_{n_k}  ] \bz   \asconv 0$ implies
\begin{align}
\label{eqn_spec_multi_eigvector_trace_4}
&\frac{1}{n} \lb b_n^2 \bar{\bone}_{n_k}^T \bL_k \bar{\bone}_{n_k} + p a_n^2  n_k^2 - p a_n^2 n_k^2 - p b_n^2 n_k  \bar{\bone}_{n_k}^T\bar{\bone}_{n_k} \rb \nonumber \\
&= \frac{1}{n} b_n^2 \Lb \bar{\bone}_{n_k}^T \lb \bL_k - p n_k \bI_{n_k} \rb \bar{\bone}_{n_k}  \Rb \nonumber \\ 
& \asconv 0. 
\end{align} 
Using $\bar{\bone}_{n_k}^T \bL_k \bar{\bone}_{n_k}=\|\bar{\bone}_{n_k}\|_2^2 \cdot \frac{\bar{\bone}_{n_k}^T}{\|\bar{\bone}_{n_k}\|_2} \bL_k \frac{\bar{\bone}_{n_k}}{{\|\bar{\bone}_{n_k}\|_2}} \geq \|\bar{\bone}_{n_k}\|_2^2 \cdot \min_{\bx \in \bbR^{n_k}:~\bx^T\bx=1,~\bx^T \bone_{n_k}=0} \bx^T \bL_k \bx = \|\bar{\bone}_{n_k}\|_2^2 \cdot  \lambda_2(\bL_k) $ and the assumption that $c_2^* = \lim_{n \ra \infty} \frac{1}{n} \min_{k \in \{1,2,\ldots,K\}}\lambda_2(\bL_k)>0$, we obtain $ \lim_{n \ra \infty}\frac{1}{n}  \bar{\bone}_{n_k}^T \bL_k \bar{\bone}_{n_k} > 0 $. Furthermore, since $\bL_k \neq p n_k \bI_{n_k}$ (the graph Laplacian matrix of a connected graph cannot be a diagonal matrix) and $\bar{\bone}_{n_k} \neq \bzero_{n_k}$, we obtain $\lim_{n \ra \infty} \frac{1}{n} | \bar{\bone}_{n_k}^T \lb \bL_k - p n_k \bI_{n_k} \rb \bar{\bone}_{n_k}  | >0$. Therefore, (\ref{eqn_spec_multi_eigvector_trace_4}) implies $\lim_{n \ra \infty} \beta_n \asconv 0$, suggesting $\bz$ is indeed a constant vector. The proof is complete by extending the analysis to $\frac{1}{n} \Lb \trace(\bY_k^T [\bL_k + p\bone_{n_k}\bone_{n_k}^T - p n_k \bI_{n_k}  ] \bY_k) \Rb$, a sum of $K-1$ terms in the form of $\frac{1}{n}   \bz^T [\bL_k + p\bone_{n_k}\bone_{n_k}^T - p n_k \bI_{n_k}  ] \bz$. 

Moreover, the condition in (\ref{eqn_spec_multi_eigvector_trace_3}) implies that in Case 1,
\begin{align}
\label{eqn_spec_multi_eigenvec_conv}
\sqrt{n_k} \bY_k \asconv \bone \bone_{K-1}^T \bV_k=\Lb v^k_1 \bone,v^k_2 \bone,\ldots,v^k_{K-1} \bone \Rb,
\end{align}
where $\bV_k=\diag(v_1^k,v_2^k,\ldots,v_{K-1}^k)$ is a diagonal matrix of constants. The scaling term $\sqrt{n_k}$ is necessary because each column in the eigenvector matrix $\bY$ has unit length.

Let $\cS=\{\bX \in \mathbb{R}^{n \times (K-1)}:~\bX^T \bX= \bI_{K-1},~\bX^T \bone_n=\bzero_{K-1}\}$.
In Case 2, since $\bY_k^T \bone_{n_k}  \asconv  \bzero_{K-1}~\forall~k$, we have 
\begin{align}
\label{eqn_S2K_lower}
&\frac{\SK(\bL)}{n} 
\asconv \lim_{n_k \ra \infty,~c>0} \frac{1}{n} \cdot \min_{\bX \in \cS}
\LB   \sum_{k=1}^K \trace(\bX_k^T \bL_k \bX_k) \right.  \nonumber \\
&~~\left. + p \sum_{k=1}^K (n-n_k) \trace(\bX_k^T\bX_k)   \RB \\
& \geq \lim_{n_k \ra \infty,~c>0} \frac{1}{n} \cdot \min_{\bX \in \cS}
\LB  \sum_{k=1}^K \trace(\bX_k^T \bL_k \bX_k)  \RB  \nonumber \\
&~~~+\lim_{n_k \ra \infty,~c>0} \frac{1}{n} \cdot \min_{\bX \in \cS}
\LB p  \sum_{k=1}^K (n-n_k) \trace(\bX_k^T\bX_k) \RB \\
& =\lim_{n_k \ra \infty,~c>0} \frac{1}{n} \cdot  \min_{k \in \{1,2,\ldots,K\}}  \SK(\bL_k)  \nonumber \\
&~~~+ (K-1)p \min_{k \in \{1,2,\ldots,K\}}(1-\rho_k) \\
\label{eqn_S2K_lower_4}
& = c^* 
+ (K-1)(1-\rho_{\max})p, 
\end{align}
where $\rho_{\max}=\max_{k \in \{1,2,\ldots,K\}}\rho_k$.

Let $\cS_k=\{\bX \in \mathbb{R}^{n \times (K-1)}:~\bX_k^T \bX_k= \bI_{K-1},~\bX_j=\bO~\forall~j \neq k,~\bX^T \bone_n=\bzero_{K-1}\}$. Since $\cS_k \subseteq \cS$, in Case 2, we have
\begin{align}
\label{eqn_S2K_upper}
&\frac{\SK(\bL)}{n} 
\asconv \lim_{n_k \ra \infty,~c>0} \frac{1}{n} \cdot \min_{\bX \in \cS}
\LB \sum_{k=1}^K \trace(\bX_k^T \bL_k \bX_k)  \right.  \nonumber \\
&~~\left.+ p \sum_{k=1}^K (n-n_k) \trace(\bX_k^T\bX_k)   \RB \\
& \leq \min_{k \in \{1,2,\ldots,K\}} \lim_{n_k \ra \infty,~c>0} \frac{1}{n} \cdot  \min_{\bX \in \cS_k}
\LB  \sum_{k=1}^K \trace(\bX_k^T \bL_k \bX_k)\right.  \nonumber \\
&~~ \left.+ p \sum_{k=1}^K (n-n_k) \trace(\bX_k^T\bX_k)   \RB \\
&= \lim_{n_k \ra \infty,~c>0} \frac{1}{n} \cdot \min_{k \in \{1,2,\ldots,K\}}
\LB \SK(\bL_k) + (K-1)p   (n-n_k) \RB \\
&\leq \lim_{n_k \ra \infty,~c>0} \frac{1}{n} \cdot \min_{k \in \{1,2,\ldots,K\}}
\LB \SK(\bL_k) + (K-1)p   (n-\nmin) \RB \\
&= \lim_{n_k \ra \infty,~c>0} \frac{1}{n} \cdot \min_{k \in \{1,2,\ldots,K\}} 
\SK(\bL_k) + (K-1) (1-\rho_{\min}) p  \\
&=c^* +(K-1)(1-\rho_{\min})p,
\label{eqn_S2K_upper_4}
\end{align}
where $\rho_{\min}=\min_{k \in \{1,2,\ldots,K\}}\rho_k$.

Comparing (\ref{eqn_partial_eig_sum}) with (\ref{eqn_S2K_lower_4}) and (\ref{eqn_S2K_upper_4}), as a function of $p$ the slope of $\frac{\SK(\bL)}{n}$ changes at some critical value $p^*$ that separates Case 1 and Case 2, and by the continuity of $\frac{\SK(\bL)}{n}$ a lower bound on $p^*$ is 
\begin{align}
\label{eqn_spec_multi_LB}
\pLB&= \lim_{n_k \ra \infty,~c>0}  \frac{\min_{k \in \{1,2,\ldots,K\}} \SK(\bL_k)}{(K-1)n_{\max}} \\
& = \frac{c^*}{(K-1)\rhomax},
\end{align}
and an upper bound on $p^*$ is
\begin{align}
\label{eqn_spec_multi_UB}
\pUB&= \lim_{n_k \ra \infty,~c>0}  \frac{\min_{k \in \{1,2,\ldots,K\}} \SK(\bL_k)}{(K-1)\nmin} \\
&=  \frac{c^*}{(K-1)\rhomin}.
\end{align}

\subsection{Proof of Theorem \ref{thm_principal_angle}}
\label{proof_thm_principal_angle}
Applying the Davis-Kahan $\sin \theta$ theorem \cite{Davis70,o2013random} to the eigenvector matrices $\bY$ and $\btY$ associated with the graph Laplacian matrices $\frac{\bL}{n}$ and $\frac{\btL}{n}$, respectively, we obtain an upper bound on the distance of column spaces spanned by $\bY$ and $\btY$, which is 
$\|\sin\mathbf{\Theta}(\bY,\btY)\|_F \leq \frac{\| \bL - \btL \|_F} {n\delta}$, where  $\delta=\inf\{|x-y|: x \in \{0\} \cup [\frac{\lambda_{K+1}(\bL)}{n},\infty),~y \in [\frac{\lambda_2(\btL)}{n},\frac{\lambda_K(\btL)}{n} ]\}$. If $p<p^*$, using the fact from (\ref{eqn_case1}) that $\frac{\lambda_j(\btL)}{n} \asconv p$ for all $j \in \{2,3,\ldots,K\}$ as $n_k \ra \infty$ and $\frac{\nmin}{\nmax} \ra c >0$, the interval 
$[\frac{\lambda_2(\btL)}{n},\frac{\lambda_K(\btL)}{n} ]$ reduces to a point $p$ almost surely. Therefore,
$\delta$ reduces to $\delta_p$ as defined in Theorem \ref{thm_principal_angle}. Furthermore, if $p_{\max} \leq p^*$, then (\ref{eqn_principal_angle_bound}) holds for all $p \leq p_{\max}$. Taking the minimum of all upper bounds in (\ref{eqn_principal_angle_bound}) for $p \leq p_{\max}$ completes the theorem.

\bibliographystyle{IEEEtran}
\bibliography{IEEEabrv,CPY_ref_20170327}

\section*{Acknowledgment}
The first author would like to thank Mr. Chun-Chen Tu from the University of Michigan Ann Arbor, USA, for his help in implementing the Newman-Reinert method.

\clearpage
\section*{{\LARGE Supplementary Material for } Phase Transitions and a Model Order Selection Algorithm for Spectral Graph Clustering
\\
Authors: Pin-Yu Chen and Alfred O. Hero III
}

\appendices

\subsection{Proof of Corollary \ref{cor_separability}}
\label{proof_separability}
Recall  the eigenvector matrix $\bY=[\bY_1^T,\bY_2^T,\ldots,\bY_K^T]^T$, where $\bY_k$ is the $n_k \times (K-1)$ matrix with row vectors representing the nodes from cluster $k$.
Since $\bY^T \bY =\sum_{k=1}^K \bY_{k}^T \bY_{k} = \bI_{K-1}$, $\bY^T \bone_n=\sum_{k=1}^K \bY_k^T \bone_{n_k}=\bzero_{K-1}$, and from (\ref{eqn_spec_multi_eigenvec_conv}) when $p<p^*$ the matrix $\sqrt{n_k} \bY_k \asconv \bone \bone_{K-1}^T \bV_k=\Lb v^k_1 \bone,v^k_2 \bone,\ldots,v^k_{K-1} \bone \Rb$ as $n_k \ra \infty$ and $\frac{\nmin}{\nmax} \ra c>0$, by the fact that $\bone_{n_k} \bone_{K-1}^T \bV_k \ra \bone \bone_{K-1}^T \bV_k$ we have 
\begin{align}
\label{eqn_spec_multi_eigvec_conv_2}
\left.
\begin{array}{ll}
\sum_{k=1}^K  \bvk \bvk^T= \bI_{K-1}; \\
\sum_{k=1}^K \bvk = \bzero_{K-1},
\end{array}
\right.
\end{align}
where $\bvk=[v_1^k,v_2^k,\ldots,v_{K-1}^k]^T$ is a vector of constants. The condition in 
(\ref{eqn_spec_multi_eigvec_conv_2}) suggests that some $\bvk$ cannot be a zero vector since $\sum_{k=1}^K  {(v^k_j)}^2=1$ for all $j \in\{1,2,\ldots,K-1\}$, and from (\ref{eqn_spec_multi_eigvec_conv_2}) we have
\begin{align}
\label{eqn_spec_multi_eigvec_coeff}
\left.
\begin{array}{ll}
\sum_{k:v^k_j>0}  v^k_j = - \sum_{k: v^k_j <0}  v^k_j,\\~~~\forall~j \in\{1,2,\ldots,K-1\}; \\
\sum_{k:v^k_i v^k_j>0}  v^k_i v^k_j = - \sum_{k: v^k_i v^k_j <0}  v^k_i v^k_j,\\~~~\forall~i,j \in\{1,2,\ldots,K-1\}, i \neq j.
\end{array}
\right.
\end{align}
Lastly, using the fact that 
\begin{align}
\label{eqn_eqn_spec_multi_eigvec_all}
\sqrt{n}\bY &=\Lb \sqrt{\frac{n}{n_1}} \sqrt{n_1} \bY_1^T,\ldots,  \sqrt{\frac{n}{n_K}} \sqrt{n_K} \bY_K^T \Rb^T \\
&\asconv \Lb \sqrt{\frac{1}{\rho_1} } \bv_1 \bone^T,\ldots,\sqrt{\frac{1}{\rho_K} } \bv_K \bone^T \Rb^T
\end{align}
as $n_k \ra \infty$ for all $k$ and $\frac{\nmin}{\nmax} \ra c>0$,
we conclude the properties in Corollary \ref{cor_separability}.

\subsection{Proof of Corollary \ref{cor_pstar}}
\label{proof_cor_pstar}
If $c_n=\Omega \lb \frac{\nmax}{n} \rb$, then by Theorem \ref{thm_spec} (c) $\pLB>0$. Therefore $p^* \geq \pLB >0$. Similarly, 
if $c_n=o\lb \frac{\nmin}{n} \rb$, then by Theorem \ref{thm_spec} (c) $\pUB=0$. Therefore $p^*=0$. Finally, since 
$\SK(\bL_k)=\sum_{i=2}^K \lambda_i(\bL_k) \geq (K-1) \lambda_2(\bL_k)$ and $\SK(\bL_k)=\sum_{i=2}^K \lambda_i(\bL_k) \leq (K-1) \lambda_K(\bL_k)$, we have $(K-1)c^*_2 \leq c^* \leq (K-1)c^*_K$.
Applying these two inequalities to Theorem \ref{thm_spec} (c) gives Corollary \ref{cor_pstar} (c).

\subsection{Proof of Corollary \ref{cor_special_graph}}
\label{proof_cor_special_graph}
If cluster $k$ is a complete graph, then $\lambda_i(\bL_k)=n_k$ for $2 \leq i \leq n_k$ \cite{Mieghem10}, which implies $c^*=\min_{k \in \{1,2,\ldots,K\}} \rho_k=\rhomin$.
Therefore,  $\pLB=\frac{\rhomin}{\rhomax}=c$, and $\pUB=1$. If cluster $k$ is a star graph, then $\lambda_i(\bL_k)=1$ for $2 \leq i \leq n_k -1$ \cite{Mieghem10}, which implies $c^*=0$ and hence $c^*=o(\rhomin)$. As a result,  by Corollary \ref{cor_pstar} (b) $p^*=0$.

\subsection{Proof of (\ref{eqn_SBM_critical})}
\label{proof_eqn_SBM_critical}
If cluster $k$ is a Erdos-Renyi random graph with edge connection probability $p_k$, then $\frac{\lambda_i(\bL_k)}{n_k} \asconv p_k$ for $2 \leq i \leq n_k$ \cite{CPY14spectral} as $n_k \ra \infty$ and $\frac{\nmin}{\nmax} \ra c>0$, where $p_k$ is a constant. Therefore,  $\pLB=\frac{\min_{k \in \{1,2,\ldots,K\}} \rho_k p_k }{ \rhomax} \geq c\cdot  \min_{k \in \{1,2,\ldots,K\}} p_k$, and $\pUB=\frac{\min_{k \in \{1,2,\ldots,K\}} \rho_k p_k }{ \rhomin} 
\leq \frac{\rhomax \cdot \min_{k \in \{1,2,\ldots,K\}} p_k }{ \rhomin}=\frac{1}{c} \cdot  \min_{k \in \{1,2,\ldots,K\}} p_k$.

\subsection{Proof of Corollary \ref{cor_K2}}
\label{proof_cor_K2}
Corollary \ref{cor_K2} (a) is a direct result from Theorem \ref{thm_spec} (a), with $K=2$ and the fact that $\min{ \{a,b\}}=\frac{a+b-|a-b|}{2}$ for all $a,b \geq 0$. Corollary \ref{cor_K2} (b) is a direct result from Theorem \ref{thm_spec} (b) and Corollary \ref{cor_separability}, with the orthonormality constraints that $\yone^T \onenone + \ytwo^T \onentwo=0$ and $\yone^T \yone + \ytwo^T \ytwo=1$. Corollary \ref{cor_K2} (c) is a direct result from Corollary \ref{cor_pstar} (c), with $\max{ \{a,b\}}=\frac{a+b+|a-b|}{2}$ for all $a,b \geq 0$.

\subsection{Proof of Corollary \ref{cor_early_breakdown}}
\label{proof_cor_early_breakdown}
We first show that when $p_{\max}<p^*$, the normalized second eigenvalue of $\bL$, $\frac{\lambda_2(\bL)}{n}$, lies within the interval $[p_{\min},p_{\max}]$ almost surely as $n_k \ra \infty$ and $\frac{\nmin}{\nmax} \ra c >0$. Consider a graph generated by the inhomogeneous RIM with parameter $\{p_{ij}\}$. In  \cite{CPY14spectral} it was established that $\frac{\|\bC_{ij}-\overline{\bC}_{ij}\|_2}{\sqrt{n_i n_j}} \asconv 0$, where $\overline{\bC}_{ij}=p_{ij} \bone_{n_i} \bone_{n_j}^T$, which means that when properly normalized by $\sqrt{n_i n_j}$ the matrices $\bC_{ij}$ and $\overline{\bC}_{ij}$ asymptotically have identical singular values and singular vectors for any cluster pair $i$ and $j$ as $n_k \ra \infty$  for all $k$ and $\frac{\nmin}{\nmax} \ra c >0$. Let $\bA(p)$ be the adjacency matrix under the homogeneous RIM with parameter $p$. Then the adjacency matrix $\bA$ of the inhomogeneous RIM can be written as $\bA=\bA(p_{\min})+ \bDelta \bA$,
and  the graph Laplacian matrix associated with $\bA$ can be written as $\bL=\bL(p_{\min})+\bDelta \bL$, where $\bL(p_{\min})$ and $\bDelta \bL$ are associated with $\bA(p_{\min})$ and $\bDelta \bA$, respectively.
Let $\overrightarrow{\Delta \bA}$, $\overrightarrow{\Delta \bL}$, and  $\overrightarrow{\bL(p)}$ denote the limit of $\frac{\bDelta \bA}{n}$, $\frac{\bDelta \bL}{n}$, and $\frac{\bL(p)}{n}$, respectively.
Since $p_{\min} = \min_{i \neq j}p_{ij}$, as $n_k \ra \infty$ and $\frac{\nmin}{\nmax} \ra c >0$,
$\overrightarrow{\Delta \bA}$ is a symmetric nonnegative matrix almost surely, and $\overrightarrow{\Delta \bL}$ is a graph Laplacian matrix almost surely. By the PSD property of a graph Laplacian matrix and Corollary \ref{cor_K2} (a), we obtain $\frac{\lambda_2(\bL)}{n} \geq p_{\min}$ almost surely as $n_k \ra \infty$ and $\frac{\nmin}{\nmax} \ra c >0$. Similarly, following the same procedure we can show that $\frac{\lambda_2(\bL)}{n} \leq p_{\max}$ almost surely as $n_k \ra \infty$ and $\frac{\nmin}{\nmax} \ra c >0$. Lastly, when $p<p^*$, using the fact from (\ref{eqn_case1}) that $\frac{\lambda_j(\bL(p))}{n} \asconv p$, and $\frac{\lambda_j(\bL(p_{\min}))}{n} \leq \frac{\lambda_j(\bL)}{n} \leq \frac{\lambda_j(\bL(p_{\max}))}{n}$ almost surely  for all $j \in \{2,3,\ldots,K\}$ as $n_k \ra \infty$ and $\frac{\nmin}{\nmax} \ra c >0$, we obtain the results.

\subsection{Proof of Theorem  \ref{thm_spec_weight}}
\label{proof_thm_spec_weight}
Similar to the proof of Theorem \ref{thm_spec}, for undirected weighted graphs under the homogeneous RIM we need to show $\frac{\|\bW_{ij}-\overline{\bW}_{ij}\|_2}{\sqrt{n_i n_j}} \asconv 0$ as $n_i,n_j \ra \infty$ and $\frac{\nmin}{\nmax} \ra c >0$, where $\bW_{ij}$ is the weight matrix of inter-cluster edges between a cluster pair ($i,j$), $\Wbar$ is the mean of the common nonnegative inter-cluster edge weight distribution, and
	 $\overline{\bW}_{ij}=p \Wbar \bone_{n_i} \bone_{n_j}^T$ when $p_{ij}=p$. Equivalently, we need to show
\begin{align}
\label{eqn_C_multi_weight_conv}
\frac{\sigma_1 \lb \bW_{ij} \rb}{\sqrt{n_i n_j}}  \asconv p  \Wbar;
~ \frac{\sigma_\ell \lb \bW_{ij} \rb}{\sqrt{n_i n_j}} \asconv 0,~\forall \ell \geq 2,
\end{align}
for all $i,j \in \{1,2,\ldots,K\}$ as $n_i,n_j \ra \infty$ and $\frac{\nmin}{\nmax} \ra c >0$. By the smoothing property in conditional expectation we have the mean of $[\bW_{ij}]_{uv}$ to be
\begin{align}
\mathbb{E} [\bW_{ij}]_{uv}&=\mathbb{E} \Lb \mathbb{E} \Lb [\bW_{ij}]_{uv} [\bC_{ij}]_{uv}|[\bC_{ij}]_{uv} \Rb \Rb
\\&=\mathbb{E} [\bC_{ij}]_{uv} \mathbb{E} \Lb [\bW_{ij}]_{uv} |[\bC_{ij}]_{uv} \Rb \\
&= p  \Wbar. \nonumber
\end{align}
Let $\bDelta=\bW_{ij}-\bWbar_{ij}$, where $\bWbar_{ij}=p \Wbar \bone_{n_i} \bone_{n_j}^T$ is a matrix whose elements are the means of entries in $\bW_{ij}$. Then
$[\bDelta]_{uv}=[\bW_{ij}]_{uv}-p \Wbar$ with probability $p$ and $[\bDelta]_{uv}=-p \Wbar$ with probability $1-p$.
The Latala's theorem \cite{Latala05} states that for any random matrix $\mathbf{M}$ with statistically independent and zero mean entries, there exists a positive constant $c_1$ such that 
\begin{align}
&\mathbb{E} \Lb \sigma_1(\mathbf{M})\Rb \leq c_1 \lb \max_u\sqrt{\sum_v \mathbb{E} \Lb [\mathbf{M}]_{uv}^2 \Rb} \right.  \nonumber \\
&~~\left.+\max_v\sqrt{\sum_u \mathbb{E} \Lb [\mathbf{M}]_{uv}^2 \Rb} + \sqrt[4]{\sum_{u,v} \mathbb{E} \Lb [\mathbf{M}]_{uv}^4 \Rb}\rb.
\end{align}
It is clear that $\mathbb{E} \Lb [\bDelta]_{uv} \Rb=0$ and each entry in $\bDelta$ is independent. Substituting $\mathbf{M}=\frac{\bDelta}{\sqrt{n_i n_j}}$ into the Latala's theorem, since $p \in [0,1]$ and the common inter-cluster edge weight distribution has finite fourth moment, by the smoothing property we have $\max_u\sqrt{\sum_v \mathbb{E} \Lb [\mathbf{M}]_{uv}^2 \Rb}=O(\frac{1}{\sqrt{n_i}})$, $\max_v\sqrt{\sum_u \mathbb{E} \Lb [\mathbf{M}]_{uv}^2 \Rb}=O(\frac{1}{\sqrt{n_j}})$, and $\sqrt[4]{\sum_{u,v} \mathbb{E} \Lb [\mathbf{M}]_{uv}^4 \Rb}=O(\frac{1}{\sqrt[4]{n_i n_j}})$.
Therefore $\mathbb{E} \Lb \frac{\sigma_1 \lb \bDelta \rb}{\sqrt{n_i n_j}}  \Rb \ra 0$
for all $i,j \in \{1,2,\ldots,K\}$ as $n_i,n_j \ra \infty$ and $\frac{\nmin}{\nmax} \ra c >0$.

\vspace{-1mm}

Next we use the Talagrand's concentration theorem stated as follows. Let $g: \mathbb{R}^k \mapsto \mathbb{R}$ be a convex and Lipschitz function.
Let $\bx \in \mathbb{R}^k$ be a random vector and assume that every element of $\bx$ satisfies
$|\bx_i|  \leq \phi$ for all $i=1,2,\ldots,k$ and some constant $\phi$, with probability one. Then there exist positive constants $c_2$ and $c_3$ such that for any $\epsilon >0$,
\begin{align}
\text{Pr}\lb \left| g(\bx)-\mathbb{E} \Lb g(\bx)  \Rb \right| \geq \epsilon\rb \leq c_2 \exp \lb \frac{-c_3 \epsilon^2}{\phi^2} \rb.
\end{align}
It is well-known that the largest singular value of a matrix $\mathbf{M}$ can be represented as $\sigma_1(\mathbf{M})=\max_{\bz^T\bz=1}||\mathbf{M} \bz||_2$ \cite{HornMatrixAnalysis} so that $\sigma_1(\mathbf{M})$ is a convex and Lipschitz function. Applying the Talagrand's theorem by substituting $\mathbf{M}=\frac{\bDelta}{\sqrt{n_i n_j}}$ and using the facts that $\mathbb{E} \Lb \frac{\sigma_1 \lb \bDelta \rb}{\sqrt{n_i n_j}}  \Rb \ra 0$ and $\frac{[\bDelta]_{uv}}{\sqrt{n_i n_j}} \leq \frac{[\bW]_{uv}}{\sqrt{n_i n_j}}$, we have
\begin{align}
\text{Pr}\lb  \frac{\sigma_1 \lb \bDelta \rb}{\sqrt{n_i n_j}} \geq \epsilon \rb \leq c_2 \exp \lb -c_3 n_i n_j \epsilon^2 \rb.
\end{align}
Since for any positive integer $n_i ,n_j >0$  $n_i n_j \geq \frac{n_i +n_j}{2}$,
$\sum_{n_i,n_j} c_2 \exp \lb -c_3 n_i n_j \epsilon^2 \rb < \infty$. By Borel-Cantelli lemma \cite{Resnick13},
$\frac{\sigma_1 \lb \bDelta \rb}{\sqrt{n_i n_j}} \asconv 0$
when $n_i, n_j \ra \infty$.
Finally, a standard matrix perturbation theory result (Weyl's inequality) \cite{HornMatrixAnalysis,weyl1912asymptotische} is
$|\sigma_\ell(\bWbar_{ij}+\bDelta)-\sigma_\ell(\bWbar_{ij})| \leq \sigma_1(\bDelta)$
for all $\ell$, and as $\frac{\sigma_1 \lb \bDelta \rb}{\sqrt{n_i n_j}} \asconv 0$, we have as $n_i ,n_j \ra \infty$,
\begin{align}
\label{eqn_Talagrand}
&\frac{\sigma_1 \lb \bW_{ij} \rb}{\sqrt{n_i n_j}}=\frac{\sigma_1 \lb \bWbar_{ij}+\bDelta \rb}{\sqrt{n_i n_j}} \asconv p \Wbar;~~  \\
&\frac{\sigma_\ell \lb \bW_{ij} \rb}{\sqrt{n_i n_j}} \asconv 0,~\forall\ell \geq 2.
\end{align}
This implies that after proper normalization by $\sqrt{n_i n_j}$,
$\bW_{ij}$ and $\bWbar_{ij}$ asymptotically have  the same singular values.
Furthermore, by the Davis-Kahan $\sin \theta$ theorem \cite{Davis70,o2013random},
the singular vectors of $\frac{\bW_{ij}}{\sqrt{n_i n_j}}$ and $\frac{\bWbar_{ij}}{\sqrt{n_i n_j}}$ are close to each other in the sense that the square of inner product of their left/right singular vectors converges to $1$ almost surely when  $\frac{\sigma_1 \lb \bDelta \rb}{\sqrt{n_i n_j}}  \asconv 0$.
Therefore, after proper normalization by $\sqrt{n_i n_j}$,
$\bW_{ij}$ and $\bWbar_{ij}$ also asymptotically have  the same singular vectors.
 Lastly, following the same proof procedure in Appendix\ref{proof_thm_spec}, we  obtain Theorem \ref{thm_spec_weight}.

\subsection{Asymptotic confidence interval for the homogeneous RIM}

Here we define the generalized log-likelihood ratio test (GLRT) under the RIM for the hypothesis $H_0:p_{ij}=p~\forall i,j,i \neq j$, against its alternative hypothesis $H_1:p_{ij} \neq p$, for at least one $i,j$, $i \neq j$. Let $f^h_{ij}(x, \theta|\{\hG_k\}_{k=1}^K)$ denote the likelihood function of observing  $x$ edges between $\hG_i$ and $\hG_j$ under hypothesis $H_h$, and $\theta$ is the edge interconnection probability. 
$\hn_k$ is the number of nodes in cluster $k$, and $\hatm_{ij}$ is the number of edges between clusters $i$ and $j$. Then under the RIM
\begin{align}
\label{eqn_prob_edge}
&f_{ij}^1(\hatm_{ij},p_{ij}|\{\hG_k\}_{k=1}^K)=\binom{\hn_i \hn_j}{\hatm_{ij}} p_{ij}^{\hatm_{ij}} (1-p_{ij})^{\hn_i \hn_j -\hatm_{ij}}; \nonumber \\
&f^0_{ij}(\hatm_{ij},p|\{\hG_k\}_{k=1}^K)=\binom{\hn_i \hn_j}{\hatm_{ij}} p^{\hatm_{ij}} (1-p)^{\hn_i \hn_j -\hatm_{ij}}.\nonumber
\end{align}
Since $\hp_{ij}$ is the MLE of $p_{ij}$ under $H_1$ and $\hp$ is the MLE of $p$ under $H_0$, the GLRT statistic is 
\begin{align}
\text{GLRT} & =2 \ln \frac{\sup_{p_{ij}}  \prod_{i=1}^{K} \prod_{j>i}^{K} f_{ij}^1(\hatm_{ij},p_{ij}|\{\hG_k\}_{k=1}^K)}
{\sup_{p_{ij}=p}  \prod_{i=1}^{K} \prod_{j>i}^{K} f_{ij}^0(\hatm_{ij},p_{ij}|\{\hG_k\}_{k=1}^K)}     \nonumber                   \\
& =2 \ln \frac{ \prod_{i=1}^K \prod_{j=i+1}^K  f_{ij}^1(\hatm_{ij},\hp_{ij}|\{\hG_k\}_{k=1}^K)}{ \prod_{i=1}^K \prod_{j=i+1}^K  f^0_{ij}(\hatm_{ij},\hp|\{\hG_k\}_{k=1}^K)}                                          \nonumber      \\
& =2 \left\{ \sum_{i=1}^K \sum_{j=i+1}^K \mathbb{I}_{\{\hp_{ij} \in (0,1)\}} \Lb \hatm_{ij} \ln \hp_{ij}  \right. \right.   \nonumber \\
&~~\left.+ (\hn_i \hn_j -\hatm_{ij} ) \ln (1-\hp_{ij}) \Rb
- \lb m-\sum_{k=1}^{K}\hatm_{i} \rb  \ln \hp   \nonumber \\
&~~\left. -  \Lb \frac{1}{2}\lb n^2-\sum_{k=1}^{K} \hn_k^2 \rb
-\lb m-\sum_{k=1}^{K}\hatm_{k} \rb \Rb \ln (1-\hp) \right\}, \nonumber
\end{align}
where we use the relations that $\sum_{i=1}^K \sum_{j=i+1}^K \hatm_{ij}=m-\sum_{k=1}^{K}\hatm_k$ and $\sum_{i=1}^K \sum_{j=i+1}^K \hn_i \hn_j=\frac{n^2-\sum_{k=1}^{K} \hn_k^2}{2}$.
By the Wilk's theorem \cite{wilks1938large}, as $n_k \ra \infty~\forall~k$,  this statistic converges in law to the  chi-square distribution, denoted by $\chi^2_\nu$, with $\nu=\binom{K}{2}-1$ degrees of freedom.
Therefore, we obtain the asymptotic $100(1-\alpha) \%$ confidence interval for $p$ in (\ref{eqn_spectral_multi_confidence_interval}).

\subsection{Phase transition tests for undirected weighted graphs}
\label{proof_phast_test_weight}
Given clusters $\{\hG_k\}_{k=1}^K$ of an undirected weighted graph obtained from spectral clustering with model order $K$,
let $\hWbar$ be the average weight of the inter-cluster edges and define 
 $\hatt_{ij}=\hp_{ij} \cdot \hWbar$, $\hatt=\hp \cdot \hWbar$, $\hatt_{\max}=\hp_{\max} \cdot \hWbar$ and  $\htLB = \frac{\min_{k \in \{1,2,\ldots,K\}} \SK( \hL_k)}{(K-1)\hn_{\max}}$.
 For undirected weighted graphs, the first phase of testing the RIM assumption in the AMOS algorithm is identical to undirected unweighted graphs, i.e., the estimated local inter-cluster edge connection probabilities $\hpij$'s are used to test the RIM hypothesis. In the second phase, if the clusters pass the homogeneous RIM test (i.e., the estimate of global inter-cluster edge probability  $\hp$ lies in the confidence interval specified in (\ref{eqn_spectral_multi_confidence_interval})), then
based on the phase transition results in Theorem \ref{thm_spec_weight}, 
 the clusters pass the homogeneous phase transition test if $\hatt < \htLB$. If the homogeneous RIM test fails, then by Theorem \ref{thm_principal_angle} the clusters pass the inhomogeneous RIM test if $\hatt_{\max}$ lies in a confidence interval $[0,\psi]$ and $\psi < t^*$. 
 Moreover, since testing $\hatt_{\max} < t^*$ is equivalent to testing $\hp_{\max} < \frac{t^*}{\hWbar}$, as discussed in Sec. \ref{subsec_phase_transition_estimator}, we can verify $\psi < t^*$ by checking the condition 
\begin{align}
\prod_{i=1}^K \prod_{j=i+1}^K F_{ij} \lb \frac{\htLB}{\hWbar},\hpij \rb \geq 1-\alpha^\prime, \nonumber
\end{align}
where $\alpha^\prime$ is the precision parameter of the confidence interval.

\subsection{Additional results of phase transition in simulated networks}
\label{appen_phase_transition}
\begin{figure}[h]
	\centering
	\begin{subfigure}[b]{0.4\textwidth}
		\includegraphics[width=\textwidth]{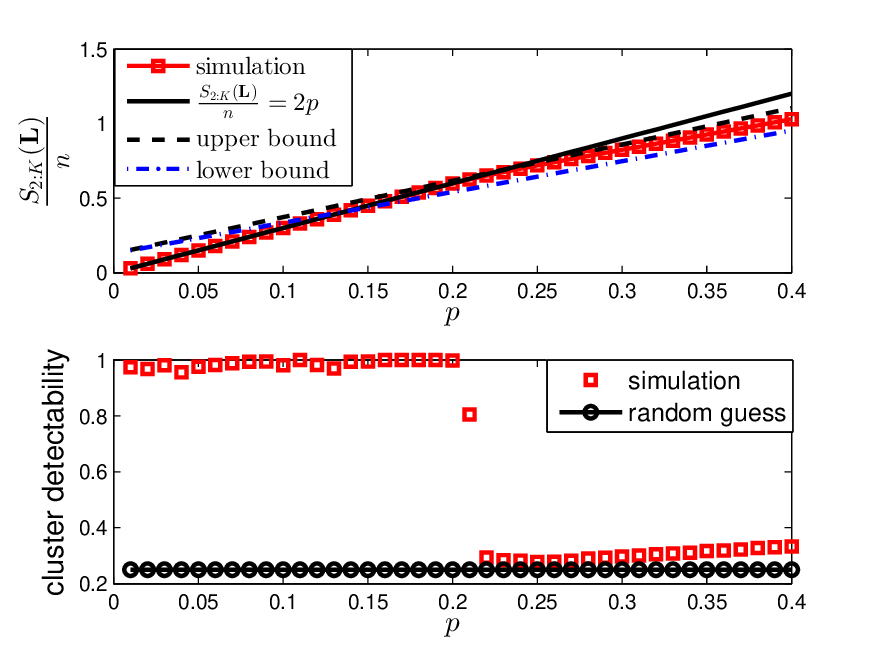}
		\caption{Phase transition in normalized partial sum of eigenvalues $\frac{\SK(\bL)}{n}$ and cluster detectability.}
		\label{Fig_SBM_6000_8000_10000}
	\end{subfigure}%
	\hspace{0.1cm}
	\centering
	\begin{subfigure}[b]{0.4\textwidth}
		\includegraphics[width=\textwidth]{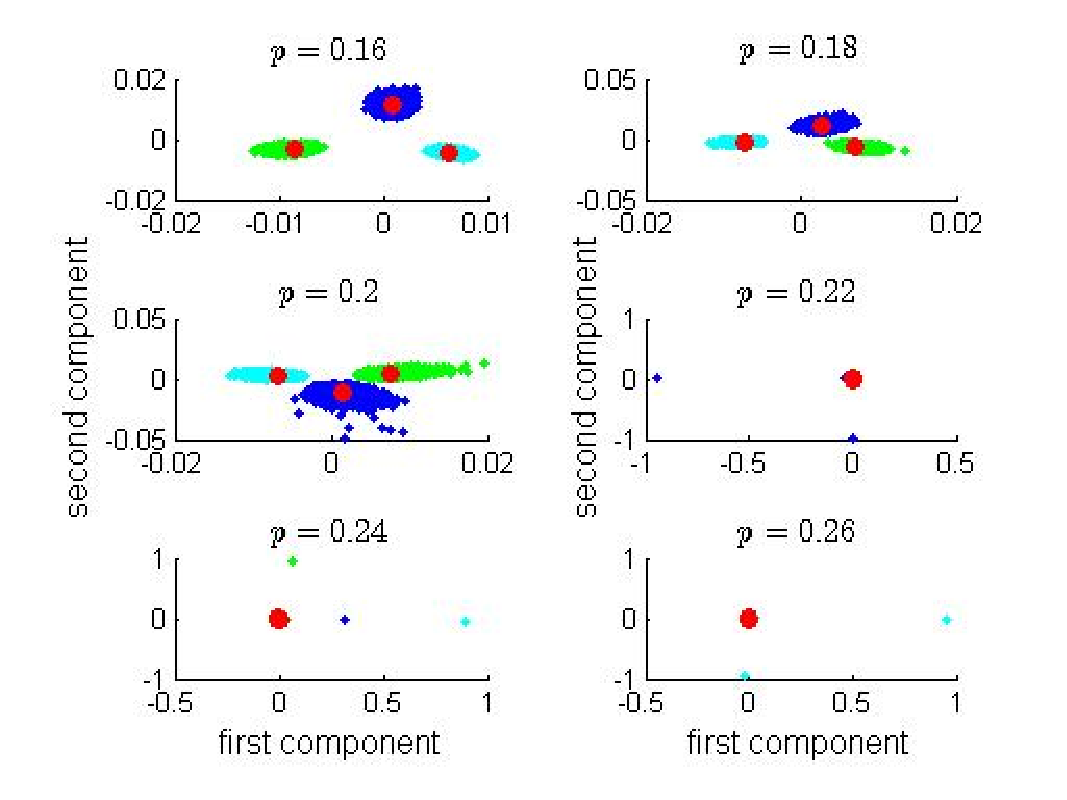}
		\caption{Row vectors in $\bY$ with respect to different $p$. Colors and red solid circles represent clusters and  cluster-wise centroids.}
		\label{Fig_6000_8000_10000_eigenvector}
	\end{subfigure}
	\caption{Phase transition of clusters generated by Erdos-Renyi random graphs. $K=3$, $(n_1,n_2,n_3)=(6000,8000,10000)$, and $p_1=p_2=p_3=0.25$.  The empirical lower bound $\pLB=0.1373$ and the empirical upper bound $\pUB=0.2288$. The results in (a) are averaged over $50$ trials.}
	\label{Fig_SBM_6000_8000_10000_all}
\end{figure}

\begin{figure}[t]
	\centering
	\begin{subfigure}[b]{0.4\textwidth}
		\includegraphics[width=\textwidth]{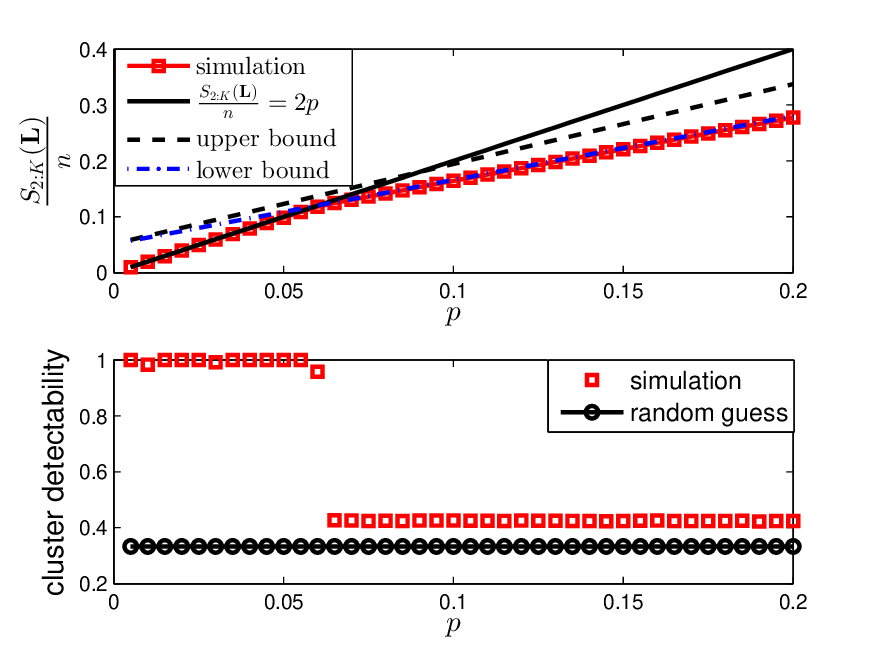}
		\caption{Phase transition in normalized partial sum of eigenvalues $\frac{\SK(\bL)}{n}$ and cluster detectability.}
		\label{Fig_spec_multi_K3_WS_1500_1000_1000}
	\end{subfigure}%
	\hspace{0.01cm}
	\centering
	\begin{subfigure}[b]{0.4\textwidth}
		\includegraphics[width=\textwidth]{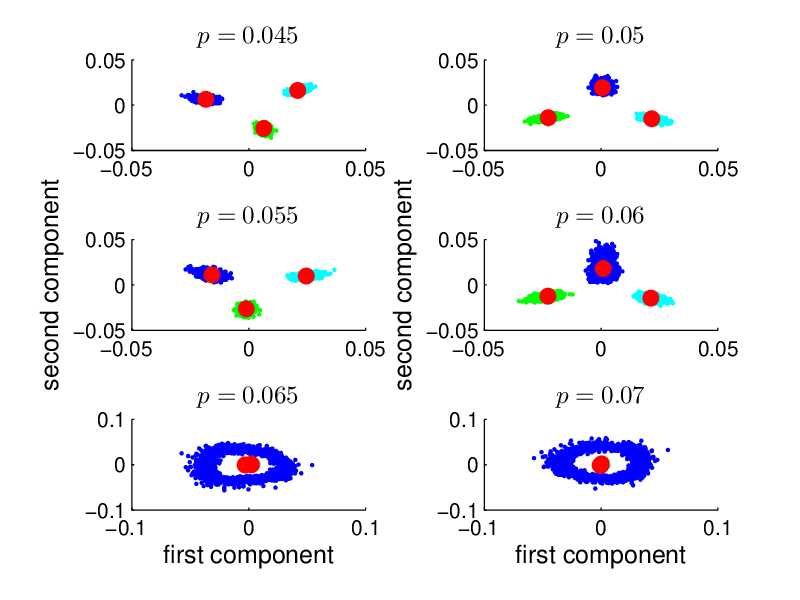}
		\caption{Row vectors in $\bY$ with respect to different $p$.  Colors and red solid circles represent clusters and  cluster-wise centroids.}
		\label{Fig_WS_K3_1500_1000_1000_eigenvector}
	\end{subfigure}
	\caption{ Phase transition of clusters generated by the Watts-Strogatz small world network model. $K=3$, $(n_1,n_2,n_3)=(1500,1000,1000)$, average number of neighbors $=200$, and rewire probability for each cluster is $0.4$, $0.4$, and $0.6$. The empirical lower and upper bounds are $\pLB=0.0602$ and $\pUB=0.0902$. The results in (a) are averaged over $50$ trials.}
	\label{Fig_spec_multi_K3_WS_1500_1000_1000_all}
\end{figure}

Fig. \ref{Fig_SBM_6000_8000_10000_all} (a) shows the phase transition in normalized partial eigenvalue sum $\frac{\SK(\bL)}{n}$ and cluster detectability for clusters generated by Erdos-Renyi random graphs with different network sizes. As predicted by Theorem \ref{thm_spec} (a), the slope of $\frac{\SK(\bL)}{n}$ undergoes a phase transition at some critical threshold value $p^*$. When $p<p^*$, $\frac{\SK(\bL)}{n}$ is exactly $2p$. When $p>p^*$, $\frac{\SK(\bL)}{n}$ is upper and lower bounded by the derived bounds. Fig. \ref{Fig_SBM_6000_8000_10000_all} (b) shows the row vectors of $\bY$ that verifies Theorem \ref{thm_spec} (b) and Corollary \ref{cor_separability}. Similar phase transition can be found for clusters generated by the Watts-Strogatz small world network model \cite{Watts98} with different cluster sizes in Fig. \ref{Fig_spec_multi_K3_WS_1500_1000_1000_all}.

\begin{figure}[t]
	\centering
	\begin{subfigure}[b]{0.24\textwidth}
		\includegraphics[width=\textwidth]{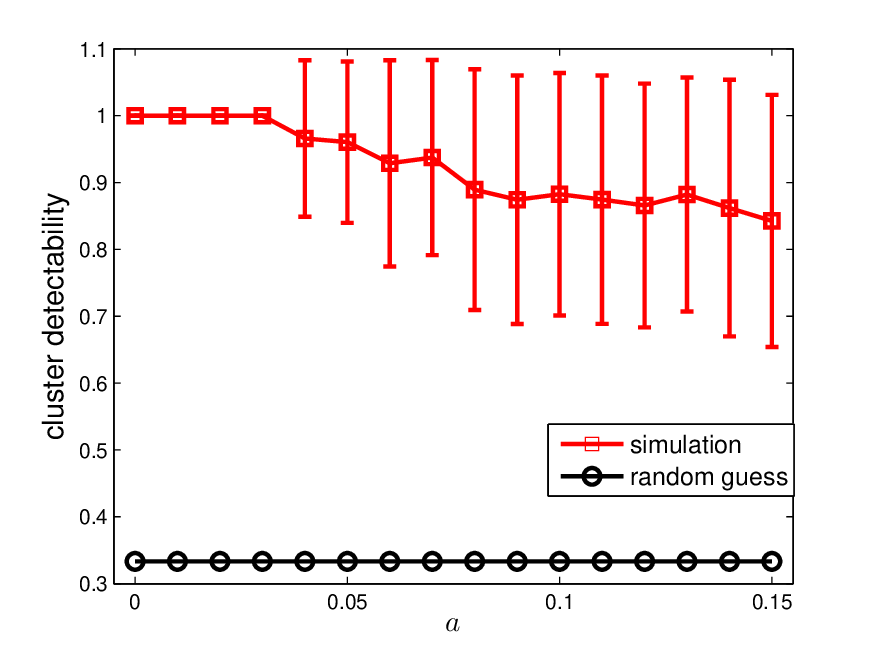}
		\caption{~}
		\label{Fig_SBM_8000_015_sensitivity}
	\end{subfigure}%
	\hspace{0.01cm}
	\centering
	\begin{subfigure}[b]{0.24\textwidth}
		\includegraphics[width=\textwidth]{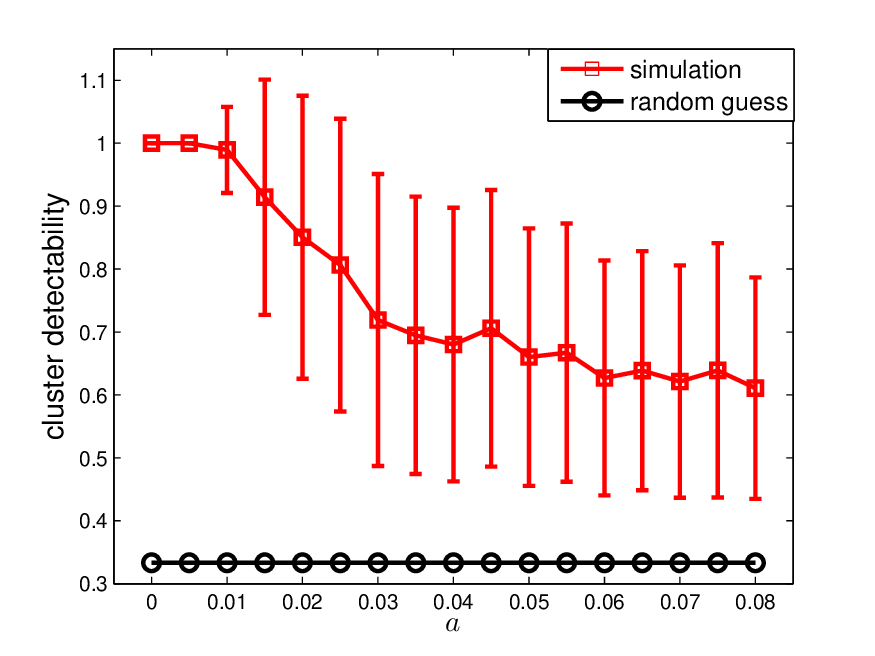}
		\caption{~}
		\label{Fig_WS_1000_008_sensitivity}
	\end{subfigure}
	\caption{Sensitivity of cluster detectability to the inhomogeneous RIM. The results are average over $50$ trials and error bars represent standard deviation. 
		(a) Clusters generated by Erdos-Renyi random graphs. $K=3$, $n_1=n_2=n_3=8000$, $p_1=p_2=p_3=0.25$, and $p_0=0.15$.
		(b) Clusters generated by the Watts-Strogatz small world network model. $K=3$, $n_1=n_2=n_3=1000$, average number of neighbors $=200$, and rewire probability for each cluster is $0.4$, $0.4$, $0.6$, and $p_0=0.08$.
	}
	\label{Fig_SGC_sensitivity}
\end{figure}

Next we investigate the sensitivity of cluster detectability to the inhomogeneous RIM. We consider the perturbation model $p_{ij}=p_0+\text{unif}(-a,a)$, where $p_0$ is the base edge connection probability and $\text{unif}(-a,a)$ is an uniform random variable with support $(-a,a)$. The simulation results in Figs. \ref{Fig_SGC_sensitivity} (a) and (b) show that almost perfect cluster detectability is still valid when $p_{ij}$ is within certain perturbation of $p_0$. The sensitivity of cluster detectability to inhomogeneous RIM also implies that if $\hp$ is within the confidence interval in (\ref{eqn_spectral_multi_confidence_interval}), then almost perfect cluster detectability can be expected.

Note that Theorem \ref{thm_impossible} also explains the effect of the perturbation model $p_{ij}=p_0+\text{unif}(-a,a)$ on cluster detectability. As $a$ increases the off-diagonal entries in $\bAt$ further deviate from $0$ and the matrix $\bAt \oplus -\bLambda$ in Appendix\ref{appen_unsuccess} gradually becomes non-singular, resulting in the degradation of cluster detectability. 
Furthermore, using Theorem \ref{thm_impossible} and the Gershgorin circle theorem \cite{HornMatrixAnalysis}, each eigenvalue of $\frac{\bAt}{n}$ lies within at least one of the closed disc centered at $\frac{[\bAt]_{ii}}{n}$ with radius $R_i$, where $R_i=\frac{n_i}{n}  \sum_{j=1,j \neq i}^{K-1} | p_{iK}-p_{ij}|$. Therefore larger inhomogeneity in $p_{ij}$ further drives the matrix $\bAt \oplus -\bLambda$ away from singularity.

\subsection{Clustering results of AMOS in the Cogent and Minnesota road datasets}
	
\begin{figure*}[t]
	\centering
	\begin{subfigure}[b]{0.49\textwidth}
		\includegraphics[width=\textwidth]{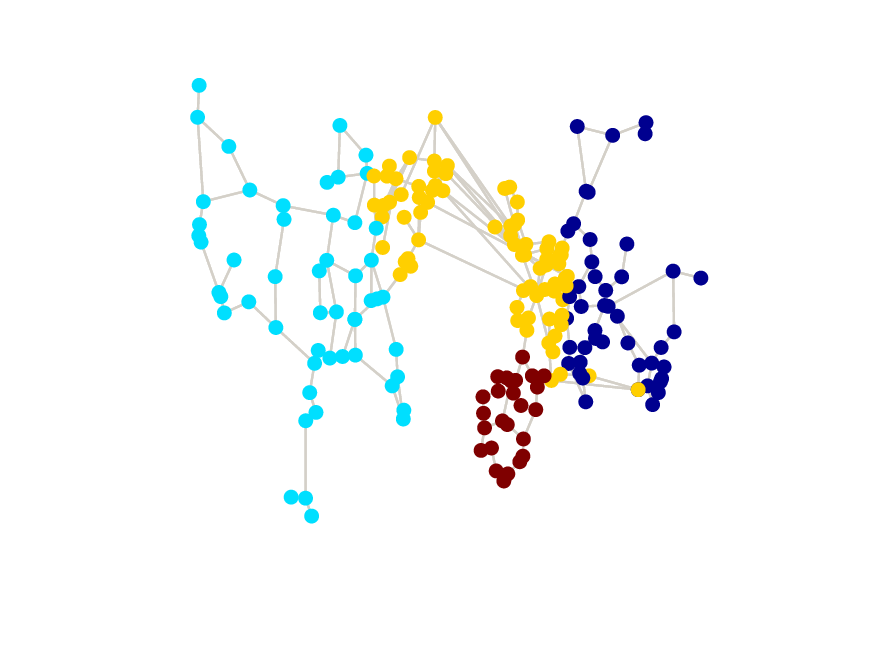}
		\vspace{-1cm}
		\caption{Proposed AMOS algorithm. The number of clusters is $4$.}
	\end{subfigure}%
	\hspace{0.01cm}
	\centering
	\begin{subfigure}[b]{0.49\textwidth}
		\includegraphics[width=\textwidth]{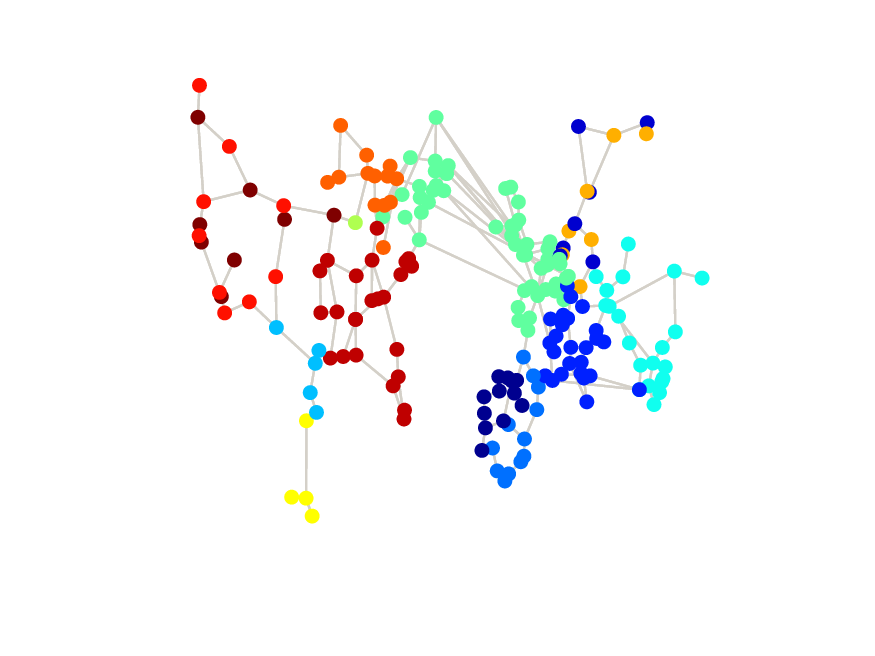}
		\vspace{-1cm}
		\caption{Self-tuning spectral clustering \cite{zelnik2004self}. The number of clusters is $14$.}
	\end{subfigure}
	\caption{The Cogent Internet backbone map across Europe and North America \cite{Knight11}. 
		Clusters from automated SGC are consistent with the geographic locations, whereas clusters from self-tuning spectral clustering are inconsistent with the geographic locations.
		Automated clusters found by AMOS, including city names, can be found in the supplementary material.}
	\label{Fig_Cogent}
\end{figure*}

	As shown in Fig. \ref{Fig_Cogent},
	the clusters of the Cogent Internet backbone map yielded by AMOS are consistent with the geographic locations except that North Eastern America and West Europe are identified as one cluster due to many transoceanic connections, wheres the clusters yielded by self-tuning spectral clustering are inconsistent with the geographic locations.

\begin{figure*}[]
	\centering
	\begin{subfigure}[b]{0.49\textwidth}
		\includegraphics[width=\textwidth]{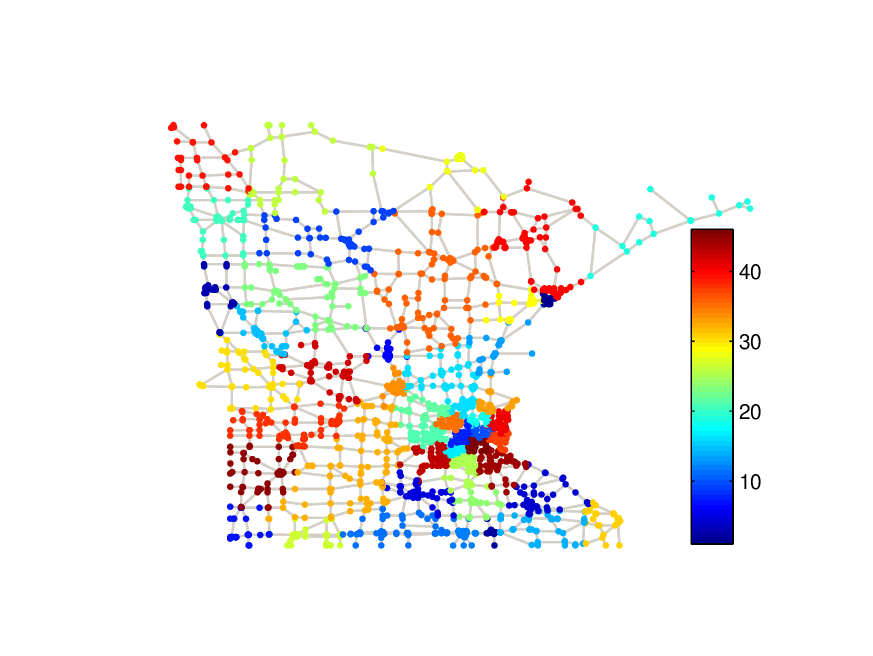}
		\vspace{-0.8cm}
		\caption{Proposed AMOS algorithm. The number of clusters is $46$.}
	\end{subfigure}%
	\hspace{0.2cm}
	\centering
	\begin{subfigure}[b]{0.49\textwidth}
		\includegraphics[width=\textwidth]{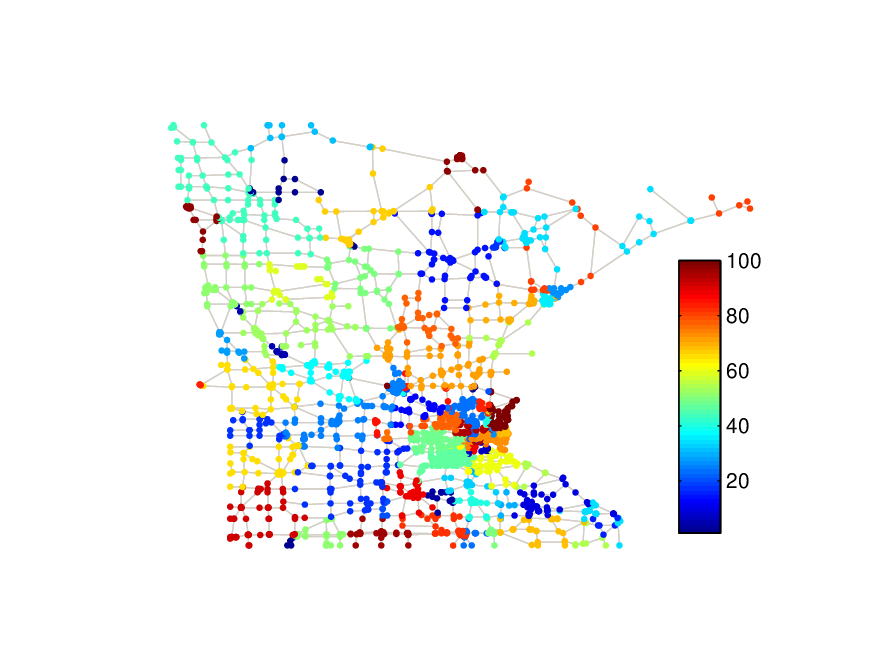}
		\vspace{-0.8cm}
		\caption{Self-tuning spectral clustering \cite{zelnik2004self}. The number of clusters is $100$.}
	\end{subfigure}
	\caption{Minnesota road map \cite{MATLAB_BGL}. Clusters from automated SGC are aligned with the geographic separations, whereas some clusters from self-tuning spectral clustering are inconsistent with the geographic separations and self-tuning spectral clustering identifies several small clusters.}
	\label{Fig_Minnesota_Road}
\end{figure*}

	Fig. \ref{Fig_Minnesota_Road} shows that the clusters of the Minnesota road map via AMOS are aligned with the geographic separations, whereas some clusters identified via self-tuning clustering are inconsistent with the geographic separations and several clusters have small sizes\footnote{For the Minnesota road map we set $K_{\max}=100$ for self-tuning spectral clustering  to speed up the computation.}.

\subsection{Performance of the Louvain method, the nonbacktracking matrix method, and the Newman-Reinert method on real-life network datasets}
\label{appen_AMOS_addition}
\begin{figure*}[t]
	\centering
	\begin{subfigure}[b]{0.49\textwidth}
		\includegraphics[width=\textwidth]{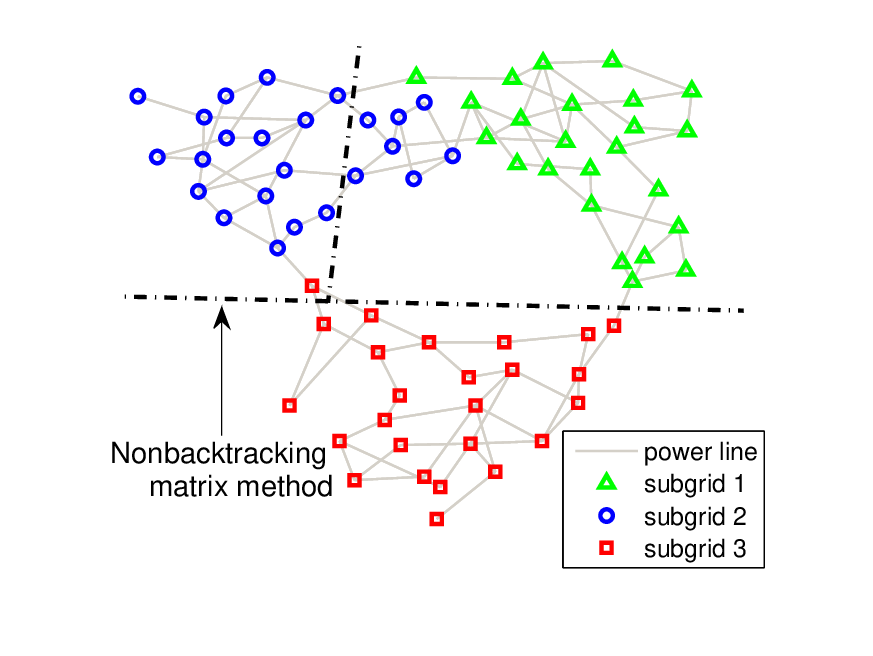}
		\caption{IEEE reliability test system. The number of clusters is $3$.}
	\end{subfigure}%
	\hspace{0.1cm}
	\begin{subfigure}[b]{0.49\textwidth}
		\includegraphics[width=\textwidth]{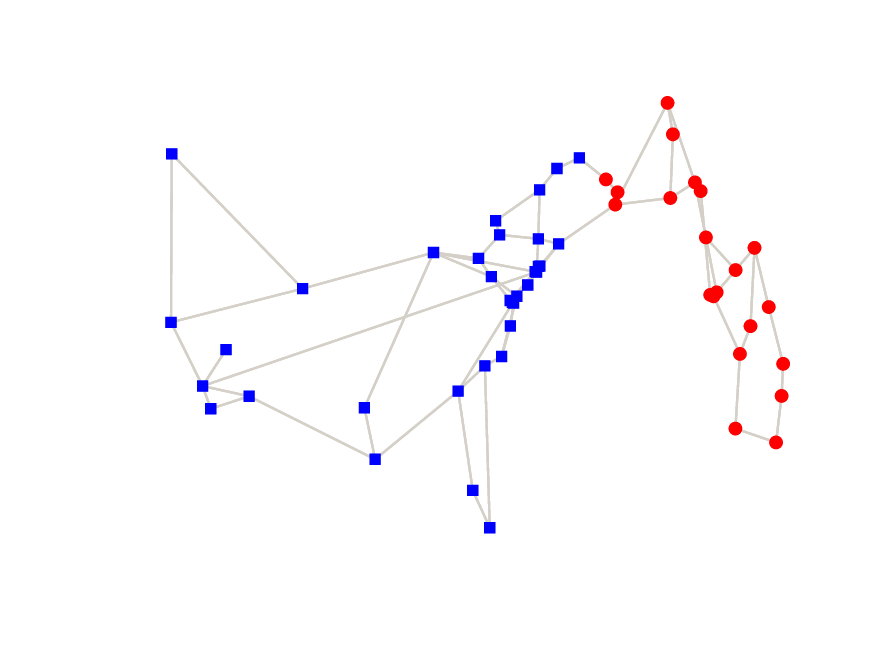}
		\caption{Hibernia Internet backbone map. The number of clusters is $2$.}
	\end{subfigure}
		\hspace{0.1cm}
		\begin{subfigure}[b]{0.49\textwidth}
			\includegraphics[width=\textwidth]{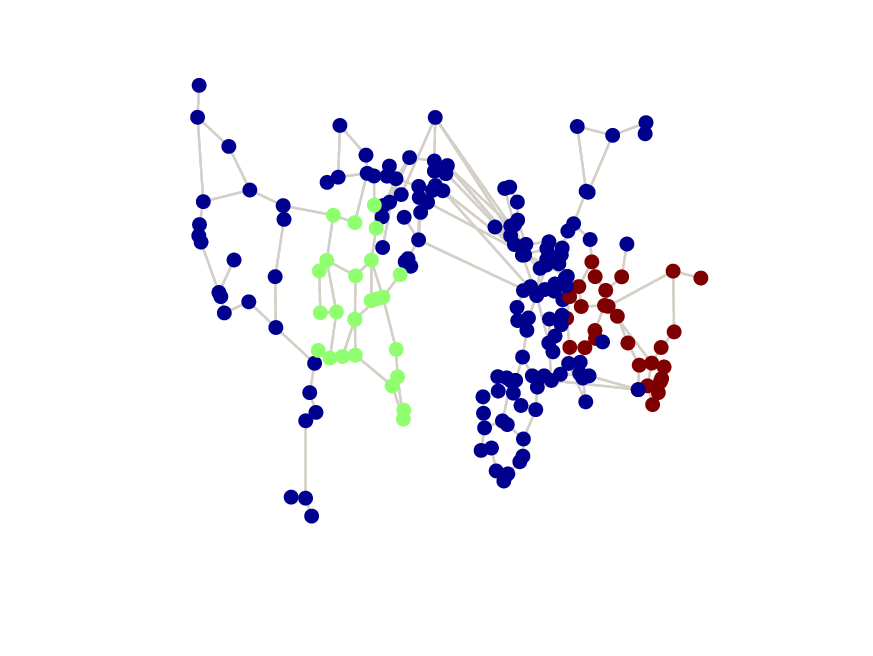}
			\caption{Cogent Internet backbone map. The number of clusters is $3$.}
		\end{subfigure}
				\hspace{0.1cm}
				\begin{subfigure}[b]{0.49\textwidth}
					\includegraphics[width=\textwidth]{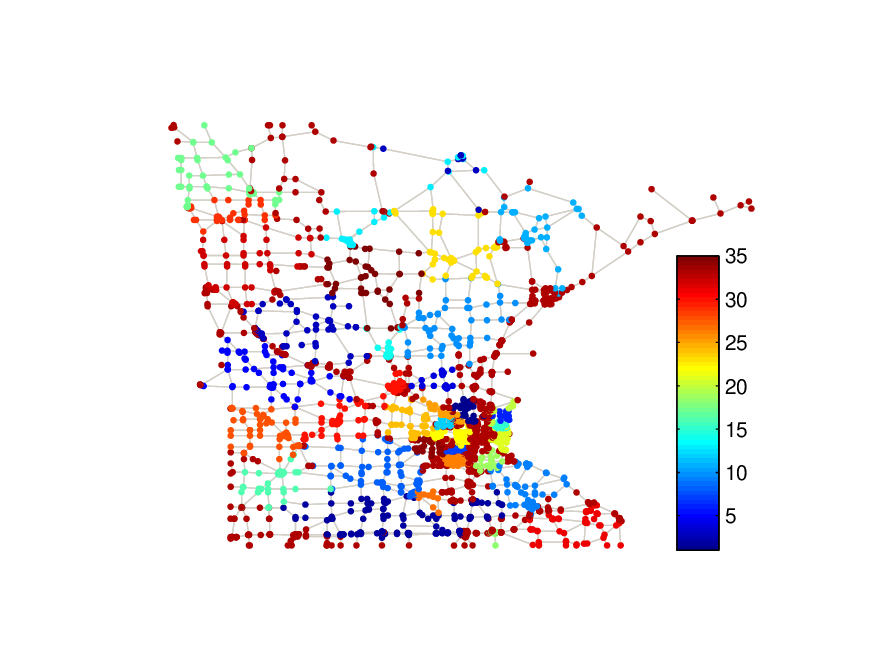}
					\caption{Minnesota road map. The number of clusters is $35$.}
				\end{subfigure}
	\caption{Clusters found with the nonbacktracking matrix method \cite{Krzakala2013,Saade2015spectral}. For IEEE reliability test system, $8$ nodes are clustered incorrectly. For Hibernia Internet backbone map, $3$ cities in the north America are clustered with the cities in Europe. For Cogent Internet backbone map, the clusters are inconsistent with the geographic locations. For Minnesota road map, some clusters are not aligned with the geographic separations.} 
	\label{Fig_NB}
\end{figure*}

\begin{figure*}[t]
	\centering
	\begin{subfigure}[b]{0.49\textwidth}
		\includegraphics[width=\textwidth]{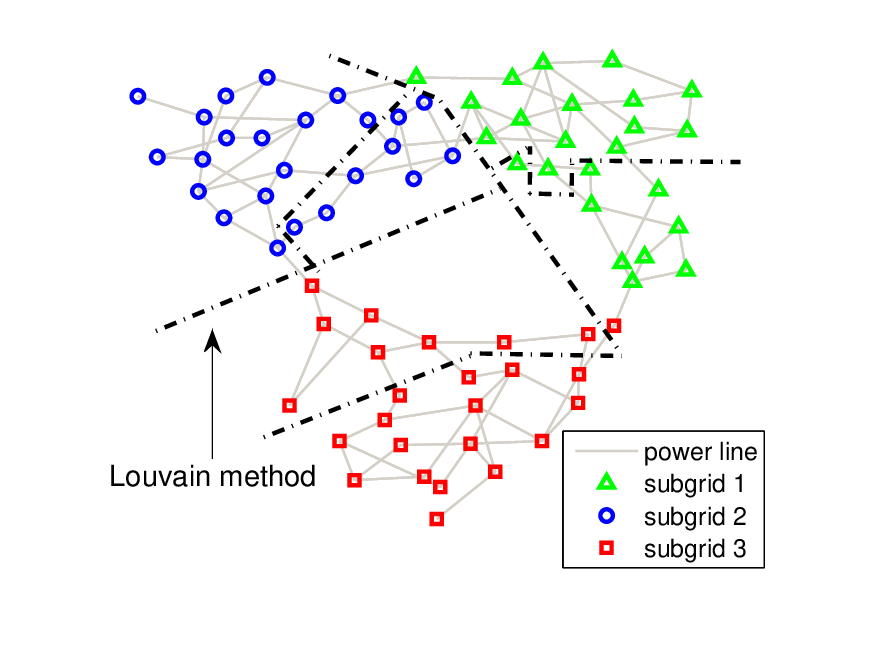}
		\caption{IEEE reliability test system. The number of clusters is $6$.}
	\end{subfigure}%
	\hspace{0.1cm}
	\begin{subfigure}[b]{0.49\textwidth}
		\includegraphics[width=\textwidth]{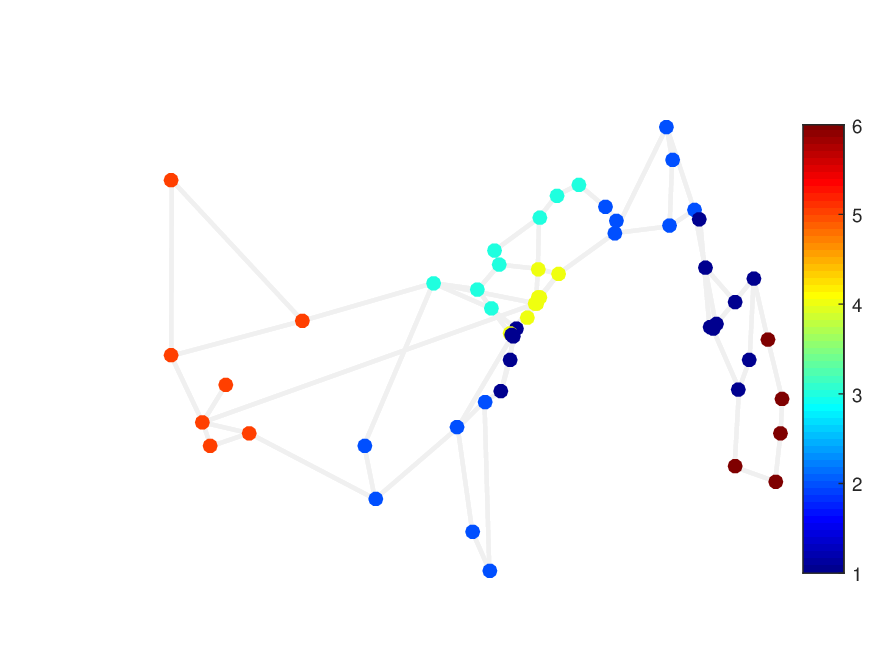}
		\caption{Hibernia Internet backbone map. The number of clusters is $6$.}
	\end{subfigure}
	\hspace{0.1cm}
	\begin{subfigure}[b]{0.49\textwidth}
		\includegraphics[width=\textwidth]{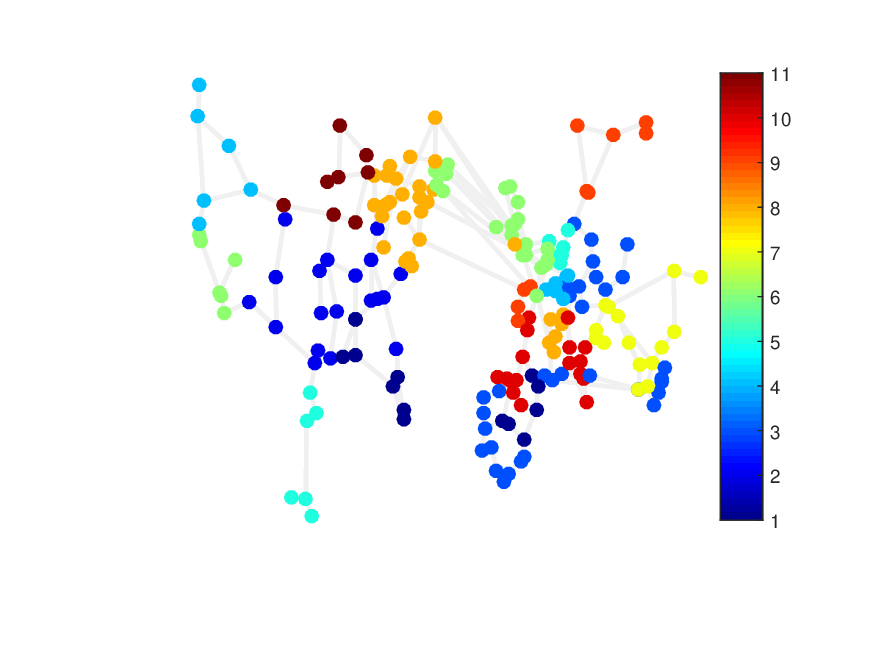}
		\caption{Cogent Internet backbone map. The number of clusters is $11$.}
	\end{subfigure}
	\hspace{0.1cm}
	\begin{subfigure}[b]{0.49\textwidth}
		\includegraphics[width=\textwidth]{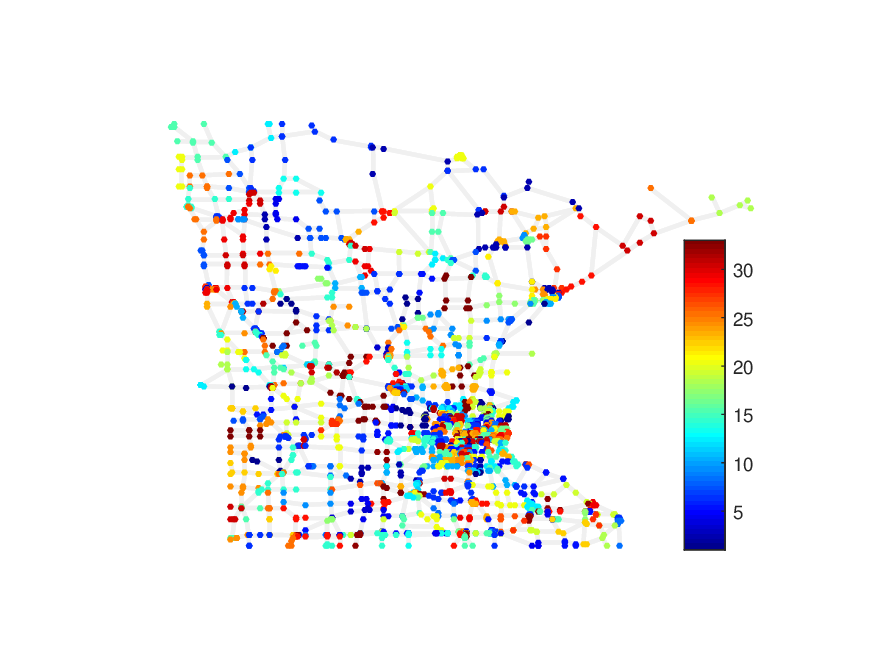}
		\caption{Minnesota road map. The number of clusters is $33$.}
	\end{subfigure}
	\caption{Clusters found with the Louvain method \cite{blondel2008fast}. For IEEE reliability test system, the number of clusters is different from the number of actual subgrids. For Hibernia and Cogent Internet backbone maps, although the clusters are consistent with the geographic locations, the Louvain method tends to identify clusters with small sizes. For Minnesota road map, the clusters are inconsistent with the geographic separations.} 
	\label{Fig_Louvain_Auto}
\end{figure*}

\begin{figure*}[t]
	\centering
	\begin{subfigure}[b]{0.49\textwidth}
		\includegraphics[width=\textwidth]{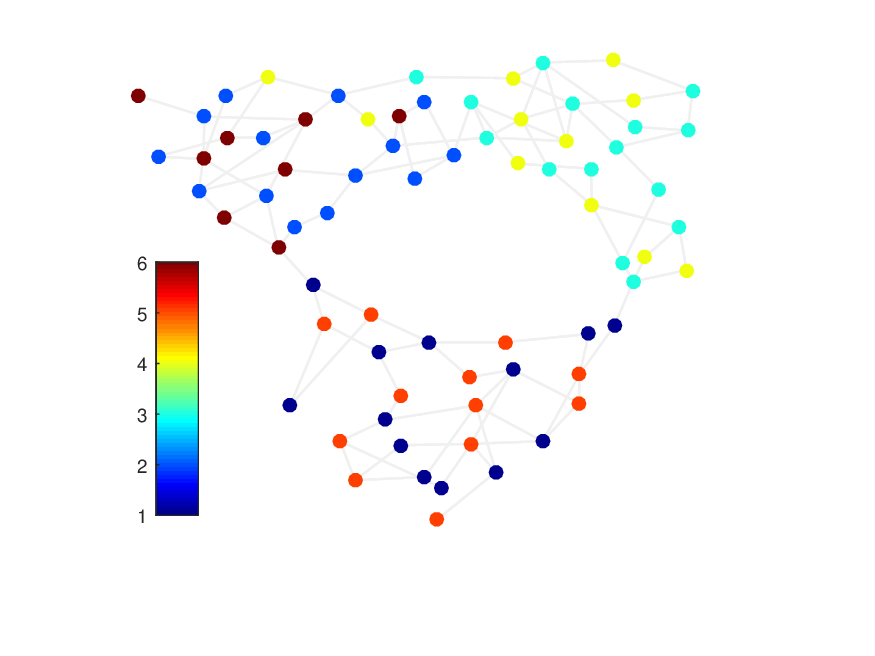}
		\caption{IEEE reliability test system. The number of clusters is $6$.}
	\end{subfigure}%
	\hspace{0.1cm}
	\begin{subfigure}[b]{0.49\textwidth}
		\includegraphics[width=\textwidth]{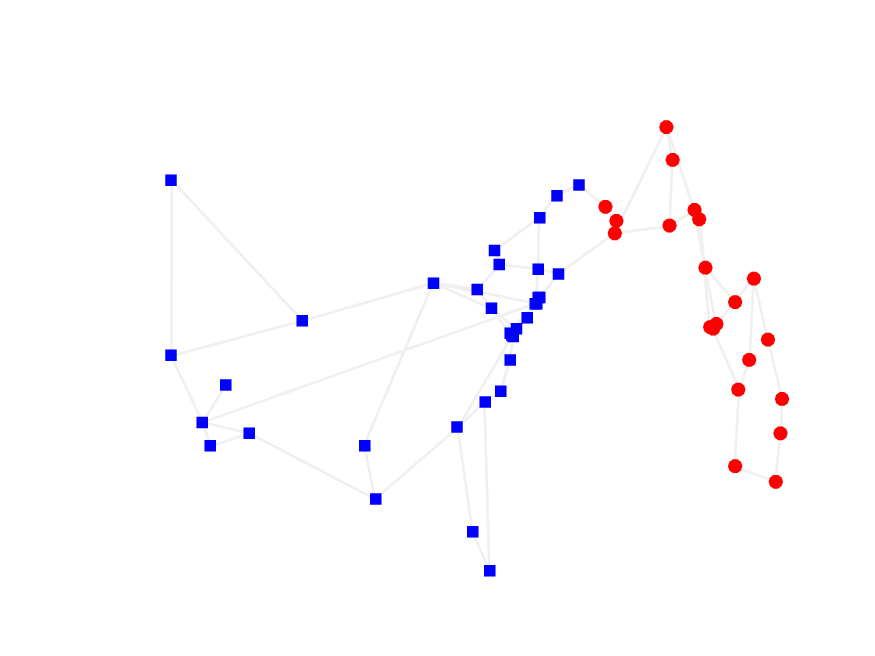}
		\caption{Hibernia Internet backbone map. The number of clusters is $2$.}
	\end{subfigure}
	\hspace{0.1cm}
	\begin{subfigure}[b]{0.49\textwidth}
		\includegraphics[width=\textwidth]{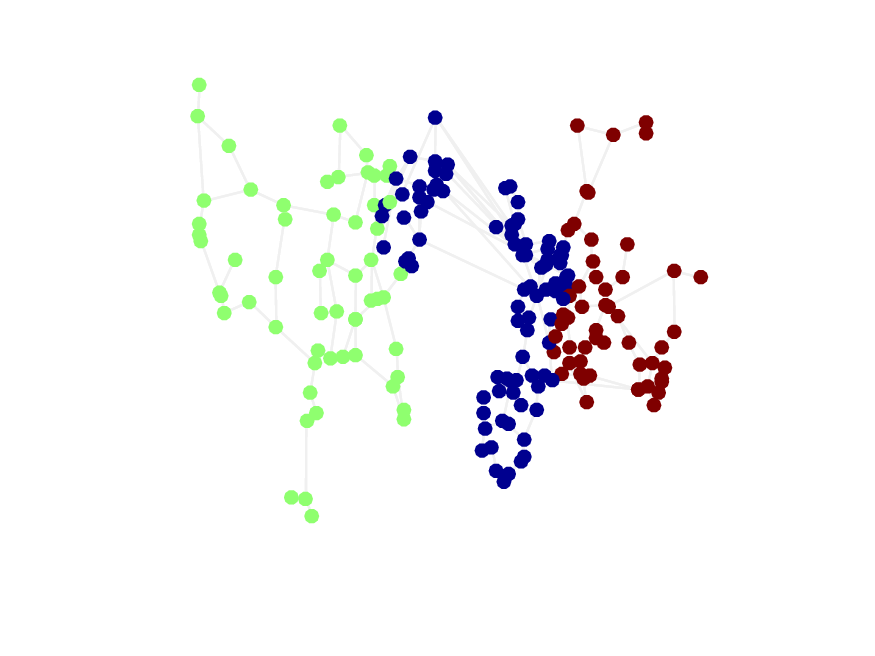}
		\caption{Cogent Internet backbone map. The number of clusters is $3$.}
	\end{subfigure}
	\hspace{0.1cm}
	\begin{subfigure}[b]{0.49\textwidth}
		\includegraphics[width=\textwidth]{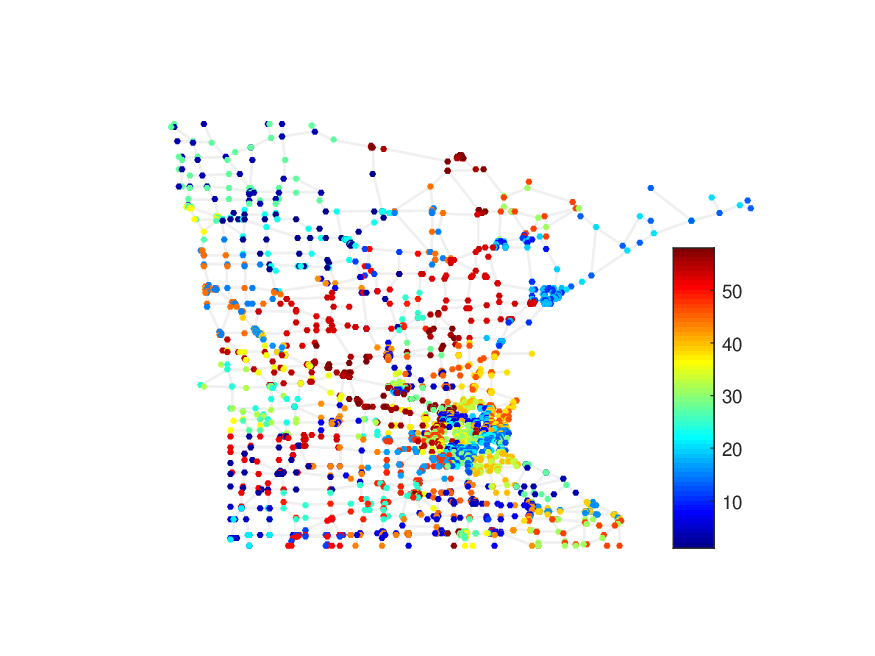}
		\caption{Minnesota road map. The number of clusters is $58$.}
	\end{subfigure}
	\caption{Clusters found with the Newman-Reinert method \cite{Newman16Estimate}. For IEEE reliability test system, the clusters are inconsistent with the actual subgrids. 
		 For Hibernia Internet backbone map, $3$ cities in the north America are clustered with the cities in Europe. 
		For Cogent Internet backbone map, the clusters are consistent with the geographic locations. For Minnesota road map, the clusters are inconsistent with the geographic separations.}
	\label{Fig_NR_Auto}
\end{figure*}

Fig. \ref{Fig_NB}, Fig. \ref{Fig_Louvain_Auto}, and Fig. \ref{Fig_NR_Auto} show the clusters of the datasets in Table \ref{table_data_SGC} identified by the nonbacktracking matrix method \cite{Krzakala2013,Saade2015spectral}, the Louvain method \cite{blondel2008fast}, and the Newman-Reinert method \cite{Newman16Estimate}, respectively. Comparing the proposed AMOS algorithm with these methods, the clusters identified by AMOS are more consistent with the ground-truth meta information provided by the datasets, except for Cogent Internet Internet backbone map. For the Cogent dataset AMOS has comparable performance to the best method (the Newman-Reinert method).

The performance of the nonbacktracking matrix method is summarized as follows. For IEEE reliability test system, $8$ nodes are clustered incorrectly. For Hibernia Internet backbone map, $3$ cities in the north America are clustered with the cities in Europe. For Cogent Internet backbone map, the clusters are inconsistent with the geographic locations. For Minnesota road map, some clusters are not aligned with the geographic separations.

The performance of the Louvain method is summarized as follows.
For IEEE reliability test system, the number of clusters is different from the number of actual subgrids. For Hibernia and Cogent Internet backbone maps, although the clusters are consistent with the geographic locations, the Louvain method tends to identify clusters with small sizes. For Minnesota road map, the clusters are inconsistent with the geographic separations.

	The performance of the Newman-Reinert method is summarized as follows. For IEEE reliability test system, $6$ clusters are identified and the clustering results are inconsistent with the ground-truth clusters.  For Hibernia Internet backbone map, $3$ cities in the north America are clustered with the cities in Europe. For Cogent Internet backbone map, the clusters are consistent with the geographic locations. For Minnesota road map, the clusters are inconsistent with the geographic separations.

\subsection{External and internal clustering metrics}
\label{subsec_clustering_metric}
	We use the following external and internal clustering metrics to evaluate the performance of different automated graph clustering methods. External metrics can be computed only when ground-truth cluster labels are known, whereas internal metrics can be computed in the absence of ground-truth cluster labels.
	In particular,  we denote the $K$ clusters identified by a graph clustering algorithm by $\{\cC_k\}_{k=1}^K$, and denote the $K^\prime$ ground-truth clusters by $\{\cC^\prime_k\}_{k=1}^{K^\prime}$.
	\\ 
	$\bullet~$\textbf{external clustering metrics} 
	\begin{enumerate}
		\item normalized mutual information (NMI) \cite{zaki2014data}: NMI is defined as 
		\begin{align}
		\textnormal{NMI}(\{\cC_k\}_{k=1}^K,\{\cC^\prime_k\}_{k=1}^{K^\prime})=\frac{2 \cdot I(\{\cC_k\},\{\cC^\prime_k\})}{|H(\{\cC_k\})+H(\{\cC^\prime_k\})|}, \nonumber
		\end{align}
		where $I$ is the mutual information between $\{\cC_k\}_{k=1}^K$ and $\{\cC^\prime_k\}_{k=1}^{K^\prime}$, and $H$ is the entropy of clusters. Larger NMI means better clustering performance.	
		\item Rand index (RI) \cite{zaki2014data}: RI is defined as 
		\begin{align}
		\textnormal{RI}(\{\cC_k\}_{k=1}^K,\{\cC^\prime_k\}_{k=1}^{K^\prime})=\frac{TP+TN}{TP+TN+FP+FN}, \nonumber
		\end{align}
		where $TP$, $TN$, $FP$ and $FN$ represent true positive, true negative, false positive, and false negative decisions, respectively. Larger RI means better clustering performance.	
		\item F-measure \cite{zaki2014data}:  F-measure is the harmonic mean of the precision and recall values for each cluster, which is defined as
		\begin{align}
		\textnormal{F-measure}(\{\cC_k\}_{k=1}^K,\{\cC^\prime_k\}_{k=1}^{K^\prime})=\frac{1}{K} \sum_{k=1}^K 	\textnormal{F-measure}_k, \nonumber
		\end{align}
		where $	\textnormal{F-measure}_k= \frac{2 \cdot PREC_k \cdot RECALL_k}{PREC_k + RECALL_k}$, and $PREC_k $ and $RECALL_k$ are the precision and recall values for cluster $\cC_k$. Larger F-measure means better clustering performance.
	\end{enumerate}		
	$\bullet~$\textbf{internal clustering metrics}	
	\begin{enumerate}
		\item conductance \cite{Shi00}: conductance  is defined as 
		\begin{align}
		\textnormal{conductance}(\{\cC_k\}_{k=1}^K)=\frac{1}{K} \sum_{k=1}^K	\textnormal{conductance}_k, \nonumber
		\end{align}
		where $\textnormal{conductance}_k=\frac{W^{out}_k}{2 \cdot W^{in}_k+W^{out}_k}$, and 
		$W^{in}_k$ and $W^{out}_k$ are the sum of within-cluster and between-cluster edge weights of cluster $\cC_k$, respectively. Lower conductance means better clustering performance.
		\item  normalized cut (NC) \cite{Shi00}:  NC is defined as 
		\begin{align}
		\textnormal{NC}(\{\cC_k\}_{k=1}^K)=\frac{1}{K} \sum_{k=1}^K	\textnormal{NC}_k, \nonumber
		\end{align}		
		where 	$\textnormal{NC}_k=\frac{W^{out}_k}{2 \cdot W^{in}_k+W^{out}_k}+\frac{W^{out}_k}{2 \cdot (W^{all}_k-W^{in}_k)+W^{out}_k}$, and 	$W^{in}_k$, $W^{out}_k$ and $W^{all}_k$ are the sum of within-cluster, between-cluster and total edge weights of cluster $\cC_k$, respectively. Lower NC means better clustering performance.
	\end{enumerate}

\clearpage

\clearpage

\clearpage

\clearpage

\begin{figure*}[!t]
	\centering
	\includegraphics[width=1\textwidth]{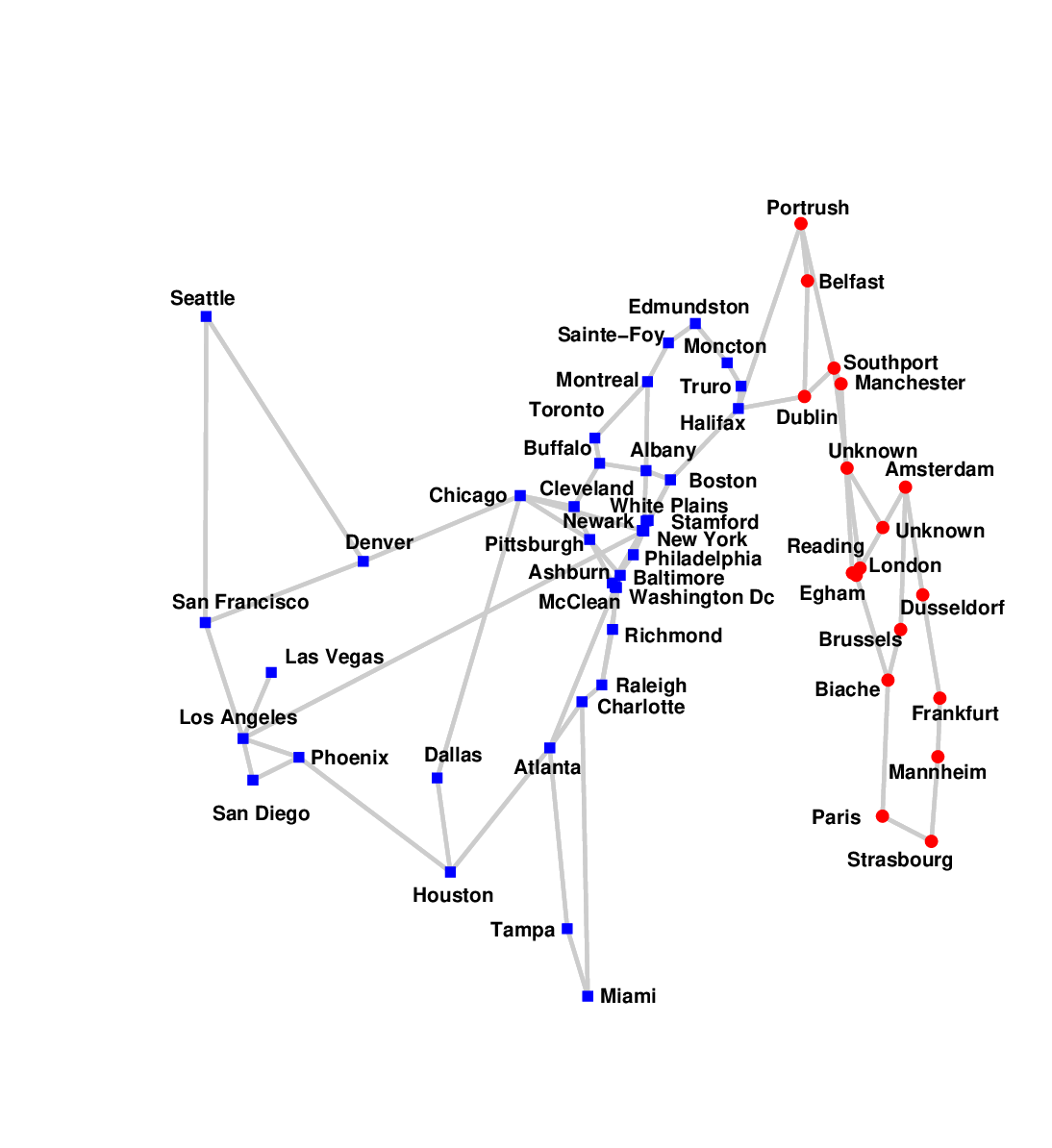}
	\caption{$2$ clusters found with the proposed automated model order selection (AMOS) algorithm for
		the Hibernia Internet backbone map with city names. The clusters are consistent with the geographic locations in the sense that one cluster contains cities in America and the other cluster contains cities in Europe. }
	\label{Fig_hibernia_name}
\end{figure*}

\begin{figure*}[!t]
	\centering
	\includegraphics[angle=90,origin=c,width=0.8\textwidth]{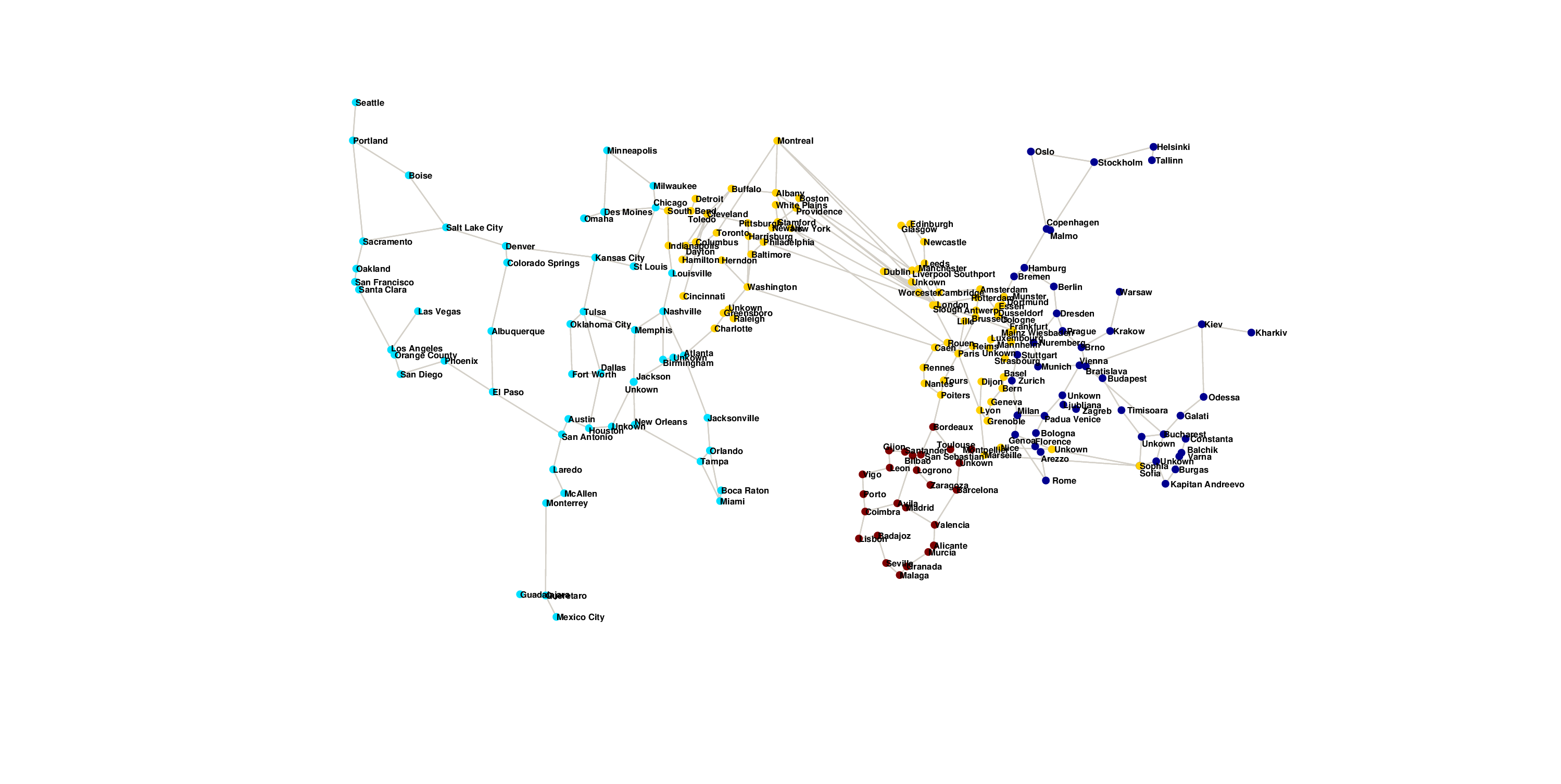}
	\vspace{-6cm}
	\caption{$4$ clusters found with the proposed automated model order selection (AMOS) algorithm
		for the Cogent Internet backbone map with city names. Clusters are separated by geographic locations except for the cluster containing cities in North Eastern America and West Europe due to many transoceanic connections.}
	\label{Fig_cogent_normalized}
\end{figure*}


\end{document}